\documentclass[twocolumn]{aastex701}

\newcommand{\poseidon}{\texttt{POSEIDON} }

\usepackage{romannum}
\usepackage{soul}
\usepackage{mhchem} 

\submitjournal{ApJL}

\shorttitle{WASP-69 b Has an Asymmetric Aerosol Distribution}
\shortauthors{L.-C. Wang et al.}

\usepackage{amsmath}
\usepackage{multirow}
\graphicspath{{./}{mainFigures/}}
\begin{document}

%\title{Dual Terminator Aerosols on WASP-69 b: Distinct Aerosol compositions, Water-rich Evening Terminator, and Escaping He Tail Revealed by JWST/NIRISS SOSS}
\title{Asymmetric Aerosol Distribution on the Terminators of the Warm Saturn WASP-69\,b Revealed by JWST NIRISS/SOSS}
%\title{JWST NIRISS Reveals Differing Aerosol Compositions and Helium Escape on the Terminators of WASP-69\,b}

% \title{WASP-69 b: Active Star or Cloudy Atmosphere?}

\correspondingauthor{Le-Chris Wang}
\email{lechris.wang@princeton.edu}

\author[0000-0002-6379-3816]{Le-Chris Wang}
\affiliation{Department of Astrophysical Sciences, Princeton University, 4 Ivy Lane, Princeton, NJ 08544, USA}
\affiliation{William H. Miller III Department of Physics \& Astronomy, Johns Hopkins University, 3400 N Charles St, Baltimore, MD 21218, USA}
\email{lechris.wang@princeton.edu}

\author[0000-0003-1622-1302]{Sagnick Mukherjee}
\altaffiliation{51 Pegasi\,b Postdoctoral Fellow}
\affiliation{School of Earth and Space Exploration, Arizona State University, Tempe, AZ 85287, USA}
\email{smukhe50@asu.edu}

\author[0000-0001-8510-7365]{Stephen P.\ Schmidt}
\altaffiliation{NSF Graduate Research Fellow}
\affiliation{William H. Miller III Department of Physics \& Astronomy, Johns Hopkins University, 3400 N Charles St, Baltimore, MD 21218, USA}
\email{sschmi42@jhu.edu}

\author[0000-0002-7352-7941]{Kevin B. Stevenson}
\affiliation{Johns Hopkins University Applied Physics Laboratory, 11100 Johns Hopkins Rd, Laurel, MD 20723, USA}
\email{kevin.stevenson@jhuapl.edu}

\author[0000-0002-1624-3360]{Mei Ting Mak}
\altaffiliation{Croucher Postdoctoral Fellow}
\affiliation{Atmospheric, Oceanic, and Planetary Physics Department, University of Oxford, Oxford OX1 3PU, UK}
\affiliation{Department of Physics and Astronomy, Faculty of Environment, Science and Economy, University of Exeter, Exeter EX4 4QL, UK}
\email{martha.mak@physics.ox.ac.uk}

\author[0000-0003-0473-6931]{Patrick McCreery}
\affiliation{William H. Miller III Department of Physics \& Astronomy, Johns Hopkins University, 3400 N Charles St, Baltimore, MD 21218, USA}
\email{pmccree2@jhu.edu}

\author[0009-0003-1097-5792]{Harry Baskett}
\affiliation{Department of Physics and Astronomy, Faculty of Environment, Science and Economy, University of Exeter, Exeter EX4 4QL, UK}
\email{hb595@exeter.ac.uk}

\author[0000-0001-5097-9251]{Carlos Gascón}
\affiliation{Space Telescope Science Institute, 3700 San Martin Drive, Baltimore, MD 21218, USA}
\email{cgascon@stsci.edu}

\author[0000-0001-6050-7645]{David K. Sing}
\affiliation{William H. Miller III Department of Physics \& Astronomy, Johns Hopkins University, 3400 N Charles St, Baltimore, MD 21218, USA}
\affiliation{Department of Earth and Planetary Science, Johns Hopkins University, 3400 N. Charles Street, Baltimore, MD 21218, USA}
\email{dsing@jhu.edu}

\author[0000-0002-9030-0132]{Katherine A. Bennett}
\affiliation{Department of Earth and Planetary Science, Johns Hopkins University, 3400 N. Charles Street, Baltimore, MD 21218, USA}
\email{kbenne50@jhu.edu}

\author[0000-0002-4997-0847]{Duncan A. Christie}
\affiliation{Max Planck Institute for Astronomy, K\"onigstuhl 17, D-69117 Heidelberg, Germany}
\email{christie@mpia.de}

\author[0000-0002-3263-2251]{Guangwei Fu}
\affiliation{William H. Miller III Department of Physics \& Astronomy, Johns Hopkins University, 3400 N Charles St, Baltimore, MD 21218, USA}
\email{guangweifu@gmail.com}

\author[0000-0003-3204-8183]{Mercedes L\'opez-Morales}
\affiliation{Space Telescope Science Institute, 3700 San Martin Drive, Baltimore, MD 21218, USA}
\email{mlopez-morales@stsci.edu}

\author[0000-0003-3667-8633]{Joshua D.\ Lothringer}
\affiliation{Space Telescope Science Institute, 3700 San Martin Drive, Baltimore, MD 21218, USA}
\email{jlothringer@stsci.edu}

\author[0000-0001-6707-4563]{Nathan J. Mayne}
\affiliation{Department of Physics and Astronomy, Faculty of Environment, Science and Economy, University of Exeter, Exeter EX4 4QL, UK}
\email{N.J.Mayne@exeter.ac.uk}

\author[0000-0002-1056-3144]{Lakeisha M. Ramos Rosado}
\affiliation{William H. Miller III Department of Physics and Astronomy, Johns Hopkins University, Baltimore, MD 21218, USA}
\email{lramosr1@jhu.edu}

\author[0000-0003-4408-0463]{Zafar Rustamkulov}
\affiliation{IPAC, California Institute of Technology, MC 100-22, 1200 E California Blvd, Pasadena, CA 91125, USA}
\email{zafar@caltech.edu}

\author[0000-0001-5761-6779]{Kevin C.\ Schlaufman}
\affiliation{William H. Miller III Department of Physics \& Astronomy, Johns Hopkins University, 3400 N Charles St, Baltimore, MD 21218, USA}
\email{kschlaufman@jhu.edu}

\author[0000-0001-7393-2368]{Kristin S. Sotzen}
\affiliation{Johns Hopkins University Applied Physics Laboratory, 11100 Johns Hopkins Rd, Laurel, MD 20723, USA}
\email{Kristin.Sotzen@jhuapl.edu}

\begin{abstract}
\noindent How aerosols form, are transported, and cycle between condensation and evaporation across exoplanet temperature regimes remains poorly understood. Recent models and observations suggest that warm giant planets near $800$--$1000$~K may span a transition between homogeneous and longitudinally heterogeneous aerosol distributions. We present a robust detection of aerosol asymmetry in a giant planet with $T_{\rm eq}\lesssim1000$~K, using the $0.86$--$2.82~\micron$ JWST NIRISS/SOSS transmission spectrum of WASP-69\,b.  The evening limb shows prominent 1.4~$\micron$ \ce{H2O} absorption ($\Delta\mathrm{BIC}_{\ce{H2O}}=+22.7$), whereas \ce{H2O} is not detected on the morning limb ($\Delta\mathrm{BIC}_{\ce{H2O}}=-8.7$). Atmospheric retrievals reveal significant aerosol opacity on both limbs, with high-altitude, optically thick clouds muting molecular features on the morning limb and lower cloud opacity allowing \ce{H2O} to emerge on the evening limb. The evening terminator is hotter by $304^{+62}_{-91}$~K, consistent with morning-limb condensates partially evaporating during transport toward the evening limb. This mechanism is independently verified with 3D general circulation models.  Stellar contamination or aerosols dominated by photochemical haze do not readily explain the asymmetry. From a limb-resolved analysis, we infer a stellar-to-superstellar atmospheric metallicity, with $\rm[M/H]=0.11^{+0.40}_{-0.46}$ from the equilibrium retrieval and [O/H]$=1.38^{+0.44}_{-0.79}$ from the free retrieval. We also detect an escaping metastable-helium tail extending to $3.08^{+0.50}_{-0.45}\,R_p$. WASP-69\,b anchors the cooler edge of the emerging population of planets with asymmetric aerosol distributions and suggests that substantial aerosol opacity may persist on both limbs across this transition.

%Atmospheric retrievals reveal significant aerosol opacity on both limbs but with different vertical distributions and particle sizes: high-altitude, optically thick clouds mute molecular features on the cooler morning limb, while lower cloud opacity on the evening limb allows \ce{H2O} to emerge.

%Atmospheric retrievals reveal significant aerosol opacity on both limbs, but the morning limb is shown to have high-altitude, optically thick clouds that mute molecular features, while the evening limb has lower cloud opacity that allows \ce{H2O} to emerge. 

%of approximately $15\times$ stellar and

% Future panchromatic analyses combining its NIRISS, NIRSpec, and MIRI transmission spectra with its dayside emission spectrum will constrain the aerosol composition and reveal how clouds, chemistry, and circulation interact throughout its three-dimensional atmosphere.
\end{abstract}


%--- Planet Formation (1241) --- Protoplanetary Disk (1300) --- Bayesian statistics (1900) --- 
\section{Introduction} \label{sec:intro}

Giant planets with equilibrium temperatures of approximately 800--1000 K may occupy a transition regime in both aerosol properties and their spatial distribution. Microphysical models predict that the dominant source of aerosol opacity shifts from hydrocarbon hazes at lower temperatures (equilibrium temperature $T_{\rm eq}\lesssim 950~\rm K$) to condensate clouds at higher temperatures ($T_{\rm eq}\gtrsim 950~\rm K$) \citep{Gao2020}. Irradiation may also determine how uniformly these aerosols are distributed around the planet. In cooler atmospheres, clouds and hazes can remain present at observable pressures across different longitudes, producing a broadly muted terminator. At higher temperatures, stronger longitudinal temperature differences can allow condensates to persist on the cooler nightside and morning limb but evaporate before reaching the evening limb, producing a cloudy morning and clearer evening \citep{Powell2019,PowellZhang2024}. The location of this transition should also depend on gravity, atmospheric circulation, and particle microphysics. 

% Planets around 1000 K may therefore probe transitions in both aerosol composition and longitudinal cloud coverage.

JWST has begun to test these predictions by resolving the morning and evening limbs separately. For WASP-39\,b, \citet{Espinoza2024} found a hotter evening limb with larger spectral features, consistent with a clearer evening terminator and a cooler, cloudier morning terminator. \citet{mukherjee2025} found an even stronger contrast for WASP-94A\,b, whose morning-limb spectrum is muted by high-altitude condensate clouds while its evening limb shows prominent \ce{H2O} absorption. At the population level, \citet{fu25} performed a uniform NIRISS/SOSS analysis of nine giant planets and identified significant muted-morning, clear-evening spectra for WASP-39\,b ($T_{\rm eq}\simeq1100~\rm K$), WASP-94A\,b ($T_{\rm eq}\simeq1500~\rm K$), and WASP-17\,b ($T_{\rm eq}\simeq1800~\rm K$). Cooler planets ($T_{\rm eq}\lesssim 1000~\rm K$) in these authors' sample did not show prominent asymmetries, although it was reported that WASP-107\,b ($T_{\rm eq}\simeq750~\rm K$) shows asymmetric aerosol opacity at longer wavelengths without correspondingly asymmetric suppression of \ce{H2O} features \citep{Murphy2024,Murphy25}. The strongest aerosol-opacity differences occur among hotter, lower-gravity planets. This is broadly consistent with theoretical expectations that larger longitudinal temperature contrasts enhance cloud evaporation toward the evening limb, while slower particle settling allows condensates to remain lofted on the morning limb \citep[see e.g.,][]{Mak2026}. The current sample does not yet establish where the transition from homogeneous to longitudinally heterogeneous aerosol coverage occurs, though. In particular, before this work, no robust detection of limb-dependent aerosol opacity had been reported for a giant planet in the $800$--$1000$~K regime.

WASP-69~b is well suited to probe this missing regime. It is an inflated, low-gravity Saturn-mass planet ($M_p\approx0.26\,M_{\rm J}$, $R_p\approx1.06\,R_{\rm J}$) with an equilibrium temperature of approximately $960$~K, orbiting an active K dwarf every 3.868 days \citep{Anderson2014}. Previous optical transmission spectra revealed a strong blueward slope that has been attributed to high-altitude aerosols \citep{Murgas2020, Estrela2021}, although stellar activity provides a competing explanation \citep{Allen2024,PR2024}. High-resolution observations have reported multiple molecular species and an extended, time-variable helium outflow \citep{Guilluy2022,Nortmann2018,Vissapragada2020,Tyler2024}. More recently, the JWST dayside emission spectrum required aerosol opacity and provided tentative evidence for a spatially inhomogeneous temperature structure \citep{schlawin2024}. Here, we present a $0.86$--$2.82\,\micron$ JWST NIRISS/SOSS transmission spectrum of WASP-69\,b. The data reveal distinct aerosol properties at the two limbs, providing a limb-resolved view of aerosol asymmetry in a giant planet at $T_{\rm eq}\lesssim 1000$~K. 

This paper is organized as follows. Section \ref{sec:obs} describes the JWST observation and data reduction methods. Section \ref{sec:lc} describes the procedures adopted to produce transmission spectra; we also analyze the helium escape that is prominent in the transmission spectrum. We update the host star characterization in Section \ref{sec:stellar}. These updated stellar parameters are then applied as inputs to the atmospheric retrievals described in Section \ref{sec:result}. The retrieval results are compared with those produced using a 3D General Circulation Model (GCM) in Section \ref{sec:gcm}. Section \ref{sec:discussion} discusses the implications of our analysis. Section \ref{sec:conclusion} summarizes our conclusions.

%To obtain accurate and precise stellar parameters for reliable transmission spectrum interpretation,

\section{Observations and Data Reductions}\label{sec:obs}

We observed a transit of WASP-69b with the James Webb Space Telescope (JWST), as part of the ``Grand Tour'' spectroscopic survey program (GO: 5924, PI: D. K. Sing), between May 4th, 2025 23:25:24 UT and May 5th, 2025 05:05:08 UT, for a total of 5.67 hours. The observations were taken with the NIRISS instrument in the Single-Object Slitless-Spectroscopy (SOSS) mode with 3 groups per integration, using the SUBSTRIP96 subarray and the NISRAPID readout setting. A total of 2064 integrations were obtained with a total exposure time of 5.08 hours. 

We used two independent %JWST
data reduction pipelines to analyze the observations of WASP-69 b - \texttt{FIREFLy} (\citealt{rustamkulov2022, rustamkulov2023}; \citealt{liu&wang2025, LCWang2026}) and \texttt{Eureka!} \citep{bell2022}. We summarize the key data reduction steps performed in each pipeline below.

\subsection{\texttt{FIREFLy} Reduction}
The \texttt{FIREFLy} reduction procedure for the NIRISS/SOSS data was similar to the reductions presented in \cite{liu&wang2025}, \cite{schmidt2025}, and \cite{LCWang2026}. To summarize, we started with uncalibrated data and followed most of the detector-level calibration steps using the \texttt{jwst} pipeline \citep{Bushouse2023}, with the addition that we subtracted the $1/f$ noise at the group level. We performed the group-level $1/f$ noise subtraction before the ramp-fitting step. This was done in three steps: we first subtracted the background from each group by scaling the NIRISS/SOSS background model\footnote{available at \href{https://jwst-docs.stsci.edu/}{https://jwst-docs.stsci.edu/}}. We then masked out the spectral trace and subtracted the median of each spectral trace-masked and background-subtracted column. Finally, we added the subtracted background back for accurate ramp fitting. After the detector-level corrections, we performed the usual bad-pixel cleaning, background subtraction, integration-level $1/f$ noise removal, and 1D spectral extraction. The background subtraction and $1/f$ removal at the integration level were performed in the same way as at the group level. We found that scaling the standard NIRISS/SOSS background model with a single flux factor did not fully remove the background. The subsequent $1/f$ correction removed the remaining background as well as the correlated noise. We also found that the partial second-order spectrum on the SUBSTRIP96 detector can dilute the measured transit depths if it is not masked properly. The white light curve was generated by summing the flux over 0.86--2.82 $\mu$m.

% \textbf{Background removal with a single scaling is not enough, as the transit depths seem to be diluted with a single background scaling! We still need the mask method to both remove 1/f noise and the remaining background.}

% \textbf{2.2 micron difference due to contamination from order 2. Should use massive mask at that region to preclude order 2, so that the background subtraction is accurate. Otherwise, transit depth would be diluted}

\subsection{\texttt{Eureka!} Reduction}
% \textcolor{red}{Kevin: Describe Eureka! Reduction Here}

We reduced the WASP-69b data using version 1.2 of the \texttt{Eureka!} pipeline \citep{bell2022} and CRDS context \texttt{pmap 1364}. We executed the standard \texttt{jwst} pipeline procedures for Stages 1 and 2. Unlike \texttt{FIREFLy}, however, we did not perform group-level background subtraction. During Stage 3, we used the curved spectral trace to straighten each order by applying integer-pixel shifts. For background subtraction, we masked the remaining spectral orders and estimated the background on a column-by-column basis using pixels located at least 22 pixels from the flattened trace. We then performed optimal spectral extraction using pixels within 17 pixels of the trace. 
We generated the white-light curve by summing the Order 1 flux over the wavelength range 0.86--2.82~$\mu m$.

% We then constructed spectroscopic light curves by binning the spectra into 49 wavelength channels, each 20~nm wide, across the same wavelength range.

\section{Light Curve Analysis}\label{sec:lc}

\begin{figure}[t]
    \centering
    \includegraphics[width=1.0\linewidth]{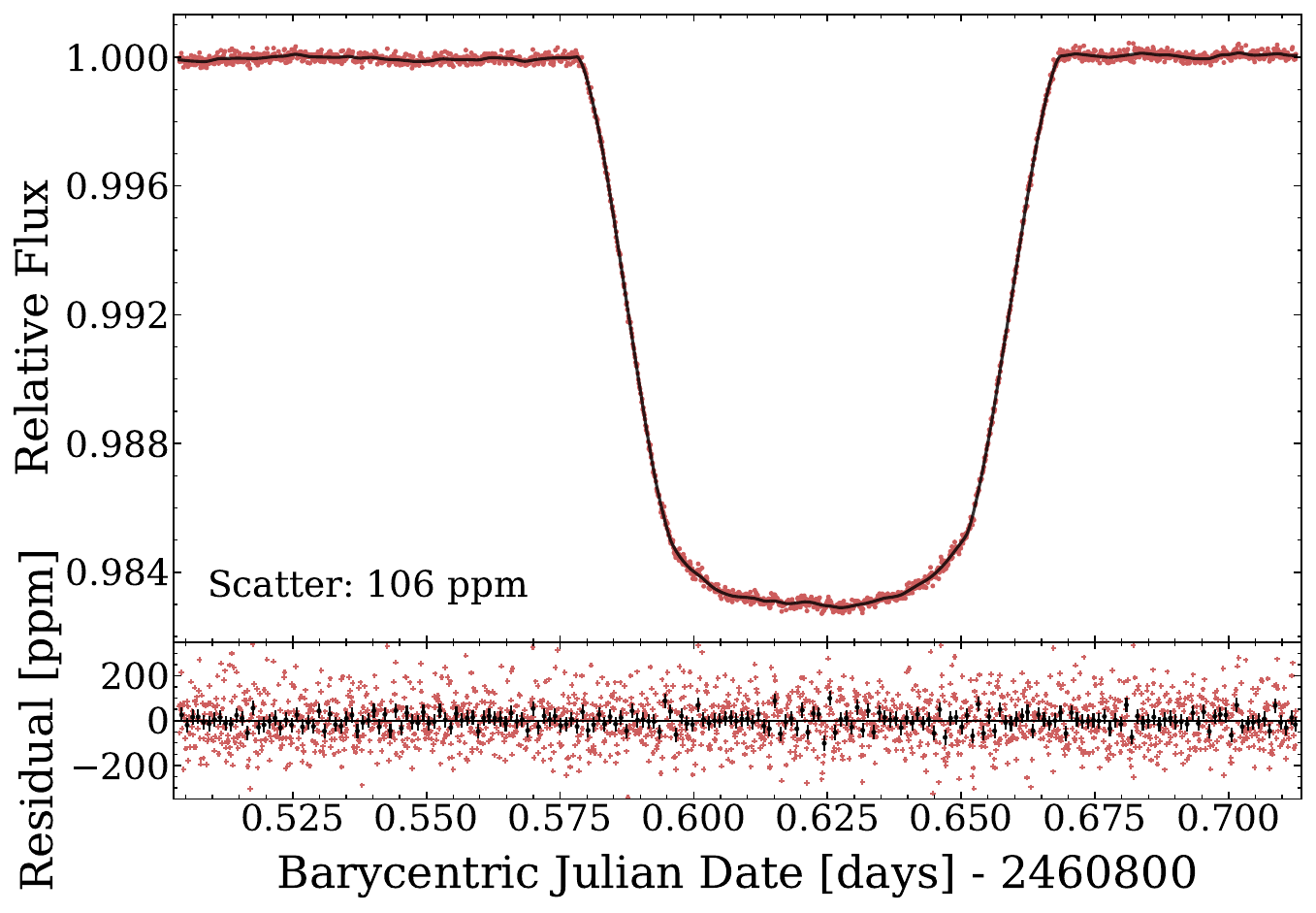}
    \caption{\texttt{FIREFLy} white light curve fit of the NIRISS transit of WASP-69~b, after applying common-mode correction.
    \textit{Top:} Relative flux measured at each integration overplotted with the best-fit light curve model. The reduced chi-square, $\chi^2_{\nu}$, is 0.31 (DoF = 2023).
    \textit{Bottom:} Residual between the measured flux and fitted model. The red points are residuals at individual integrations and the black are binned at 10 integrations. The scatter of the residual is 106 ppm.
    }
    \label{fig:wlc}
\end{figure}

% We fit the white light curve using \texttt{batman} \citep{kreidberg2015} and the Markov Chain Monte Carlo (MCMC) algorithm implemented by \texttt{emcee}, with burn-in of 1000 and a length of 2000. We trim out the first thirty integrations that exhibit non-linearity.  We use the linear term in time and spectral trace position shifts in the y-direction as the initial systematics vector, as preferred by the Bayesian Information Criterion (BIC). 

\begin{deluxetable}{lcc}
\centering
\tablewidth{0pt}
\tablecaption{Best-fit Orbital Parameters for WASP-69 b from NIRISS/SOSS data}
\tablehead{
Model Setting & \colhead{\texttt{FIREFLy}} & \colhead{\texttt{Eureka!}}
}
\startdata
$T_0$ (BJD) & $2460800.623360 \pm 0.00001$ & $2460800.62339 \pm 0.00005$ \\
$a/R_\ast$ & $12.274 \pm 0.02 $ & $12.30 \pm 0.05$ \\
$b$ & $0.670 \pm 0.002$ & $0.678\pm0.007$ \\
$i$ (degrees) & $86.87\pm0.01$ & $86.84 \pm 0.03$ \\
$\left(R_p/R_\ast\right)^2$ & $0.01595 \pm 0.00006$  & $0.01650 \pm 0.00006$ \\
$u_1$ & $0.29\pm0.01$ & 0.2314 \\
$u_2$ & $0.17\pm0.02$ & 0.2479 \\
\enddata
\tablecomments{$T_0$ is the time of transit in BJD, $a/R_\ast$ is the unitless scaled semi-major axis, $b$ is the unitless impact parameter, $i$ is the inclination in degrees, $\left(R_p/R_\ast\right)^2$ is the squared planet-to-star radius ratio, $u_1$ is the first quadratic limb darkening coefficient, and $u_2$ is the second quadratic limb darkening coefficient. The limb-darkening coefficients from \texttt{Eureka!} were fixed from the \texttt{mps2} grid \citep{mps2}.}
\label{tab:orbparams}
\end{deluxetable}

We fitted the white light curve from \texttt{FIREFLy} with \texttt{batman} \citep{mandel2002,kreidberg2015} and the Markov Chain Monte Carlo (MCMC) sampler \texttt{emcee} \citep{emcee}, using a burn-in of 1000 steps and a chain length of 2000\footnote{We found the number of steps used is sufficient for convergence for both white light curve and spectroscopic light curve fits.}. We discarded the first 30 integrations, which are affected by detector non-linearity.  We fitted the systematics vector together with the \texttt{batman} transit model. We fitted for the quadratic limb darkening coefficients $u_1$ and $u_1$, the planet's impact parameter $b$, scaled semi-major axis $a/R_*$, transit depth $R_p/R_\ast$, and mid-transit time $T_0$. The orbital period is fixed at 3.868 days \citep{Anderson2014}. The final white light curve fit is shown in Figure~\ref{fig:wlc}. The final best-fit parameters are shown in Table~\ref{tab:orbparams}. The fit was performed in three iterations, for reasons we explain below.

In the first iteration, we allowed the system parameters and the baseline systematics vector to vary freely. The baseline systematics model included a linear trend in time and the spectral-trace shift in the $y$ direction, as preferred by the Bayesian Information Criterion (BIC). This fit yielded unphysical limb darkening coefficients that suggest limb brightening, with $(u_1+u_2)=-0.25\pm0.04$. As a check, we compared these values with the limb darkening fitted from the NIRSpec G395H data (P. E. Cubillos et al., Priv. Comm.) and found that the NIRSpec result is consistent with the Stagger 3D stellar-atmosphere model \citep{Magic2015}. We therefore interpret the apparent limb brightening in the NIRISS/SOSS white light curve as evidence for residual unmodeled correlated noise, rather than a physical limb-darkening profile. We found the correlated noise to be a common-mode structure across all spectroscopic bins, whose amplitude is not correlated with atmospheric spectral features. Intriguingly, such a structure was also observed in other NIRISS/SOSS SUBSTRIP96 observations (e.g., \citealt{schmidt2026}, E. M. May et al 2026, in preparation). In the second iteration, we fixed the limb darkening to the Stagger 3D values and refit the light curve. We then smoothed the residuals with a Gaussian filter and adopted the smoothed trend as a common-mode systematic. In the third and final iteration, we added this common-mode term to the systematics vector and repeated the fit with free limb darkening, orbital parameters, and systematics vector.

%and $u_1-u_2=0.87\pm0.07$

We then proceeded to fit the spectroscopic light curves. We fixed the limb darkening coefficients to the Stagger 3D stellar-atmosphere model, as we found the fitted limb darkening coefficients to be well-described by the model values. We fixed orbital parameters to the fitted white light curve values. We fitted for the transit depth, baseline flux, and systematics vector in each spectroscopic bin using the Levenberg-Marquardt algorithm from \texttt{lmfit}. 

\begin{figure*}[t]
    \centering
    \includegraphics[width=1.0\linewidth]{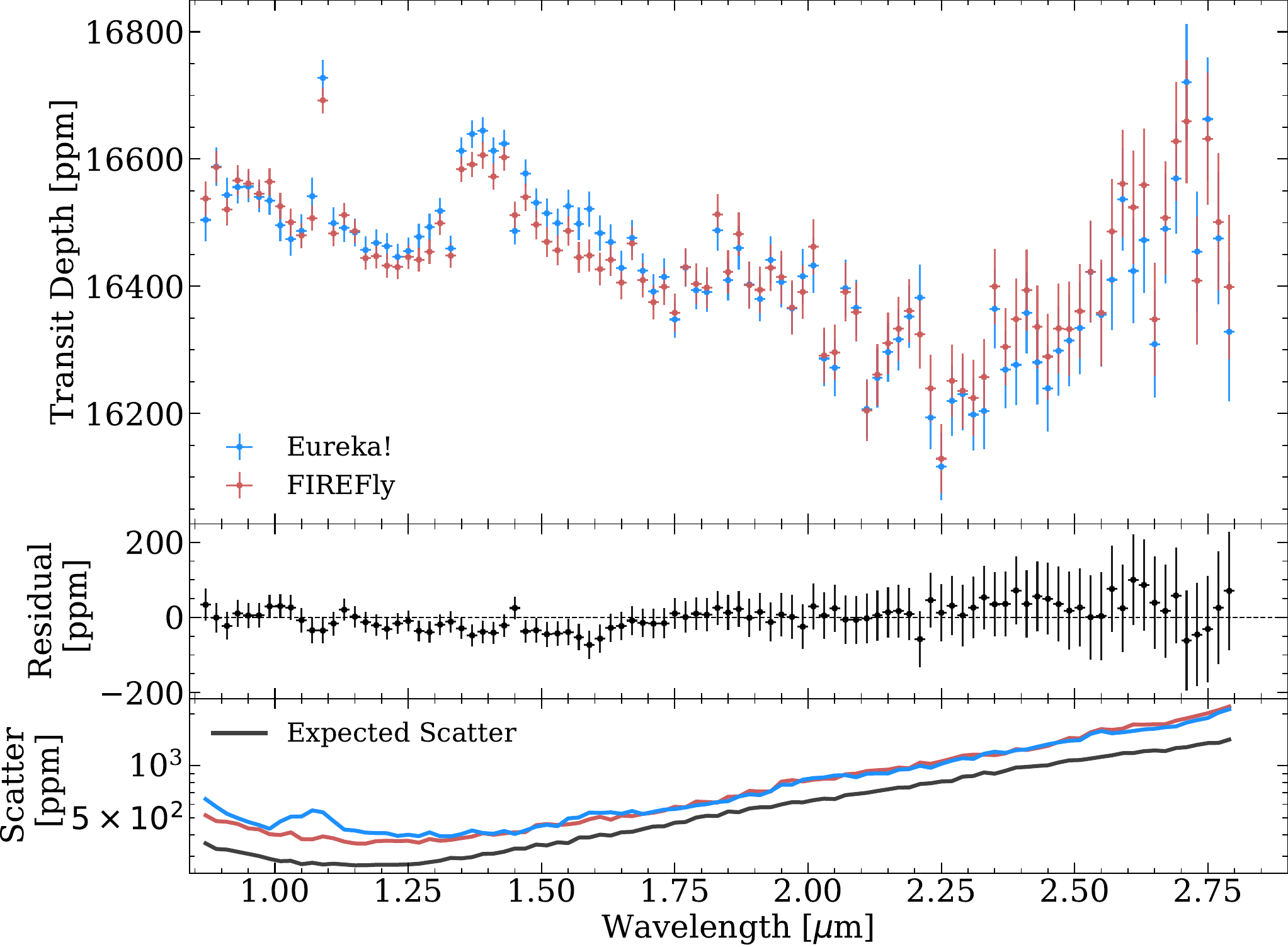}
    \caption{Comparison of limb-averaged NIRISS/SOSS transmission spectra of WASP-69\,b binned to 0.02 $\micron$ from two independent reductions. \textit{Top:} Transmission spectrum from \texttt{Eureka!} (blue) and \texttt{FIREFLy} (red). \textit{Middle:} Difference between the \texttt{FIREFLy} reduction and \texttt{Eureka!} reduction. \textit{Bottom:} Measured uncertainty per spectral channel for each of the two reductions, compared with the expected photon noise (black).
    }
    \label{fig:spectrum}
\end{figure*}

The final transmission spectrum binned by 0.02 $\mu m$ is shown in Figure \ref{fig:spectrum}. We also performed white light curve and spectroscopic light curve fitting with \texttt{Eureka!} following similar procedures. We compare the resulting transmission spectrum from \texttt{FIREFLy} and \texttt{Eureka!} in Figure \ref{fig:spectrum}.

\subsection{Limb-resolved Transmission Spectrum}\label{sec:limb_spectra}
\begin{figure*}[t]
    \centering
    \includegraphics[width=1.0\linewidth]{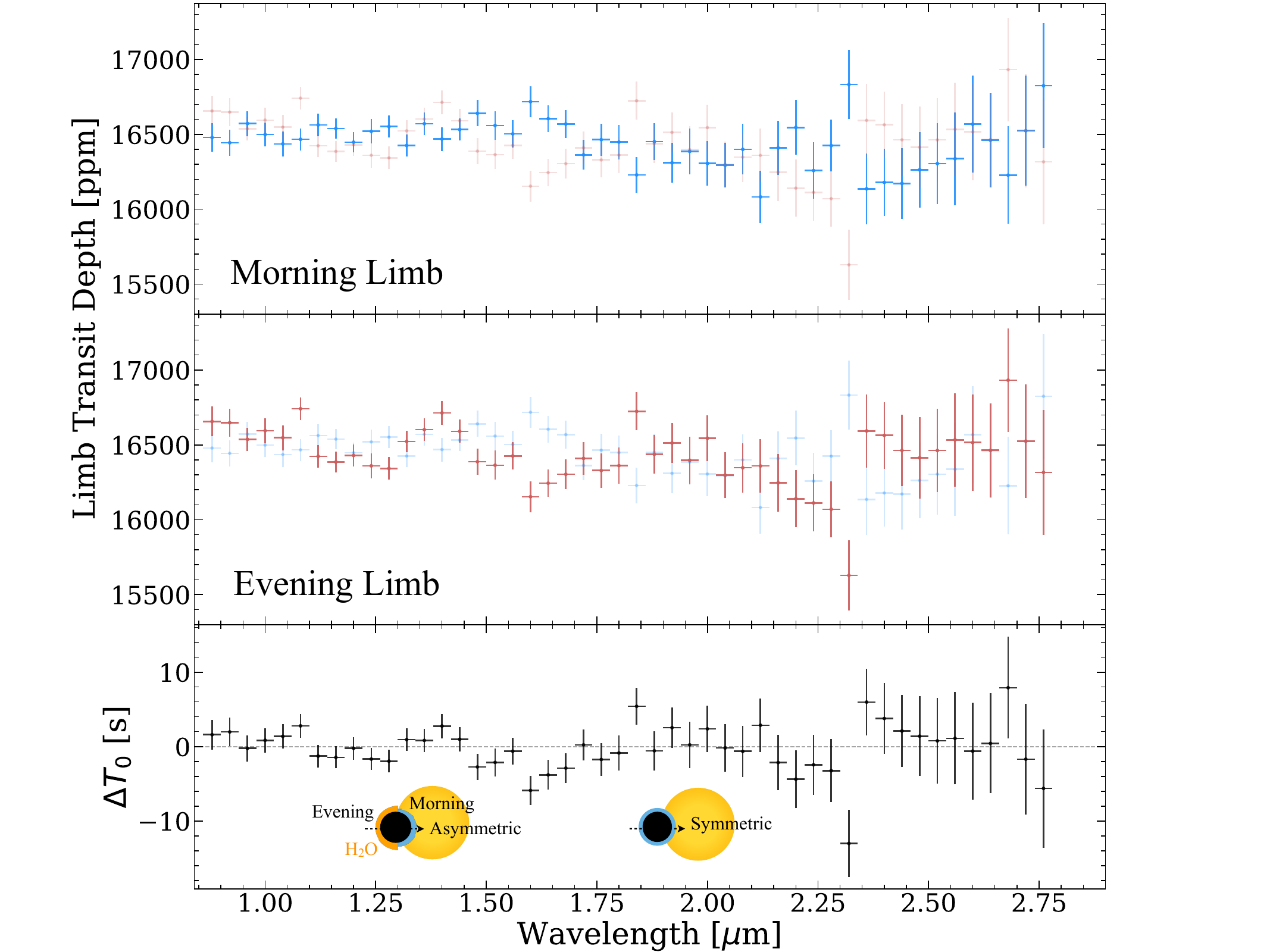}
    \caption{Evidence for limb asymmetry in WASP-69\,b. The upper and middle panels show the morning- and evening-limb transmission spectra, respectively, binned to 0.04~$\micron$; the opposite limb is overplotted in a lighter shade for comparison. The evening limb exhibits a prominent 1.4~$\micron$ \ce{H2O} feature that is muted on the morning limb. The bottom panel shows the chromatic transit-time offset, $\Delta T_0$, defined as the fitted mid-transit time in each spectroscopic bin minus the fitted mid-transit time from the white-light curve. The structure in $\Delta T_0$, particularly near the 1.4~$\micron$ \ce{H2O} band, provides an independent indication of limb asymmetry. The schematics in the bottom illustrate the origin of this timing signal: wavelength-dependent differences in the effective radii of the morning and evening limbs shift the apparent transit midpoint, whereas a symmetric planet produces no such offset.
    }
    \label{fig:limb_limb}
\end{figure*}

We extracted limb-resolved transmission spectra by repeating the above spectroscopic light curve fitting procedures with \texttt{catwoman} \citep{catwoman2020, catwoman2021}. We show the fitted morning limb and evening limb spectra from \texttt{FIREFLy} in the upper two panels of Figure \ref{fig:limb_limb}. The limb-resolved spectra from \texttt{Eureka!} are consistent with the \texttt{FIREFLy} spectra. The presence of limb asymmetry is evidenced by the presence of 1.4 $\mu m$ water feature in the evening limb and the lack of such a feature in the morning limb. 

% Approximating the leading (morning) limb and trailing (evening) limb as two semi-circles, \texttt{catwoman} allows us to examine whether there are limb asymmetries in WASP-69~b.

As an independent test, we also examined the chromatic variation of the fitted mid-transit time, a commonly used diagnostic of limb asymmetry \citep{rustamkulov2023,Espinoza2024,murphy24T0,LCWang2026, Radica2026}. We fitted each spectroscopic light curve with a spherical-planet \texttt{batman} model, allowing the mid-transit time, $T_0$, to vary, and compared it with the value obtained from the white-light curve. To zeroth order, a difference in the sizes of the morning and evening limbs shifts the apparent transit earlier or later, so limb asymmetry can appear as a wavelength-dependent offset in $T_0$. The resulting $\Delta T_0$ spectrum (bottom panel of Figure \ref{fig:limb_limb}) shows structure similar to the evening-limb spectrum, particularly near the 1.4~$\micron$ \ce{H2O} band. This provides an independent indication of limb asymmetry.

The presence of limb asymmetry motivates us to focus on analyzing limb-resolved spectra, rather than the limb-averaged spectrum, in Section \ref{sec:result}, as ignoring limb asymmetries will lead to biased inferences of atmospheric compositions and metallicities \citep{mukherjee2025}.

\subsection{Escaping Helium}\label{sec:He}

% it is a bloated, low-gravity Saturn-mass planet on a short 3.868-day orbit around an active K dwarf from which it receives substantial high-energy irradiation.
WASP-69 b is an especially compelling target to search for photoevaporative atmospheric loss because of its low density and close proximity to its active K-dwarf host star. Indeed, WASP-69 b was one of the first exoplanets to have been detected with escaping helium \citep{Nortmann2018}.  The helium excess was subsequently confirmed by \citet{Vissapragada2020}, and later homogeneous higher-resolution analyses with SPIRou and GIANO-B reinforced the helium detection and showed that the signal can vary from transit to transit \citep{Allart2023,Masson2024,Guilluy2024,Levine2024}. Keck/NIRSPEC observations further showed that the post-transit absorption can persist for at least 1.28 hr, implying a tail extending to at least $7.5~R_p$ \citep{Tyler2024}. Recently, \citet{Allart2025} resolved helium absorption to $\sim50$ min after egress and modeled the data with a thermosphere-plus-exosphere outflow, finding a mass-loss rate of order $2.25\times10^{11}$ g/s. Together, these studies indicate that WASP-69 b has a long, likely time-variable helium tail shaped by ongoing atmospheric loss and interaction with the stellar environment. 

In line with previous observations, WASP-69\,b's NIRISS/SOSS transmission spectrum reveals a prominent metastable helium triplet absorption feature at 10830\,\AA\ (Figure \ref{fig:spectrum}). Here, we characterize the helium escape with the NIRISS/SOSS data.

\subsubsection{Helium Outflow Morphology}

% \textcolor{red}{Carlos: Describe escaping helium characterization with Harmonica here. A figure from Harmonica analysis.}

% \begin{figure*}[t]
%     \centering
%     \includegraphics[width=0.49\linewidth]{mainFigures/W69_helium_lc.pdf} 
%     \includegraphics[width=0.49\linewidth]{mainFigures/W69_helium_envelope.pdf}
%     \caption{Main results of the WASP-69\,b helium envelope fit. The left panel shows the pixel level light curve containing the helium triplet at 1.083 $\micron$ in red, together with the model white light curve in black and the fitted model helium light curve in blue. The lower panel shows the residuals of the helium light curve fit.  The right panel shows the median retrieved envelope in blue, with the shaded region showing the 1$\sigma$ uncertainty region. The gray line represents the star, and the black circle illustrates the planet.}
%     \label{fig:outflow_shape}
% \end{figure*}

\begin{figure*}[t]
    \centering
    \includegraphics[width=0.49\linewidth]{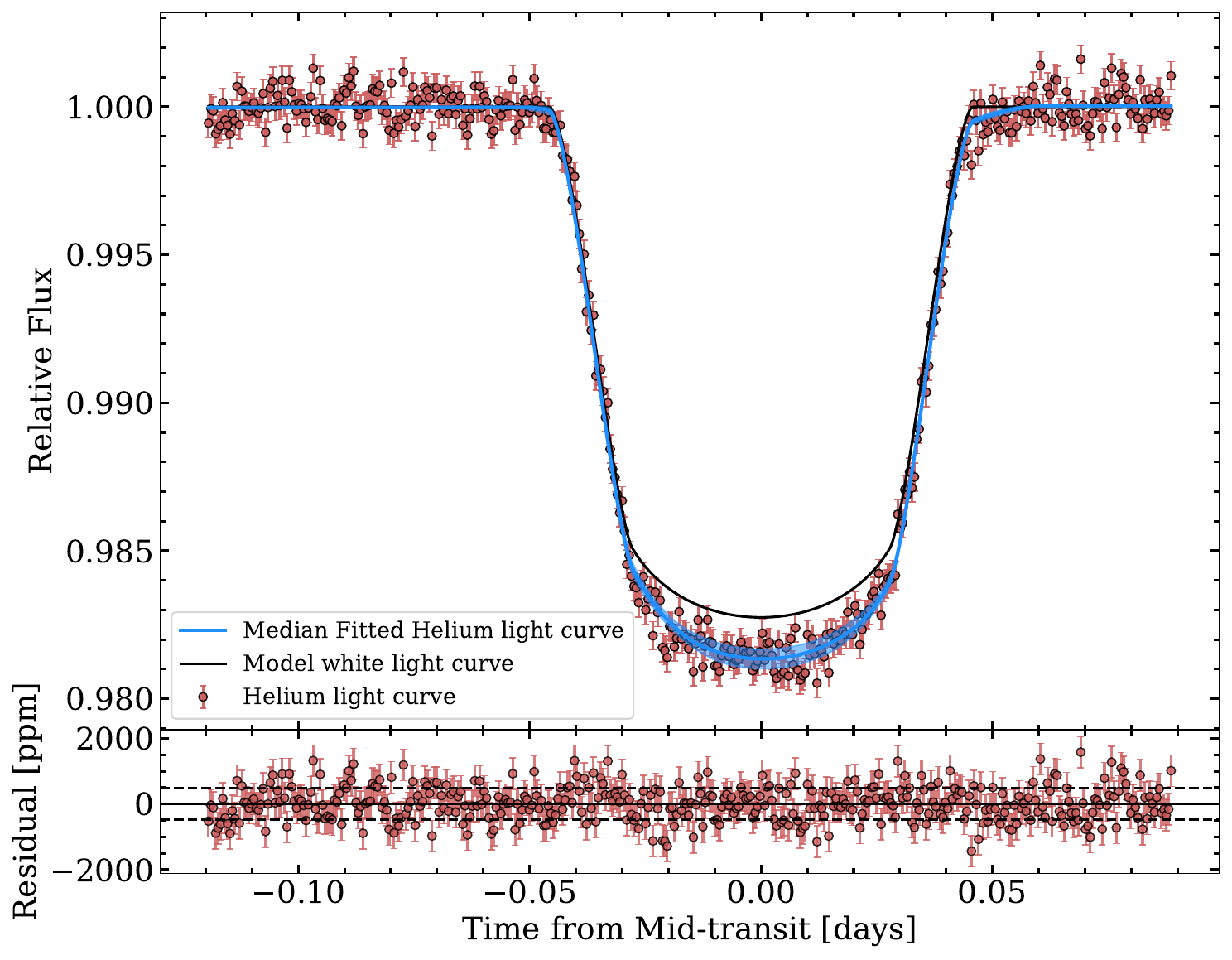} 
    \includegraphics[width=0.49\linewidth]{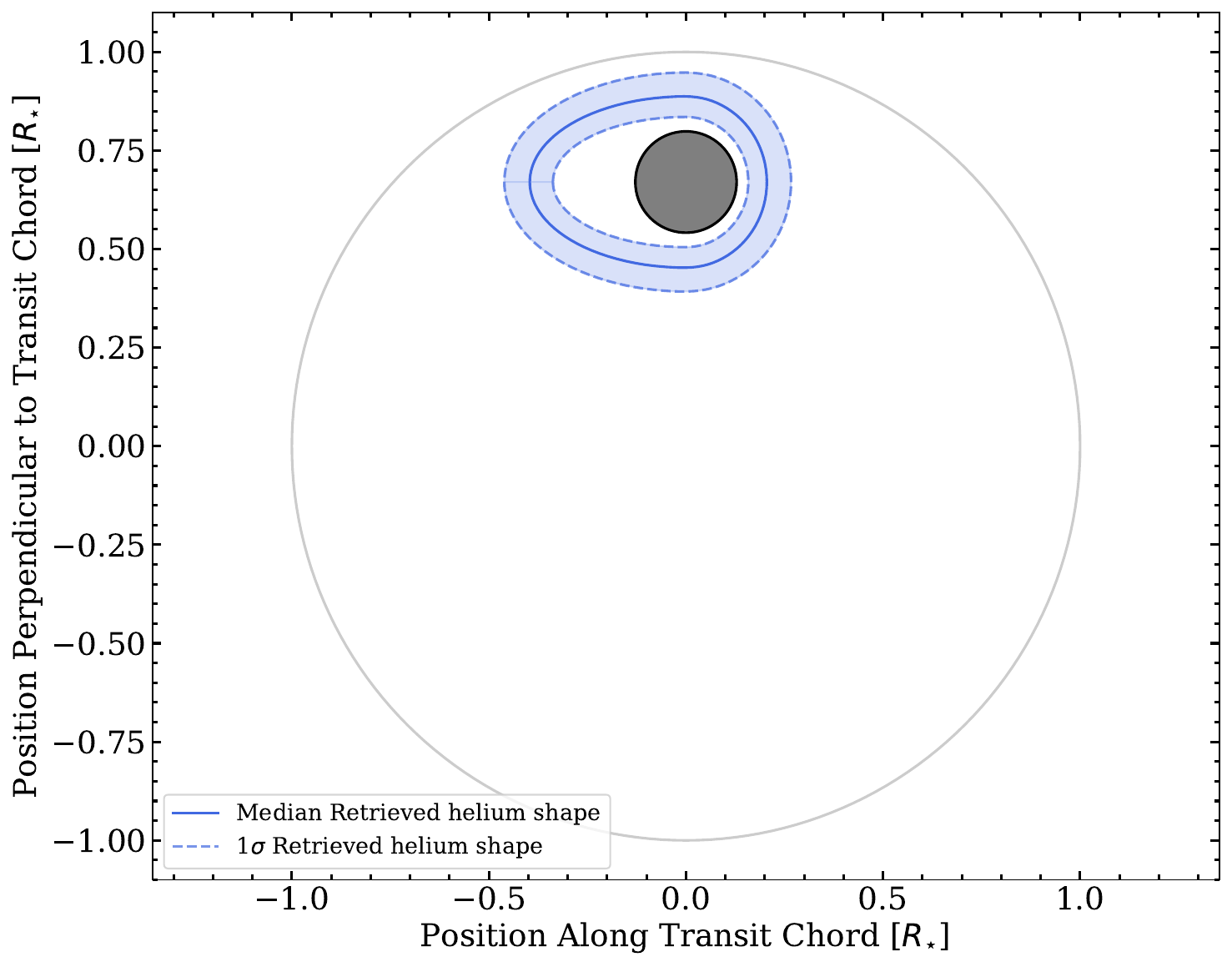}
    \includegraphics[width=0.72\linewidth]{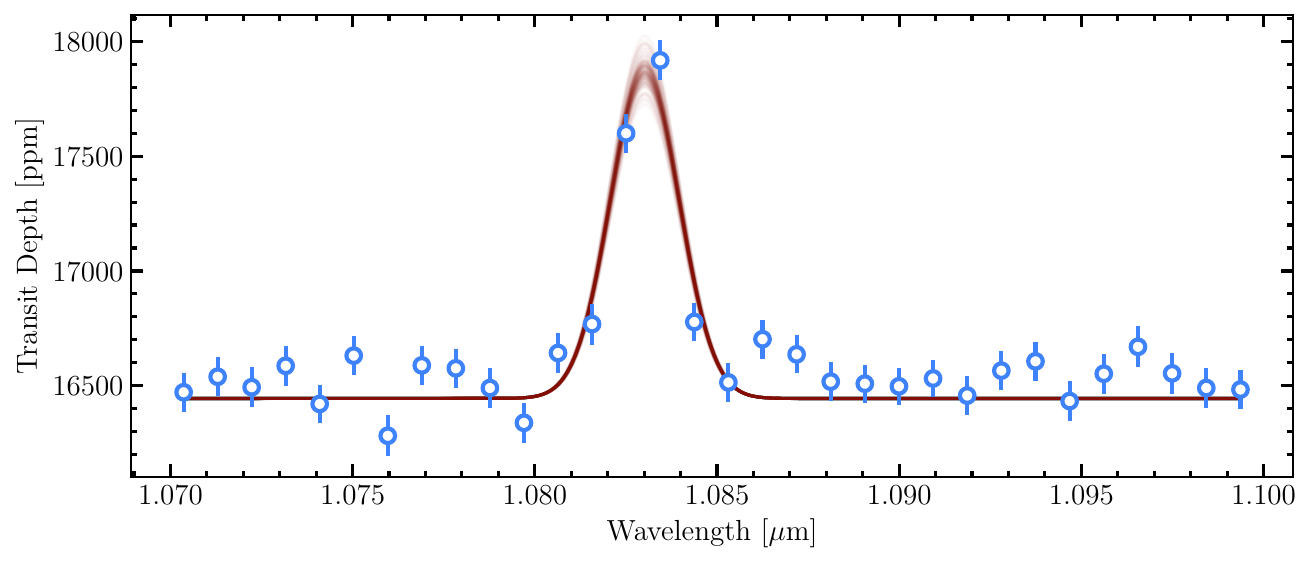}
    \caption{Helium outflow constraints for WASP-69\,b. \textit{Upper left}: pixel-level light curve containing the helium triplet at 1.083~$\micron$ in red, together with the white-light transit model in black and the fitted helium light curve from \texttt{Harmonica} in blue; the lower subpanel shows the fit residuals. \textit{Upper right}: Helium outflow morphology constrained with \texttt{Harmonica}. The median retrieved helium envelope is in blue, with the shaded region denoting the 1$\sigma$ uncertainty; the gray line marks the stellar limb and the black circle represents the planet. \textit{Bottom}: \texttt{p-winds} model fits to the metastable helium triplet absorption feature in the WASP-69\,b NIRISS/SOSS transmission spectrum.  The blue points show the pixel-level NIRISS/SOSS transmission spectrum over the wavelength range 1.070--1.100\,$\mu$m and the red curves show 100 random model draws from the posterior distribution.}
    \label{fig:outflow_shape}
\end{figure*}

% \begin{figure*}
%     \centering
%     \includegraphics[width=\linewidth]{mainFigures/helium-spectrum_fit_90.pdf}
%     \caption{\texttt{p-winds} model fits to the metastable helium triplet absorption feature in the WASP-69\,b NIRISS/SOSS transmission spectrum. The blue points show the NIRISS/SOSS transmission spectrum over the wavelength range 1.070--1.100\,$\mu$m. Red curves show 100 random model draws from the posterior distribution. The agreement between the observed feature and the posterior model realizations demonstrates that NIRISS/SOSS robustly constrains the absorption depth of the helium triplet, despite not resolving the detailed line profile.}
%     \label{fig:helium-fit}
% \end{figure*}

Figure \ref{fig:outflow_shape} shows in red the \texttt{FIREFLy} pixel-level light curve containing the helium triplet at 1.083 $\micron$, compared to the white light curve model in black computed with the best fit values from the \texttt{FIREFLy} reduction (Table \ref{tab:orbparams}). By visual inspection, we observed that the helium light curve exhibited not only a larger transit depth, but also an apparent post-transit absorption spanning $\sim0.5$ hours. In order to set constraints on the outflow morphology, we fitted the pixel-level helium light curve using the model and methodology presented in \cite{gascon25}. In short, we modeled the helium envelope as two semi-ellipses with semi-major axes $l_1$ and $l_2$, a shared semi-minor axis $h$, and a uniform envelope opacity $\alpha$ defined as the fraction of stellar light blocked by the escaping gas where its envelope overlaps the stellar disk. The transit light curves of both the planet and the helium envelope were calculated with \texttt{Harmonica} \citep{grant22}, which represents arbitrary shapes as sums of circular harmonics. While this model does not take into account any of the physical processes involved in the onset of helium escape, it allows us to place constraints on the outflow's morphology solely based on the shape of the helium transit light curve.

We fixed the system parameters to the best-fit values obtained with the \texttt{FIREFLy} reduction (Table \ref{tab:orbparams}), and fitted for the four envelope parameters ($l_1$, $l_2$, $h$, $\alpha$), as well as a linear trend to account for systematics. We performed the fits using the MCMC sampler \texttt{emcee} \citep{foreman13}, running 30 walkers, each with 500 steps as burn-in and 2500 steps. We considered uniform priors for all variables.  The fitted morphology is shown in the upper right panel of Figure \ref{fig:outflow_shape}, with the blue solid lines representing the median fitted model, and the shaded blue representing the 1$\sigma$ uncertainty region. Our model is able to %successfully 
fit the helium light curve, including the apparent post-transit absorption, as shown in the upper left of Figure \ref{fig:outflow_shape}. We obtain a prominent trailing tail of $l_1 = 3.08^{+0.50}_{-0.46} R_p$, and no clear evidence for a leading tail in WASP-69\,b ($l_1 = 1.60^{+0.47}_{-0.40} R_p$). The envelope opacity is also constrained to $1/\alpha = 35.6^{+13.6}_{-12.1}$. However, we note that $\alpha$ and $h$ are slightly degenerate, given that both parameters imprint similar features in the transit light curve \citep{gascon25}. While the detection of post-transit absorption agrees with previous observations \citep{Tyler2024, Allart2025}, our derived tail sizes are relatively smaller. This could be either attributed to NIRISS/SOSS' low resolution ($R \sim 700$) compared to ground-based observations or the time-variable nature of WASP-69\,b's helium tail. Still, JWST's unprecedented spectrophotometric precision and stability are essential to studying the morphology of atmospheric escapes like the one presented in this section.
%which is consistent with previous studies. We also constraint. We note that, given the low resolution of NIRISS/SOSS compared to high resolution spectrographs, the

% , which significantly dilutes the helium signal compared to high resolution ground-based observatories

\subsubsection{Mass Loss}

The resolving power of NIRISS/SOSS ($R\sim700$) is insufficient to resolve the metastable helium triplet, and we therefore cannot simultaneously constrain both the mass-loss rate and the outflow temperature \citep{dossantos22}. Nevertheless, the unresolved 10833\,\AA\ absorption provides a useful consistency check against ground-based high-resolution measurements that resolve the helium line profiles. For a direct comparison, we adopted the same one-dimensional Parker-wind framework used by \citet{mccreery2025} to reanalyze the CARMENES spectrum of \citet{Nortmann2018}. Their \texttt{p-winds} analysis inferred mass-loss rates of $2.2^{+0.2}_{-0.2}\times10^{11}$ and $5.9^{+0.6}_{-0.5}\times10^{10}$\,gs$^{-1}$ for H/(H+He) = 0.99 and 0.90, respectively, with corresponding outflow temperatures of $4700^{+110}_{-110}$ and $6200^{+140}_{-140}$\,K.

We fitted the NIRISS/SOSS spectrum with \texttt{p-winds}, which predicts the metastable-helium absorption for a specified mass-loss rate, outflow temperature, H/He composition, stellar high-energy spectrum, and planetary parameters \citep{dossantos22}. We sampled broad priors in mass-loss rate and temperature with \texttt{dynesty} \citep{dynesty} and interpreted the resulting joint posterior rather than the marginalized constraints on either parameter alone. For the stellar XUV spectrum, we adopted the MUSCLES spectrum of HD~40307, scaled to the orbital distance of WASP-69\,b \citep{musclesI,musclesII,musclesIII,musclesIV,musclesV}. Motivated by hydrodynamic models that favored sub-solar helium abundances in escaping atmospheres \citep{Salz2016a,Rumenskikh2022,Yan2024}, we considered both a solar-like composition, H/(H+He) = 0.90, and a helium-poor case, H/(H+He) = 0.99.

For both assumed H/He compositions, the joint posteriors from NIRISS/SOSS overlap the corresponding CARMENES-based constraints from \citet{mccreery2025}, indicating consistency between the low- and high-resolution measurements. Representative \texttt{p-winds} models drawn from these posteriors also reproduce the amplitude and spectral shape of the unresolved helium feature, as shown in the bottom panel of Figure~\ref{fig:outflow_shape}. From the posterior model draws, we measure a maximum excess absorption of $7.91^{+0.29}_{-0.32}$\%. Independently, integrating the excess transit depth relative to the fitted continuum over 1.070--1.081 and 1.086--1.100~$\mu$m yields an equivalent width of $29\pm2$\,m\AA.

Despite the limited spectral resolution of NIRISS/SOSS, these results provide an independent consistency check that the unresolved helium absorption is compatible with the mass-loss rates and outflow temperatures inferred from ground-based high-resolution spectroscopy.

\section{Stellar Characterization}\label{sec:stellar}
% Precise and accurate stellar parameters are required for accurate planet atmospheric inference. Since the publication of Gaia Data Releases (DR) 2 and 3, precision parallax measurements allow for significantly improved precisions of stellar parameters inferred with isochrone fitting, which derives stellar parameters by fitting multi-wavelength, broadband photometry spanning the ultraviolet through the infrared to theoretical stellar evolution models. In tandem with Gaia DR3 parallax measurements, archival stellar photometric data and three-dimensional (3D) local extinction and reddening maps, we present constraints on the stellar fundamental and photospheric parameters of WASP-69 using the \texttt{isochrones} Python package \citep{mor15}. \texttt{isochrones} performs simultaneous Bayesian fits of the Modules for Experiments in Stellar Evolution \citep[MESA;][]{pax11,pax13,pax18,pax19,jer23} Isochrones \& Stellar Tracks \citep[MIST;][]{dot16,cho16} v1.2 isochrone grid to a curated collection of data for the star using \texttt{MultiNest} \citep{fer08,fer09,fer19}. This methodology has been extensively used in the literature \citep[e.g.,][McCreery et al. in preparation]{Reggiani22, Reggiani24, Hamer22} and has become the state-of-the-art when inferring stellar parameters. 

Precise and accurate stellar parameters are essential for reliable inference of planetary atmospheres. Gaia parallaxes, particularly those from Data Release 3 (DR3), have substantially improved stellar characterization by providing strong distance constraints for isochrone fitting. We determined the fundamental and photospheric properties of WASP-69 using the \texttt{isochrones} Python package \citep{mor15}, combining its Gaia DR3 parallax with archival ultraviolet-to-infrared broadband photometry and constraints from three-dimensional extinction and reddening maps. The \texttt{isochrones} package simultaneously fits these data to the MESA Isochrones and Stellar Tracks \citep[MIST;][]{pax11, pax13, dot16,cho16, pax18,pax19,jer23} v1.2 grid, and samples the posterior distribution using \texttt{MultiNest} \citep{fer08,fer09,fer19}. Similar methods have been widely applied to infer stellar parameters in the literature \citep[e.g.,][Z. Rustamkulov et al. in review, McCreery et al., in preparation]{Reggiani22,Reggiani24,Hamer22}.

We fitted the MIST grid to the following data:
\begin{enumerate}
\item Galaxy Evolution Explorer \citep[GALEX;][]{mart05}
near-ultraviolet (NUV) photometry from the GUVcat\_AIS \citep{bia17} including in quadrature a zero-point uncertainty
of 0.02 mag;
\item SkyMapper Southern Survey DR4 $uvr$ photometry including in quadrature
their zero-point uncertainties (0.03,0.02,0.01) mag \citep{skymapper};
\item Gaia EDR3 $G$ photometry including in quadrature its zero-point uncertainty
\citep{Gaia_mission2016, GaiaDR32023, GaiaEDR3Validation, EDR3photometry, row21, tor21};
\item Two-micron All-sky Survey (2MASS) $JHK_{s}$ photometry including their zero-point uncertainties \citep{skr06};
\item Wide-field Infrared Survey Explorer (WISE) All-sky \citep{wri10} $W1$ and $W2$ photometry in quadrature their zero-point uncertainties (0.032,0.037) mag\footnote{\url{https://wise2.ipac.caltech.edu/docs/release/allsky/expsup/sec4\_4h.html\#PhotometricZP}}.
\item We also utilized a zero point-corrected Gaia EDR3 parallax \citep{GaiaDR32023, lin21a, lin21b, row21, tor21} and
\item an estimated extinction value based on the 3D local extinction maps using \texttt{G-Tomo} Python module \citep{lal22, ver22}.
\end{enumerate}
Furthermore, we verified the quality of the input photometry using the data quality flags described in \citet{Hamer22}.  

An extinction prior informed by 3D local extinction and reddening maps as well as a distance prior informed by the \citet{bai21} geometric distance were enforced. We show the results of our model fit in Table \ref{tab:stellar_params}.

\begin{deluxetable}{lcc}
\label{tab:stellar_params}
\centering
\tablewidth{0pt}
\tablecaption{Derived stellar fundamental and photospheric parameters of WASP-69}
\tablehead{
Parameter & Value 
}
\startdata
Mass ($M_\star$) & $0.85^{+0.01}_{-0.01} M_\odot$ \\
Radius ($R_\star$) & $0.81^{+0.01}_{-0.01} R_\odot$ \\
Log Age ($\log_{10}\tau_\star$) & $9.86^{+0.10}_{-0.17}$ \\
Effective Temperature ($T_\text{eff}$) & $4920^{+60}_{-40} \text{ K}$ \\
Surface Gravity ($\log g$) & $4.54^{+0.02}_{-0.02} \text{ dex}$ \\
Metallicity ([Fe/H]) & $0.28^{+0.03}_{-0.02} \text{ dex}$ \\
Distance ($d$) & $50.2^{+0.1}_{-0.1} \text{ pc}$ \\
Extinction ($A_V$) & $0.08^{+0.01}_{-0.01} \text{ mag}$ \\ 
\enddata
\tablecomments{Quoted values are median and the central 68\% credible interval from the model fit, and thus do not include uncertainties from model systematics.}
\end{deluxetable}

\section{Retrieval Analysis}\label{sec:result}
We next performed atmospheric retrievals for the limb-resolved transmission spectra shown in Section \ref{sec:limb_spectra}, adopting the updated stellar parameters presented in Section \ref{sec:stellar}. Specifically, \texttt{PICASO} was used to simultaneously retrieve the morning and evening limb spectra with assumptions of chemical equilibrium, and \texttt{POSEIDON} was used to independently retrieve the two spectra with free chemistry.

\begin{deluxetable*}{lccc}
    \tablewidth{0pt}
    \tablecaption{\texttt{PICASO} Retrieval Priors \& Fitted Values}
    \label{tab:picaso}
    \tablehead{
        \colhead{Parameter} & \colhead{Description} &
        \colhead{Prior} & \colhead{Fit Value}
    }
    \startdata
    % \multicolumn{4}{c}{\textbf{Retrieved Parameters}} \\
    % \hline
    $\mathrm{[M/H]}$ (dex)
    & Atmospheric Metallicity
    & $\mathcal{U}(-1,3)$
    & $0.11^{+0.40}_{-0.46}$ \\
    $\mathrm{C/O}$ (unitless)
    & Atmospheric C/O Ratio
    & $\mathcal{U}(0.0458,0.916)$
    & $0.43^{+0.55}_{-0.30}$ \\  
    $\log f_{\mathrm{sed}}$ (unitless)
    & Sedimentation Factor
    & $\mathcal{U}(-4,1)$
    & $-2.7^{+1.2}_{-0.8}$ \\
    $\log K_{\mathrm{zz}}~(\mathrm{cm^2\,s^{-1}})$
    & Vertical Eddy Diffusion Coefficient
    & $\mathcal{U}(8,13)$
    & $10.6^{+1.0}_{-1.3}$ \\
    $x_{R_p}$ (unitless)
    & Radius Adjustment Factor
    & $\mathcal{U}(-0.1,0.1)$
    & $-0.0096^{+0.0065}_{-0.0052}$ \\
    $T_{\rm eq}$ (K)
    & Evening Limb Temperature
    & $\mathcal{U}(600,1800)$
    & $983^{+138}_{-186}$ \\
    $\delta T$ (K)
    & Limb--limb Temperature Contrast
    & $\mathcal{U}(-400,0)$
    & $-304^{+91}_{-62}$ \\
    $T_{\rm int}$ (K)
    & Intrinsic Temperature
    & $\mathcal{U}(30,600)$
    & $311^{+180}_{-170}$ \\
    $\log \kappa_{\rm IR}$
    & Guillot Profile Parameter
    & $\mathcal{U}(-2,+0.5)$
    & $-1.16^{+0.56}_{-0.50}$ \\
    $\log \gamma$
    & Guillot Profile Parameter
    & $\mathcal{U}(-2,+0.5)$
    & $-0.30^{+0.52}_{-0.76}$ \\
    $\alpha$
    & Guillot Profile Parameter
    & $\mathcal{U}(0,1)$
    & $0.48^{+0.32}_{-0.30}$ \\
    $\delta R_p^2/R_\ast^2$
    & Transit Depth Offset
    & $\mathcal{U}(0,6\times10^{-4})$
    & $(1.64^{+0.98}_{-0.83})\times10^{-4}$ \\
    \hline
    %\multicolumn{4}{c}{\textbf{Equilibrium Abundances}} \\
    %\hline
    %$\log_{10}(\mathrm{H_2O})$
    %    & Equilibrium \ce{H2O} Abundance
    %    & Derived
    %    & \\
    %$\log_{10}(\mathrm{CO_2})$
    %    & Equilibrium \ce{CO2} Abundance
    %    & Derived
    %    & \\
    \enddata
\end{deluxetable*}

\subsection{\texttt{PICASO} 1.5D Limb-limb Retrieval}

% \begin{deluxetable*}{lccc}\label{tab:picaso}
%     \centering
%     \tablewidth{0pt}
%     \tablecaption{\texttt{PICASO} retrieval Priors \& Fitted Values}
%     \tablehead{Parameter & Description & Prior & Fit Value}
%     \startdata
%     \hline
%     $\text{T}_{\text{int}}/\text{K}$ & Intrinsic Temperature & $\mathcal{U}(500,600)$ &  \\
%     $\text{[M/H]}$ (dex) & Atmospheric Metallicity & $\mathcal{U}(0,2.5)$ &  \\
%     $\text{C/O}$ (unitless) & Atmospheric C/O Ratio & $\mathcal{U}(0.25,2.5)$ &\\
%     $\text{f}_{\text{rfacv}}$ (unitless) & Recirculation Efficiency & $\mathcal{U}(0.3,0.7)$ &  \\
%     $\log \text{f}_{\text{sed}}$ (unitless) & Sedimentation Factor & $\mathcal{U}(-4,1)$ & \\
%     $\log \text{K}_{\text{zz}}~\text{(cm$^2$ s$^{-1}$)}$ & Vertical Eddy Diffusion Coefficient & $\mathcal{U}(8,11)$ & \\
%     $x_{\text{R}_{\text{p}}}$ (unitless) & Radius Adjustment Factor & \\
%     $\log \text{P}_{\text{ref}}~\text{(bar)}$ & Reference Pressure & $\mathcal{U}(-6, 0.9)$ & \\
%     $\delta T$ (K) & Limb-limb temperature contrast & & \\    $\delta R_p^2/R_\ast^2$ & Transit depth offset& &
%     \enddata
% \end{deluxetable*}

We used \texttt{PICASO} \citep{Mukherjee2023,batalha19,mang2026} and \texttt{VIRGA} \citep{batalha2026} to jointly model the morning and evening limb spectra, following \citet{mukherjee2025}. We modeled the $T(P)$ profiles of the morning and evening limbs as two parametric profiles that follow the parameterization from \citet{Guillot2010}. These profiles were allowed to have different values for the $T_{\rm eq}$ parameter but share the other parameters between them, including $\log(\kappa_{\rm IR})$, $\log{\gamma}$, and $a$. Following \citet{mukherjee2025}, we also forced the two $T(P)$ profiles to share the same $T_{\rm int}$ because the thermal profiles are expected to converge to the same adiabat at pressures deeper than the radiative--convective boundary. 

We assumed thermochemical equilibrium for both planet limbs. WASP-69~b's atmosphere can show substantial deviations from thermochemical equilibrium, owing to its equilibrium temperature \citep{moses2013,Mukherjee2025chem}. However, the effect of such disequilibrium chemistry processes on the abundance of H$_2$O is relatively low ($\sim{0.1}$ dex; \citealt{Mukherjee2025chem}) compared to their effects on other gases (e.g., CH$_4$, CO, NH$_3$, and CO$_2$) at WASP-69~b's equilibrium temperature \citep{Bangera2026}. As our observations covered wavelengths primarily sensitive to H$_2$O absorption, thermochemical equilibrium was a justified physical assumption.

We used \texttt{VIRGA} \citep{batalha2026} to model condensed clouds in each planet limb. \texttt{VIRGA} is based on the cloud model presented in \citet{ackerman2001}. It simulates the vertical distribution of cloud particles assuming a balance between the lofting of cloud particles and condensible vapor due to vertical transport and the subsequent settling of cloud particles under gravity. The vertical transport was parametrized by the 1D eddy diffusion coefficient $K_{\rm zz}$, whereas the settling of cloud particles was parameterized through the $f_{\rm sed}$ parameter. The eddy diffusion coefficient profile was held constant, i.e., independent of pressure/altitude. $f_{\rm sed}$ is a ratio between the settling speed of the cloud particles and the convection speed.  We allowed both limbs to have the same $f_{\rm sed}$ and $K_{\rm zz}$ parameters following \citet{mukherjee2025}. We included clouds composed of MnS, KCl, and Na$_2$S in our models, as they have been predicted to form in this temperature regime from thermochemical equilibrium calculations \citep{morley2012,visscher2006}. 

This forward modeling framework had 12 parameters-- $T_{\rm eq}$, $\delta{T_{\rm eq}}$, $\log(\kappa_{\rm IR})$, $\log{\gamma}$, $a$, $T_{\rm int}$, [M/H], C/O, $\log(f_{\rm sed})$, $\log(K_{\rm zz})$, $x_{R_p}$, and an offset. $\delta{T_{\rm eq}}$ represented the difference in the temperatures between the evening and morning limb. We scaled the radius of the planet at 1~mbar reference pressure with the $x_{R_p}$ parameter, where the radius at 1~mbar was given by $R=R_p(1+x_{R_p})$. We also included an offset parameter between the morning and evening limb spectra, as this offset is dependent on the uncertainty in the measured mid-transit time from the white light curve \citep{fu25,mukherjee2025}. We used the PyMultiNest Bayesian sampler \citep{buchner16} to constrain these parameters by fitting this forward model jointly to the measured morning and evening limb spectra.

\begin{figure*}[t!]
    \centering
    \includegraphics[width=1.0\linewidth]{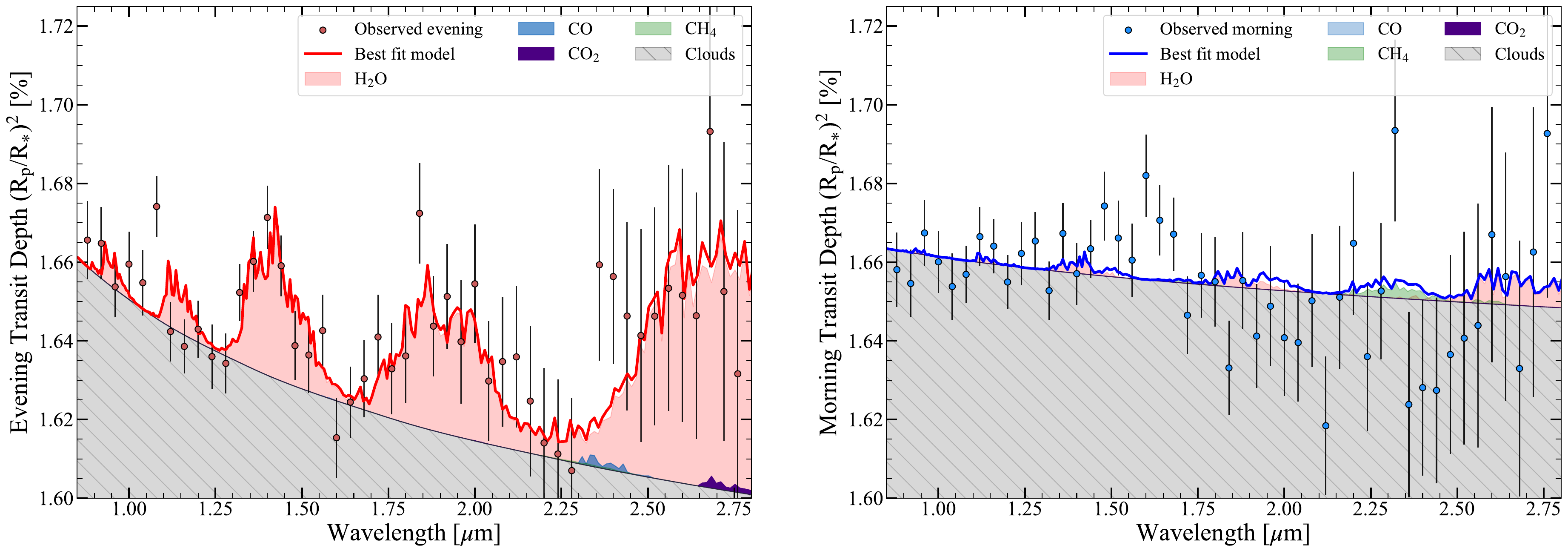}
    \caption{Best-fitting spectra from the joint \texttt{PICASO} limb--limb retrieval for the evening limb (left) and morning limb (right). Points with error bars show the measured spectra, and the solid curves show the best-fitting models. The shaded regions indicate the contributions from \ce{H2O}, CO, \ce{CO2}, \ce{CH4}, and clouds. Clouds introduce a blueward slope on both spectra. The evening-limb spectrum retains prominent molecular structure dominated by \ce{H2O}, whereas cloud opacity strongly mutes the molecular features on the morning limb.
    }
    \label{fig:spec_picaso}
\end{figure*}

\begin{figure*}[t!]
    \centering
    \includegraphics[width=1.0\linewidth]{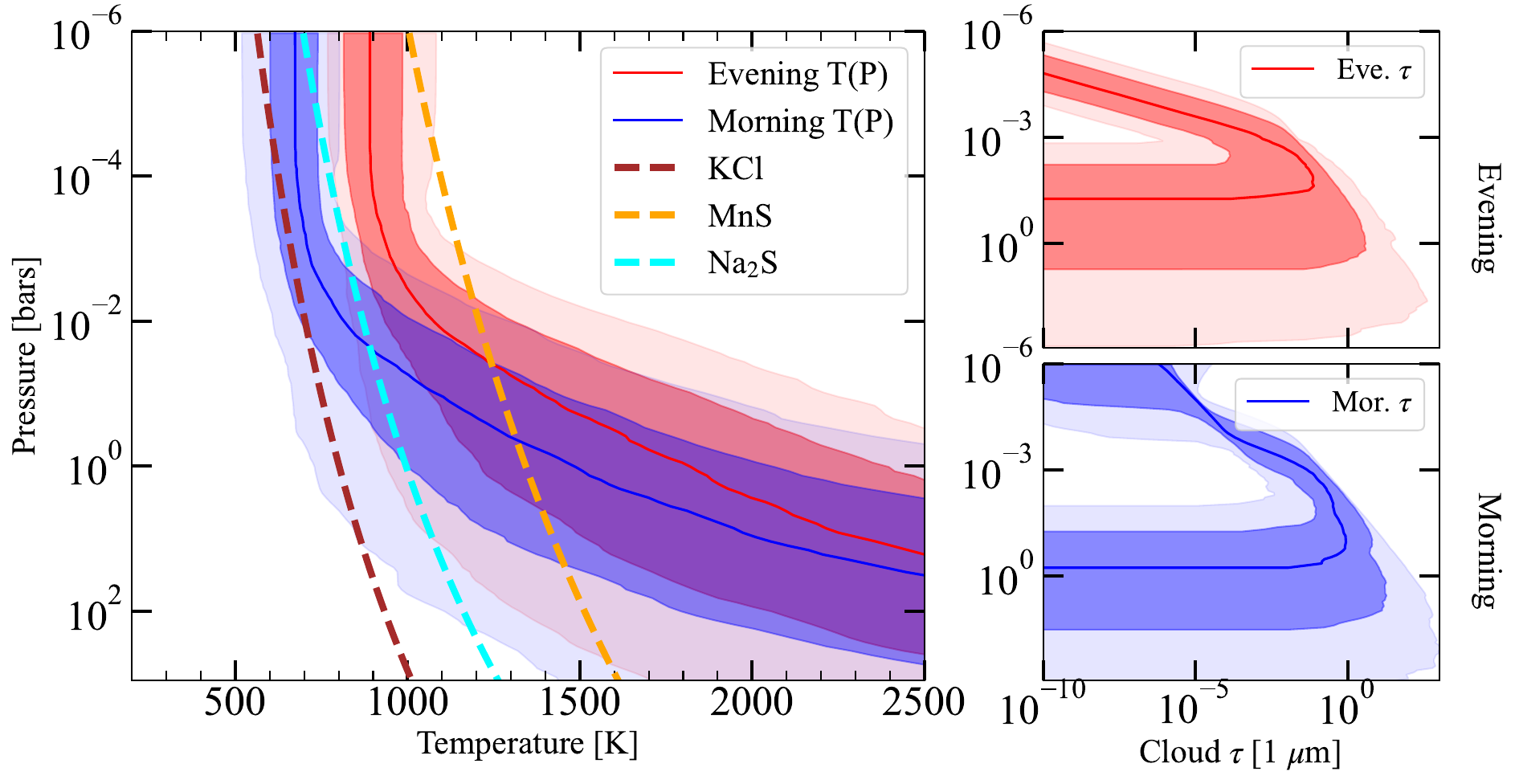}
    \caption{Retrieved atmospheric structures from the \texttt{PICASO} limb--limb retrieval. The left panel shows the temperature--pressure profiles of the evening (red) and morning (blue) limbs, with shaded regions denoting the $1\sigma$ and $2\sigma$ posterior uncertainties. Dashed curves mark the condensation temperatures of \ce{KCl}, MnS, and \ce{Na2S}. The right panels show the corresponding cloud optical-depth profiles at 1~$\micron$ for the evening (top) and morning (bottom) limbs.
    }
    \label{fig:tp_picaso}
\end{figure*}

We summarize the priors and posteriors for the fitted parameters in Table \ref{tab:picaso}. The retrieved best-fit model are shown in Figure \ref{fig:spec_picaso} along with contributions from key absorbers. We find that gaseous \ce{H2O} is the dominant gaseous absorber in both limbs. However, the aerosol contribution to the best-fit spectrum varies between the limbs. The morning limb spectra is dominated by aerosol absorption whereas the evening limb is dominated by \ce{H2O} absorption. The corresponding retrieved temperature-pressure ($T(P)$) profiles for the morning and evening limbs are shown in Figure \ref{fig:tp_picaso} along with cloud optical depth profiles. We discuss the interpretations of these results in Section \ref{sec:retrieval_results}.

% \textcolor{red}{Sagnick: Describe PICASO here.\begin{itemize}
%     \item Description of the retrieval setup (Table \ref{tab:picaso})
%     \item Description of the retrieved spectrum (Figure \ref{fig:spec_picaso})
%     \item Description of the TP profile (Figure \ref{fig:tp_picaso})
%     \item takeaway: global cloud with different opacities in the morning and evening limb, causing scattering blueward slope and muted morning limb.
% \end{itemize}}

\begin{deluxetable*}{lc}
\centering
\tablewidth{0pt}
\tablecaption{\poseidon Atmospheric Retrieval Priors and Settings}
\tablehead{
\colhead{Component} & \colhead{Setting} }
\startdata
    Planetary Surface Gravity ($\log_{\rm{10}} g_{\rm{p}}$) &  5.76 (fixed)  \\
    Reference Radius ($\text{R}_{\mathrm{p, \, ref}}$) & $\mathcal{N}(1.057, 0.21^2)\,\text{R}_{\text{Jup}}$ \\
    \hline
    Atmospheric Temperature ($T_{\mathrm{ref}}$) & $\mathcal{U}(750, 1500)$\,K  \\
    $T(P)$ Profile Curvature 1 ($\alpha_{1}$) & $\mathcal{U}(0.3, 2.00)$\,K$^{-\frac{1}{2}}$  \\
    $T(P)$ Profile Curvature 2 ($\alpha_{2}$) & $\mathcal{U}(0.3, 2.00)$\,K$^{-\frac{1}{2}}$ \\
    $T(P)$ Profile Region 1 ($\log_{10} (P_{1}$ / bar)) & $\mathcal{U}(-6, 0)$ \\
    $T(P)$ Profile Region 2 ($\log_{10} (P_{2}$ / bar)) & $\mathcal{U}(-6, 0)$ \\
    $T(P)$ Profile Region 3 ($\log_{10} (P_{3}$ / bar)) & $\mathcal{U}(-2, 2)$ \\
    \hline
    Rayleigh Enhancement Factor ($\log_{\rm{10}} a$) & $\mathcal{U}(-4, 8)$ \\
    Haze Scattering Slope ($\gamma$) & $\mathcal{U}(-20, 1)$ \\
    Cloud Top Pressure ($\log_{\rm{10}} (P_{\rm{cloud}}$ / bar)) & $\mathcal{U}(-5, 0.5)$ \\
    \hline
    Stellar Photosphere Temperature $T_{\rm{phot}}$ & $\mathcal{N}(4920, 60^{2})$\,K  \\
    Stellar Heterogeneity Temperature $T_{\rm{het}}$ & $\mathcal{U}(2710, 5310)$\,K   \\
    Stellar Heterogeneity Fraction $f_{\rm{het}}$ & $\mathcal{U}(0, 0.5)$   \\
    \hline
    Free Chem Molecule Volume Mixing Ratio ($\log_{\rm{10}} X_i$) & $\mathcal{U}(-12, -0.3)$ \\
    \hline
\enddata
\tablecomments{Gaussian priors are summarized as $\mathcal{N}(\mu, \sigma^2)$, where $\mu$ and $\sigma$ are the mean and standard deviation, respectively. 
$T_{\mathrm{ref}}$ refers to the top-of-atmosphere temperature.
}
\label{tab:poseidon_priors}
\end{deluxetable*}

\begin{deluxetable*}{lccc}

\tablecaption{Retrieved Molecular Volume Mixing Ratios and the adopted opacity sources from \texttt{POSEIDON}, without stellar contamination
\label{tab:poseidon}}

\tablehead{
\colhead{Species} &
\colhead{Morning Limb} &
\colhead{Evening Limb} &
\colhead{Opacity Source}
}

\startdata
$\log_{10}\mathrm{H_2O}$  & $< -3.20$ & $-1.70^{+0.44}_{-0.78}$ & \citet{polyansky18} \\
$\log_{10}\mathrm{CH_4}$   & $< -3.19$ & $< -3.48$ & \citet{Yurchenko2024} \\
$\log_{10}\mathrm{CO_2}$   & $< -2.59$ & $< -2.07$ & \citet{yurchenko20} \\
$\log_{10}\mathrm{CO}$     & $< -1.16$ & $< -1.21$ & \citet{li15} \\
$\log_{10}\mathrm{NH_3}$   & $< -3.44$ & $< -3.59$ & \citet{Coles19} \\
$\log_{10}\mathrm{SO_2}$   & $< -1.15$ & $< -1.61$ & \citet{Underwood2016} \\
$\log_{10}\mathrm{HCN}$    & $< -2.35$ & $< -1.69$ & \citet{barber14} \\
$\log_{10}\mathrm{H_2S}$   & $< -2.03$ & $< -1.82$ & \citet{azzam16exomol} \\
$\log_{10}\mathrm{OCS}$    & $< -3.51$ & $< -2.55$ & \citet{Owens2024} \\
$\log_{10}\mathrm{CS_2}$   & $< -1.12$ & $< -1.54$ & \citet{Gordon22} \\
\enddata
\tablecomments{
Abundances are reported as $\log_{10}(\mathrm{VMR})$.
Only $\mathrm{H_2O}$ on the evening limb is detected and constrained; its
abundance is reported as the posterior median with the 16th--84th percentile
credible interval. All other values are $3\sigma$ upper limits.
}

\end{deluxetable*}

\subsection{\texttt{POSEIDON} 1D Limb-independent Free Retrieval}
We used the \poseidon Python package \citep{MacDonald2017,MacDonald2023} to perform free chemistry atmosphere retrievals on the individual limb spectra of WASP-69\,b's NIRISS transmission spectrum.
Our atmosphere model assumed background gas consisting of H$_2$ and He at Solar He/H$_2$ = 0.17 \citep{asplund09}.
We constructed atmospheres with 150 layers uniformly spaced in log pressure, ranging from 10$^{-7}$-10$^2$ bar, under hydrostatic equilibrium.
We solved for hydrostatic equilibrium with the assumption that the atmosphere has a pressure of 10$^{-3}$ bar at the reference radius.
We freely retrieved the six $T(P)$ profile parameters following the prescription of \citet{Madhusudhan2009}.
Our free chemistry atmosphere model considered the following trace species: H$_2$O, CH$_4$, CO$_2$, CO, NH$_3$, SO$_2$, HCN, H$_2$S, OCS, and CS$_2$.
We retrieved the log volume mixing ratios (VMRs) for each of the trace species in our model for our free chemistry retrievals.
For clouds and aerosols, we used the three-parameter species-agnostic parameterization of \citet{MacDonald17}, with a Rayleigh enhancement factor $\log a$, a haze slope $\gamma$, and a cloud-top pressure $\log P_{\text{cloud}}$.
As we examined cloud coverage on each limb individually, we did not allow for patchy clouds in these retrievals.
In retrievals that considered impacts on the transmission spectra from the transit light source effect \citep[TLSE;][]{Rackham18}, we marginalized over possible stellar contamination from unocculted starspots or faculae by interpolating PHOENIX stellar atmosphere models \citep{husser2013phoenix} using the \texttt{PyMSG} package \citep{Townsend2023} and the three-parameter stellar contamination prescription from \citet{Rathcke2021}.
In total, our free retrievals had 20 free parameters: the reference radius, 6 $T(P)$ profile parameters, 3 cloud/aerosol parameters, and 10 trace species parameters; models with stellar contamination added three parameters for a total of 23.
The priors and additional settings for our model are listed in Table \ref{tab:poseidon_priors}.

We calculated model spectra in our \poseidon retrievals by solving the equation of radiative transfer in a cylindrical coordinate system for 100 incident stellar rays that are attenuated according to the atmospheric opacity along line-of-sight.
Opacities were precomputed at $R=20,000$ across a grid of temperatures and pressures using the \texttt{Cthulhu} Python package \citep{Agrawal2024,Cthulu}.
The opacity sources for the species we included are listed in Table \ref{tab:poseidon}. We also included continuum collision-induced absorption from H$_2$-H$_2$ and H$_2$-He pairs \citep{Karman2019} and Rayleigh scattering for all gases \citep{MacDonald2022}. 
We additionally performed six nested retrievals, three for each limb: one with stellar contamination, one without H$_2$O, and one without CO$_2$ to calculate Bayes factors and assess the statistical significance of these components of our model. 
We calculated model transmission spectra using opacity sampling on a wavelength grid from 0.8--2.9 $\mu$m, and we performed nested sampling via the \texttt{MultiNest} algorithm using the \texttt{PyMultiNest} \citep{fer08,fer09,fer19} Python package.
This \texttt{MultiNest} run used 1000 live points to constrain the parameter space for each of our retrievals.

% \textcolor{red}{Takeaway:\begin{itemize}
%     \item potential degeneracy between aerosol scattering and stellar heterogeneity, but the stellar contamination scenario requires high temperature contrast
%     \item fit for the specific haze and cloud species to argue against haze
%     \item metallicity and C/O from evening limb vs limb averaged retrieval
% \end{itemize}}

\begin{figure*}[t!]
    \centering
    \includegraphics[width=1.0\linewidth]{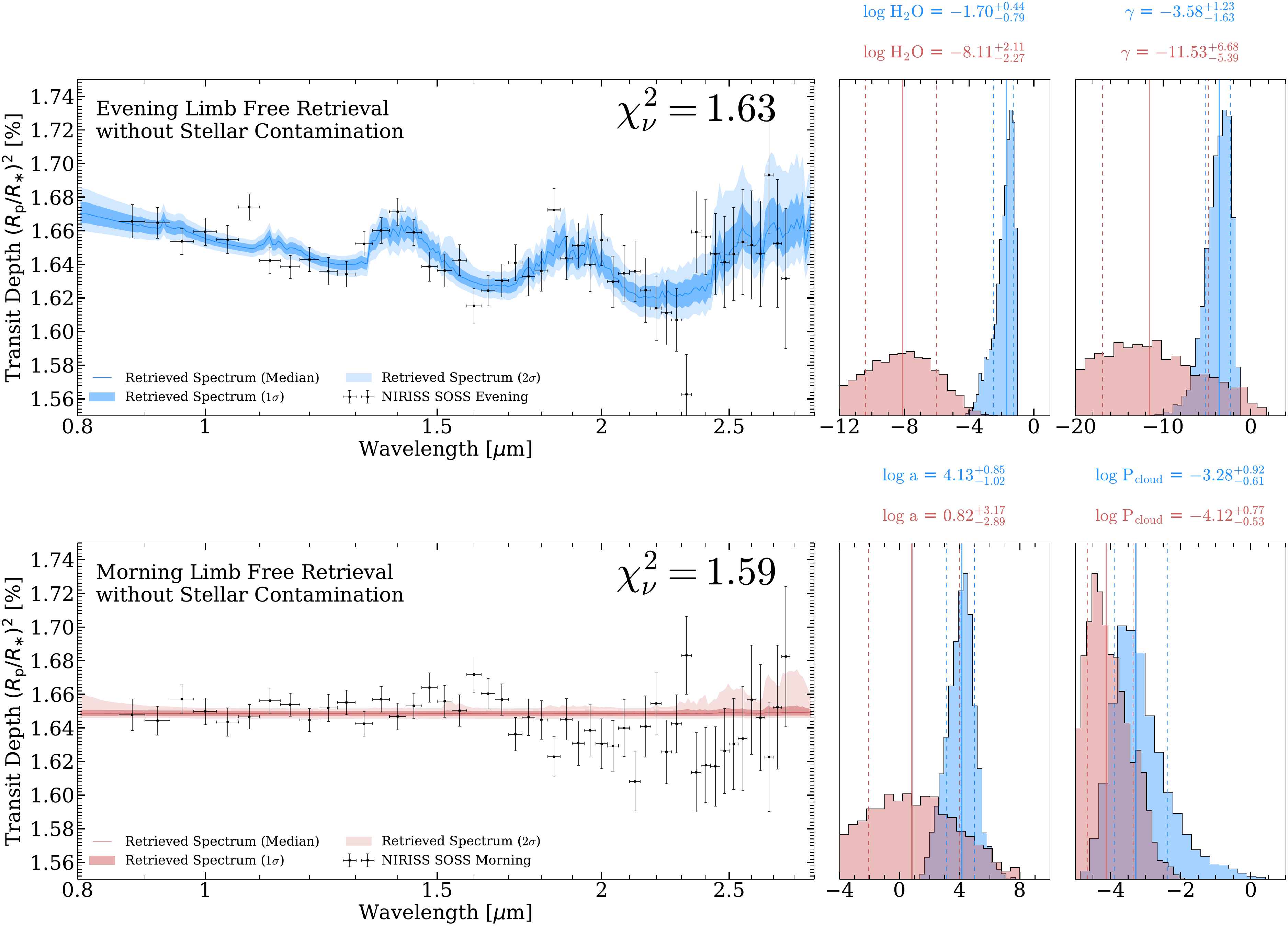}
    \caption{Limb-independent \texttt{POSEIDON} free retrievals of the evening and morning limb spectra without stellar contamination. The left panels show the measured evening (top) and morning (bottom) spectra, median retrieved models, and 1$\sigma$ and 2$\sigma$ credible intervals. The right panels compare the marginalized posteriors for selected atmospheric parameters, with blue and red denoting the evening and morning limbs, respectively; solid and dashed lines mark the posterior medians and 1$\sigma$ intervals.
    }
    \label{fig:poseidon_noTLSE}
\end{figure*}

\begin{figure*}[t!]
    \centering
    \includegraphics[width=1.0\linewidth]{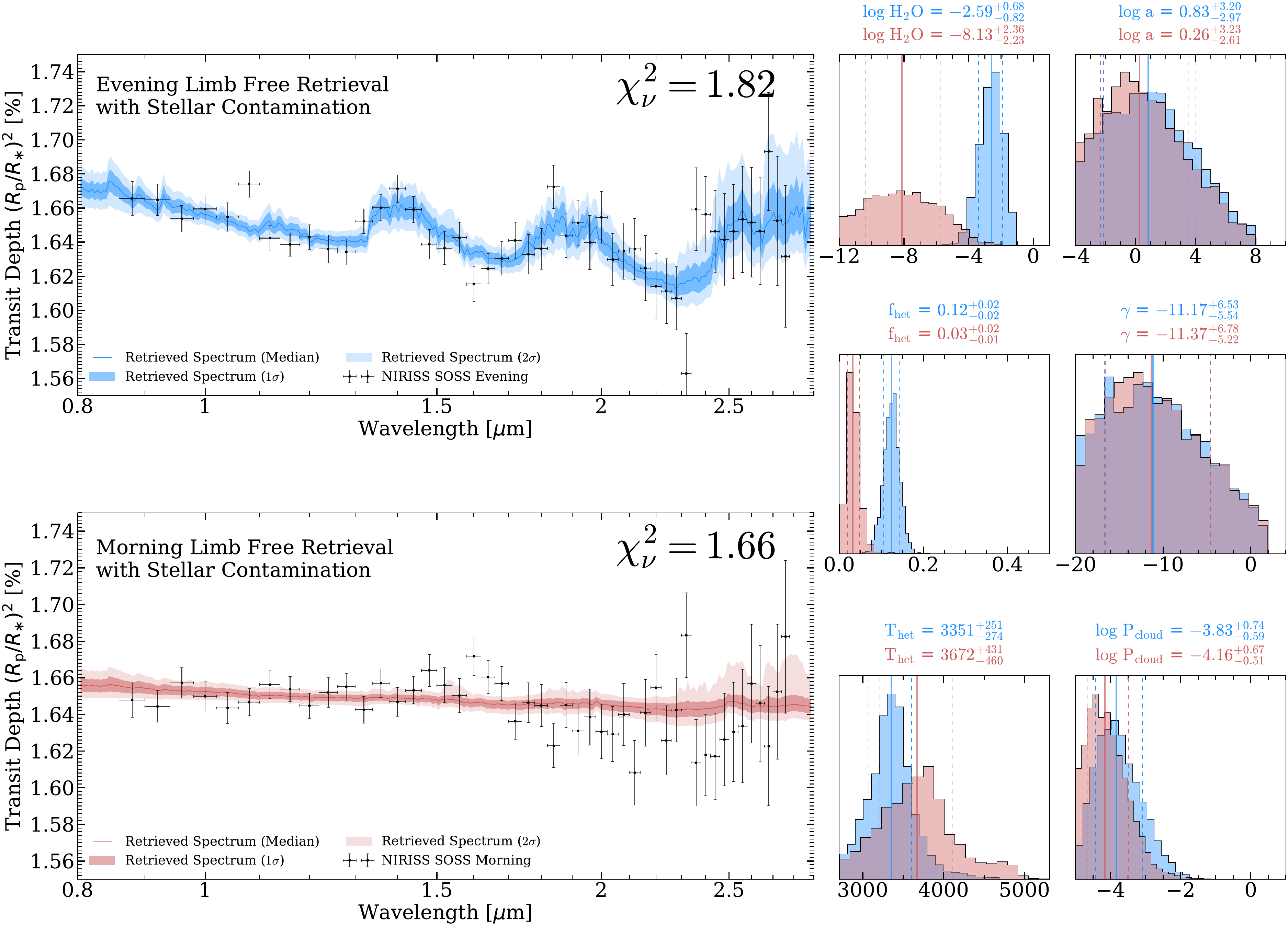}
    \caption{Same as Figure \ref{fig:poseidon_noTLSE}, but with stellar contamination included in the retrieval setup. The TLSE lowers the retrieved H$_2$O VMR in the evening limb slightly as less aerosol-driven muting of the feature is needed. 
    }
    \label{fig:poseidon_TLSE}
\end{figure*}

\subsection{Retrieval Results}\label{sec:retrieval_results}

% \textcolor{red}{Chris: Rough Skeleton now, to be expanded}
\subsubsection{Detection of Water and a Stellar-to-superstellar Metallicity}
% \textcolor{red}{Pending POSEIDON results here}

As shown in Figure \ref{fig:spec_picaso} and Table \ref{tab:poseidon}, our retrievals indicate the presence of \ce{H2O} in the evening limb of WASP-69\,b's atmosphere. Our limb-limb equilibrium retrievals additionally showed some \ce{CH4} opacities in the morning limb towards longer wavelengths (Figure \ref{fig:spec_picaso}) expected for equilibrium chemistry at that temperature, but it was not constrained by the free retrievals. Instead, we place 3$\sigma$ upper limits on species other than \ce{H2O} from limb-independent free retrievals in Table \ref{tab:poseidon}.

% \ce{CO2} agrees between the limb-limb equilibrium chemistry retrieval and limb-independent free retrieval, although the free retrievals only provide constraints on upper limits as \ce{CO2} does not have significant opacities in the SOSS band.

% We calculate the detection significance of \ce{H2O} and \ce{CO2} by performing the full free retrievals without either species on morning limb and evening limb spectra and comparing the resulting Bayesian Information Criterion to the full free retrieval  ($\Delta\rm BIC$, following \citealt{thorngren2026}). Specifically, $\Delta\mathrm{BIC}_{\rm X} = \mathrm{BIC}_{\rm\ce{X}} -  \mathrm{BIC}_{\rm ref}$, where $\mathrm{BIC}_{\rm\ce{X}}$ is the BIC for the best-fit model from the retrieval without species $\rm X$ and $\mathrm{BIC}_{\rm ref}$ is the BIC for the best-fit model from the full retrieval. A positive $\Delta \rm BIC_{X}$ therefore represents the case where species $\rm X$ is favored. We find that \ce{H2O} is not favored ($\Delta\mathrm{BIC}_{\ce{H2O}} = -8.7$ for the morning limb, while it is significantly detected in the evening limb ($\Delta\mathrm{BIC}_{\rm\ce{H2O}} = +22.7$). We discuss the implications of this result in Sections \ref{sec:retrieval_asymmetry}. In either limbs, \ce{CO2} is not favored ($\Delta \rm BIC_{\ce{CO2}} = -4.3$ in the morning limb and $\Delta \rm BIC_{\ce{CO2}} = -4.2$ in the evening limb).

We calculate the detection significance of \ce{H2O} by performing the full free retrievals without \ce{H2O} on morning limb and evening limb spectra and comparing the resulting BIC to the full free retrieval  ($\Delta\rm BIC$, following \citealt{thorngren2026}). Specifically, $\Delta\mathrm{BIC}_{\rm X} = \mathrm{BIC}_{\rm\ce{X}} -  \mathrm{BIC}_{\rm ref}$, where $\mathrm{BIC}_{\rm\ce{X}}$ is the BIC for the best-fit model from the retrieval without species $\rm X$ and $\mathrm{BIC}_{\rm ref}$ is the BIC for the best-fit model from the full retrieval. A positive $\Delta \rm BIC_{X}$ therefore represents the case where species $\rm X$ is favored. We find that \ce{H2O} is not favored ($\Delta\mathrm{BIC}_{\ce{H2O}} = -8.7$ for the morning limb, while it is significantly detected in the evening limb ($\Delta\mathrm{BIC}_{\rm\ce{H2O}} = +22.7$). We discuss the implications of this result in Sections \ref{sec:retrieval_asymmetry}.

The limb-limb equilibrium retrieval and evening limb free retrieval give consistent constraints on the planet's metallicity. Interpreting the retrieved evening-limb \ce{H2O} abundance from \texttt{POSEIDON} as a proxy for the atmospheric oxygen abundance yields an enrichment of
$24^{+45}_{-20}\times$ solar, or
$[\mathrm{O/H}]=1.38^{+0.44}_{-0.79}$. The equilibrium retrieval from \texttt{PICASO} constrains an atmospheric metallicity [M/H] $= 0.11^{+0.40}_{-0.46}$. The two estimates agree within 1.4 $\sigma$, and suggest WASP-69\,b has approximately stellar to superstellar metallicity ([Fe/H]$_\ast =0.28^{+0.03}_{-0.02}$ and [O/H]$_\ast = 0.26 \pm 0.08$). Panchromatic analysis incorporating NIRSpec/G395H will be needed to constrain additional metallicity-sensitive species such as CO and \ce{CO2}, thereby refining the atmospheric metallicity measurement and providing an accurate constraint on the C/O ratio.

% The uncertainty is too broad due to SOSS's negligible sensitivity to metallicity-sensitive species such as CO and \ce{CO2}. We are therefore unable to meaningfully constrain the planet's metal enrichment.

% Both estimates suggest that WASP-69\,b's atmosphere is more metal-rich than its host star, for which we measure $[\mathrm{Fe/H}]=0.28^{+0.03}_{-0.02}$ (Section~\ref{sec:stellar}). This superstellar enrichment is consistent with core-accretion scenarios in which the planet's envelope was enriched through the accretion and dissolution of heavy-element-rich solids \citep[e.g.,][]{Mordasini2016, Venturini2016}. The metal enrichment is also qualitatively consistent with the giant-planet mass--metallicity relation \citep{Thorngren2016}, in which Saturn-mass planets retain larger heavy-element abundances relative to their smaller H/He envelopes.

%The metallicity constraint from the limb-independent free retrieval comes from the \ce{H2O} constraints.

\subsubsection{Cloudy Morning and Less Cloudy Evening: Evidence against High Haze Fractions and Stellar Contamination} \label{sec:retrieval_asymmetry}

The muted morning-limb spectrum is best explained by an optically thick, high-altitude cloud layer. As shown by the limb-limb \texttt{PICASO} retrieval (Figure \ref{fig:spec_picaso}), cloud opacity dominates the morning-limb spectrum and suppresses the molecular absorption bands across the NIRISS/SOSS wavelength range. By contrast, the evening-limb spectrum retains prominent \ce{H2O} absorption, although clouds still contribute to the overall scattering slope. The limb-independent \texttt{POSEIDON} retrievals reach the same qualitative conclusion (Figure \ref{fig:poseidon_noTLSE}): the morning limb requires a high-altitude cloud deck ($\log_{10} P_{\rm cloud} < -2$), whereas the evening limb is characterized by enhanced aerosol scattering ($\log a>1$). The retrieved temperature structure provides a natural explanation for this opacity contrast. The evening-terminator $T(P)$ profile is hotter than the morning-terminator profile by $304^{+62}_{-91}$~K. As in WASP-94A\,b \citep{mukherjee2025}, condensate particles can form and remain lofted to low pressures on the cooler morning limb, but partially evaporate as they are transported into the hotter evening terminator. The resulting reduction in high-altitude cloud opacity allows the evening-limb \ce{H2O} bands to emerge. Unlike the nearly clear evening limb of WASP-94A\,b, however, WASP-69\,b appears to retain appreciable evening-limb aerosol opacity manifested as a scattering slope extending to the reddest wavelength of NIRISS/SOSS. This may reflect the more efficient atmospheric recirculation predicted at lower equilibrium temperatures \citep{Roth2024}. The resulting reduction in longitudinal temperature contrast limits the difference in condensation and evaporation between the two terminators, allowing clouds to persist on both limbs while their vertical distributions remain different.
Is the aerosol scattering mostly due to clouds or photochemically-produced haze? With NIRISS/SOSS data alone, it seems difficult to distinguish in this case. This contrasts with WASP-94A\,b, where a cloudy morning limb and nearly clear evening limb more directly favor condensates formed on the cooler nightside \citep{mukherjee2025}. 3D GCMs show that comparably muted spectral features require a very high haze concentration, which would also produce a thermal inversion in the atmospheric $T(P)$ profile (M. T. Mak et al. 2026b, in preparation). The absence of such an inversion in the dayside emission spectrum reported by \citet{schlawin2024} therefore provides evidence against substantial dayside haze on WASP-69\,b.

Host-star activity, such as unocculted star spots, can contaminate a transmission spectrum by a chromatic slope that can be confused with planetary signals \citep{Rackham18}. Such contamination, however, cannot readily explain the limb-dependent suppression of molecular features observed here: the evening-limb spectrum shows prominent \ce{H2O} absorption, whereas the same bands are strongly muted on the morning limb. Our limb-independent retrievals that explicitly included stellar contamination infer heterogeneous-region temperatures of $3351^{+251}_{-274}$ and $3672^{+431}_{-460}$~K for the evening and morning spectra, respectively. These regions are approximately $1570$ and $1250$~K cooler than the measured photospheric temperature of $4920^{+60}_{-40}$~K. More importantly, the inferred spot covering fractions are $0.12\pm0.02$ and $0.03\pm0.02$ for the evening and morning spectra (Figure \ref{fig:poseidon_TLSE}), differing by more than $3\sigma$. The retrieval therefore requires statistically different stellar-contamination slopes for the two limb spectra. The additional stellar parameters are also not justified by the improvement in fit: the models without stellar contamination are favored by $\Delta\mathrm{BIC}=10.3$ and 8.2 for the evening and morning limbs, respectively. Stellar contamination may contribute a common-mode slope, but the inconsistent retrieved spot covering fractions due to different slopes and poorer model preference indicate that it is unlikely to be the primary contributor to the observed spectra.

\section{Comparison to 3D Circulation Model}\label{sec:gcm}

To test whether the limb-limb aerosol distribution difference we found through retrievals emerges naturally from the atmospheric circulation of WASP-69\,b, we simulated the planet's three-dimensional thermal and chemical structure with a GCM. The resulting atmospheric profiles were used to predict the cloud distributions and transmission spectra of both limbs.

\subsection{GCM Setup}

\begin{deluxetable}{lcc}
\label{tab:input_parameter_UM}
\centering
\tablewidth{0pt}
\tablecaption{Stellar and planetary parameters for GCM input}
\tablehead{
Parameter & Value 
}
\startdata
Stellar irradiance [W\,m$^{-2}$] & 3.509$\times$10$^{6}$ \\
Stellar constant at 1\,AU [W\,m$^{-2}$] & 471.192\\
Inner radius [m] &  6.851$\times$10$^{8}$\\
Domain height [m] & 2.5$\times$10$^{7}$ \\
Semi-major axis [AU] & 4.661$\times$10$^{-2}$ \\
Orbital period [Earth day]  & 3.868 \\
Rotation rate [rad\,s$^{-1}$] &  1.88$\times$10$^{-5}$ \\
Surface gravity [m\,s$^{-2}$] & 5.76 \\
Specific gas constant [J\,K$^{-1}$\,kg$^{-1}$] & 3160.575 \\
Specific heat capacity [J\,K$^{-1}$\,kg$^{-1}$] & 11165.036 \\
Metallicity & 10$\times$solar\\
Filtering constant & 0.16\\
Depth of sponge layer & 0.75\\
\enddata
\end{deluxetable}

We simulated WASP-69\,b using the Met Office Unified Model (UM) 3D GCM, assuming a clear-sky atmosphere with kinetics chemistry. The UM has been used to simulate the atmosphere over a range of gaseous exoplanets \citep{Christie21,Zamyatina23,Zamyatina24,Christie24,Mak25}. It consists of a dynamical core, ENDGame, which solves the non-hydrostatic, full deep-atmosphere equations of motion with varying gravity \citep[see][for discussion]{Wood14,Mayne14a,Mayne14b,Mayne17}. To maintain numerical stability across the evolution of simulation, we set the filtering constant $K$ of $\sim$0.16, the depth of the sponge layer of 0.75 relative to the model domain height, with the vertical damping coefficient of 0.25 \citep[see discussions on treatment of longitudinal filtering and sponge layers in ][]{Christie24}. The dynamical core is coupled to a 2-stream radiative transfer module from the \texttt{Socrates} radiative transfer code \citep{Edwards_and_Slingo_1996} which solves the gaseous absorption from \ce{H2O}, \ce{CH4}, \ce{CO}, \ce{CO2}, \ce{HCN}, \ce{NH3}, \ce{Li}, \ce{Na}, \ce{K}, \ce{Rb}, and \ce{Cs}, as well as collision-induced absorption from \ce{H2}-\ce{H2} and \ce{H2}-\ce{He}, and Rayleigh scattering due to \ce{H2} and \ce{He} using the correlated-\textit{k} method \citep[see details of line list opacity sources in Appendix A in][]{Zamyatina23}. 

The kinetics chemistry scheme consists of the C/H/N/O reduced chemical network from \citet{Venot19}. The scheme solves for the stiff ordinary differential equations which calculate the production and loss of 30 chemical species with 181 reversible thermochemical reactions \citep[see][for further details]{Drummond20,Zamyatina23,Zamyatina24}. This chemistry scheme does not take into account any photochemical reactions. We summarize the stellar and planetary input parameters and the setup of UM in Table~\ref{tab:input_parameter_UM}. We performed our simulation using a horizontal grid spacing of 2.5$^\circ$ in longitude and 2$^\circ$ in latitude, with a vertical grid of 80 equally spaced model levels. Throughout the simulation runtime, we varied both the radiative and dynamical time steps between 20 and 30\,s before increasing to 60\,s once the simulation became numerically stable. The chemistry timescale is fixed at 3750\,s throughout the simulation time. We ran the simulation for 2000\,Earth days to reach a pseudo-equilibrium, determined by the plateauing of changes in the globally averaged top-of-atmosphere net flux, the maximum vertical and horizontal wind speeds, and the volume mixing ratio of \ce{CH4}, \ce{CO}, \ce{H2O}, \ce{CO2}, \ce{NH3} and \ce{HCN} with respect to their initial states. We present the temporally averaged simulation results from the last 50 Earth days. To model the condensed clouds using \texttt{VIRGA}, we calculated the GCM output global-averaged $K_{\rm zz}$ profile using the mixing length theory ($K_{\rm zz}=wH$ where $w$ is the global-averaged vertical wind and $H$ is the scale height) which results in $K_{\rm zz}$ varying between $10^9$--$10^{11}$\,cm$^2$\,s$^{-1}$ across the atmosphere, agreeing with the retrieved $K_{\rm zz}$ profile from our limb-limb chemical equilibrium retrieval (Table~\ref{tab:picaso}). We also adopted log($f_{\rm sed}$)$=-2.65$ according to the same retrieval (Table~\ref{tab:picaso}. We adopted GCM output $T(P)$ profiles of both limbs. We further calculated the clear-sky and cloudy transmission spectra with \texttt{PICASO}, using the GCM output $T(P)$ limb profiles as well as the species abundance limb profiles of \ce{H2O}, \ce{CO2}, \ce{CO}, \ce{CH4}, \ce{NH3}, and \ce{HCN} as input files. We note that the cloudy transmission spectrum was calculated using the GCM output $T(P)$ profiles post-processed with \texttt{VIRGA}, and that the GCM simulation itself was cloud-free. Without explicitly coupling cloud microphysics, atmospheric dynamics, chemistry, and radiative transfer, the estimated cloud abundance might be inaccurate, as the clouds were assumed to remain locally static, whereas in reality clouds can form through processes such as nucleation, condensation, coagulation etc \citep{Gao2020}. Cloud particles can also be transported across longitudes by atmospheric circulation \citep{Christie_2022,Powell_and_Zhang_2024}. This decoupled treatment also neglects the radiative effects of clouds on the atmospheric thermal structure. However we emphasize that this is an exploratory calculation rather than an attempt to calculate the detailed cloud formation and evolution. Our aim is to assess, to first order, what cloud species are preferentially expected to condense under the simulated atmospheric conditions, and hence how their presence may affect the limb transmission spectra.

\begin{figure*}
    \centering
    \includegraphics[width=0.95\linewidth]{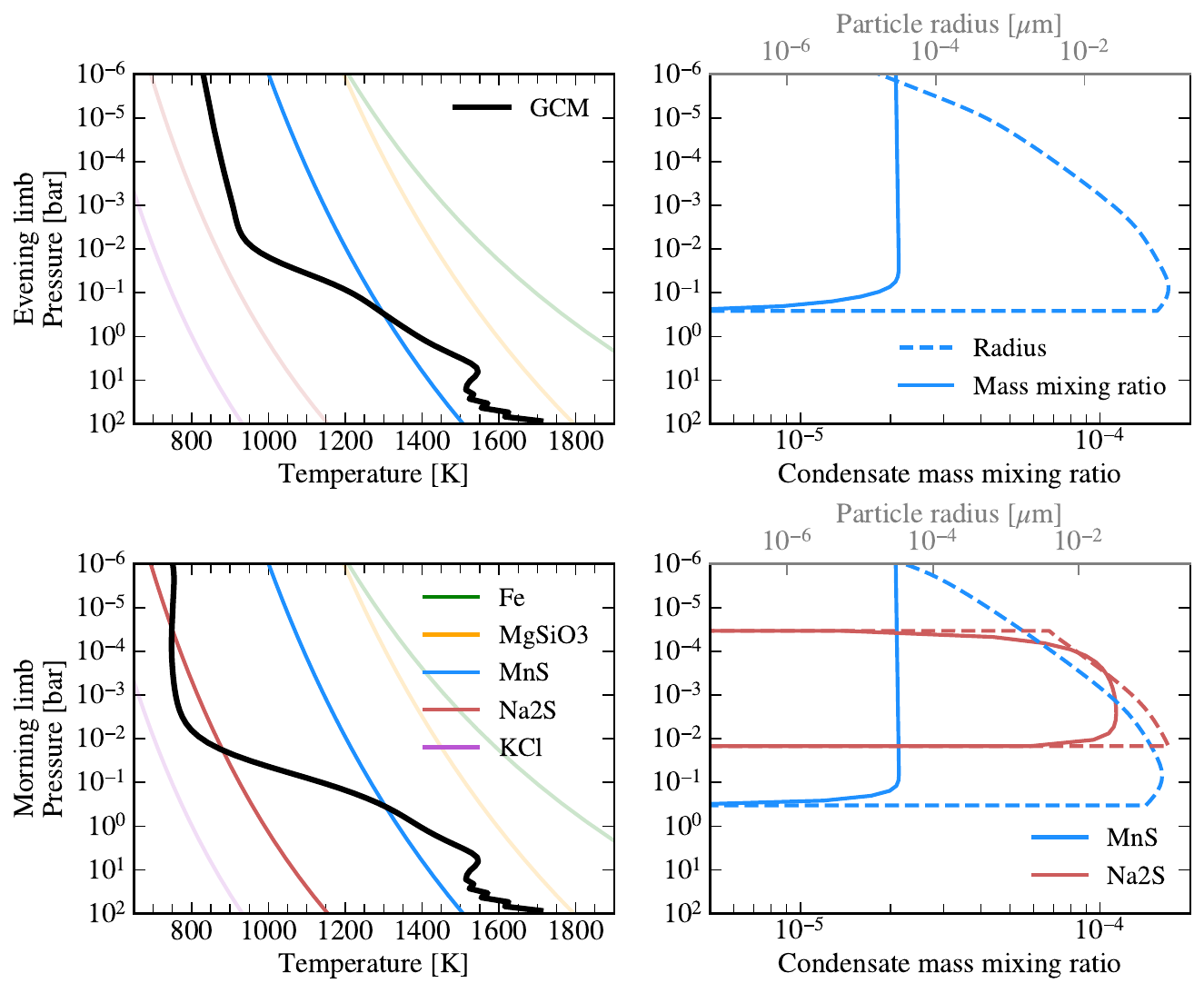}
    \caption{\textit{Left}: Evening (top) and morning limb (bottom) temperature-pressure profiles of the clear-sky atmosphere of WASP-69b simulated with kinetic chemistry (black). The condensation temperature of \ce{Fe}, \ce{MgSiO3}, \ce{MnS} and \ce{Na2S} are overlaid. Species that condense are represented in bright colors, whereas species that do not condense are represented in faded colors. The hotter evening limb favors only the condensation of \ce{MnS} cloud, whereas the colder morning limb favors the condensation of both \ce{MnS} and \ce{Na2S} clouds. \textit{Right}: Evening (top) and morning limb (bottom) limb with condensate mass mixing ratio and particle radius [$\mu$m] of \ce{MnS} and \ce{Na2S} clouds. The morning limb is dominated by \ce{Na2S} clouds in the observable region.}
    \label{fig:GCM_cloud_n_r}
\end{figure*}

\subsection{GCM Results}

\begin{figure*}[t!]
    \centering
    \includegraphics[width=0.95\linewidth]{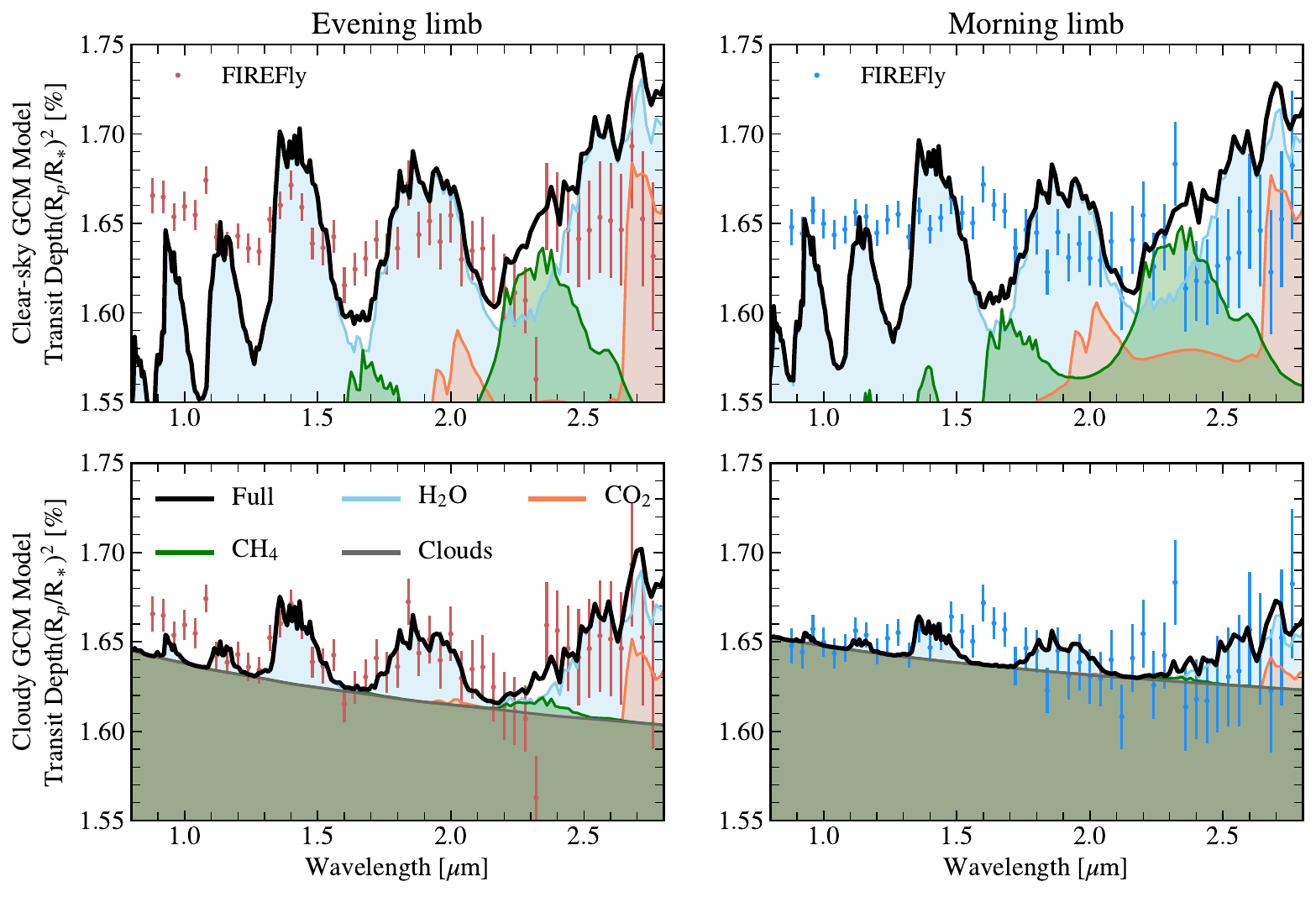}
    \caption{Clear-sky (top) and cloudy (bottom) transmission spectra of WASP-69b for the evening limb (left) and morning limb (right) plotted against the \texttt{FIREFLy} reduction. The cloudy spectra is post-processed with \texttt{VIRGA} using the GCM-output atmospheric output. The evening limb is assumed to have \ce{MnS} clouds, whereas the morning limb is assumed to have both and \ce{Na2S} clouds. The contributions from \ce{H2O}, \ce{CO2}, \ce{CH4} and clouds are overlaid. The cloud opacity strongly mutes the spectral features of \ce{H2O} in the atmosphere of both limbs.}
    \label{fig:GCM_clear_cloudy}
\end{figure*}

The left panels of Figure~\ref{fig:GCM_cloud_n_r} show the evening and morning limb temperature-pressure profiles assuming a clear-sky atmosphere, averaged across latitudes and overlaid with the condensation curves of multiple cloud species, including \ce{Fe}, \ce{MgSiO3}, \ce{MnS} \ce{Na2S}, and \ce{KCl}. Matching the retrieved thermal structure, the GCM calculated thermal profile of the evening limb is hotter than the morning limb, with only \ce{MnS} condensing over the evening limb and both \ce{MnS} and \ce{Na2S} condensing over the morning limb. The right panels of Figure~\ref{fig:GCM_cloud_n_r} show the condensate mass mixing ratio and particle radius distribution between the two limbs. \ce{MnS} clouds share a similar distribution between the two limbs, expected from the $T(P)$ profile and the shared $f_{\rm sed}$ in our initial conditions; \ce{Na2S} clouds significantly dominate in the observable region over the morning limb.

%, as well as the GCM-derived $K_{\rm zz}$ averaged across the whole atmosphere

Postprocessing the formation of clouds with \texttt{VIRGA} using the GCM-output atmospheric profiles, Figure~\ref{fig:GCM_clear_cloudy} compares the clear-sky and cloudy transmission spectra of both limbs, showing that the clear-sky atmosphere is strongly dominated by \ce{H2O} features in the transmission spectra of both limbs. Such prominent spectral features result in a mismatch between the theoretical clear-sky prediction and the observed data, agreeing with the retrieval results that WASP-69b should possess a cloudy atmosphere. Fig.~\ref{fig:GCM_clear_cloudy} further demonstrates that the inclusion of clouds significantly mutes the \ce{H2O} spectral features on both limbs. Consistent with the retrieval results, the cloud opacity is slightly weaker over the evening limb than over the morning limb, allowing the \ce{H2O} absorption band to remain visible in the evening spectrum, while it is strongly suppressed in the morning spectrum. Overall, the inclusion of clouds in the GCM spectra provides a better fit to the observed transmission spectra compared with the corresponding clear spectra, suggesting that aerosol opacity is likely present on both limbs in WASP-69\,b.

\section{Discussion}\label{sec:discussion}
% \subsection{Atmosphere of WASP-69~b}
% We confirmed the escaping atmosphere of WASP-69 b. We found haze probably can’t explain the blue part of the spectrum, in tension with pre-JWST results, because we see muted morning terminator. This is supported by our \texttt{PICASO} retrievals. We found degeneracy in spectral interpretation between stellar contamination and aerosol scattering: there could be components of both. Stellar heterogeneity can explain the blueward slope, but it requires high temperature contrast, as seen in \texttt{POSEIDON}; aerosols in the form of clouds can explain both the slope and limb asymmetry. In the cloud-scattering scenario, the cloud will be global, but with the morning limb having higher opacity. The MIRI data will provide an independent test of this hypothesis.

% \textcolor{red}{Chris: I will expand the above summary after the retrievals are finalized.}

\subsection{Differentiate between Aerosols vs Stellar Heterogeneity}

Previous observations of WASP-69\,b have shown that it is difficult to distinguish stellar contamination from aerosols. \citet{Murgas2020}, \citet{Ouyang2023}, and \citet{Allen2024} measured steep optical slopes consistent with aerosol scattering, but could not rule out activity from the host star as their origin. More recently, \citet{PR2024} claimed detection of a facular crossing and found that unocculted faculae can dominate the spectrum at the shortest wavelengths, while still requiring a high-altitude cloud deck in the planetary atmosphere. 

Our study shows that limb-resolved spectra could break the degeneracy between stellar contamination and aerosols. Because the morning and evening spectra are measured within the same transit, they share the common-mode contamination produced by unocculted stellar regions. Such contamination may manifest as a common blueward slope, but it cannot readily explain why prominent \ce{H2O} absorption is present on the evening limb and suppressed on the morning limb. Such a wavelength-dependent difference between the two spectra therefore localizes at least the longitude-dependent component of the planetary atmosphere. The retrieval tests presented in Section~\ref{sec:retrieval_asymmetry} reinforce this interpretation, as the retrieved spot properties differ statistically between the two limbs and the additional model complexity from stellar contamination is not justified by the fit improvement.

% Retrievals that include stellar contamination require spot covering fractions that differ by more than 3$\sigma$ between the two limbs and are disfavored by $\Delta\mathrm{BIC}=10.3$ and 8.2 compared to retrievals without stellar contamination for the evening and morning spectra, respectively.

% In this case, limb-resolved spectroscopy breaks the degeneracy between stellar contamination and aerosols, and we can more confidently identify asymmetric aerosol distribution as the primary source of the morning--evening difference.

% We therefore cannot exclude a common-mode stellar contribution, particularly at the shortest wavelengths, but stellar heterogeneity cannot account for the observed difference in molecular-feature amplitudes.

% In this case, limb-resolved spectroscopy breaks the stellar-contamination--aerosol degeneracy for the asymmetric component of the transmission spectrum and identifies heterogeneous planetary aerosols as the primary source of the morning--evening difference.

\subsection{Aerosol Properties}

The presence of high-altitude aerosols, primarily in the form of photochemical haze, has been consistently invoked to explain the optical slope of WASP-69\,b's transmission spectrum \citep{Murgas2020,Estrela2021,Khalafinejad2021,Ouyang2023, Allen2024}. Recently, \citet{schlawin2024} found that dayside aerosol opacity is required to explain the 2--12~$\micron$ JWST emission spectrum, but did not uniquely identify the aerosol composition. A high-altitude \ce{MgSiO3} cloud model reproduced the spectrum but required extreme vertical lofting from deep atmospheric layers; \ce{Na2S} and MnS could condense at the inferred dayside pressures, but inefficient nucleation may inhibit their formation; KCl nucleates more readily but would require transport from cooler regions \citep{Gao2020}.

The NIRISS/SOSS bandpass alone does not uniquely identify the aerosol composition, because different aerosol can produce similarly smooth opacity slopes and muting effects over 0.9--2.8~$\micron$. Nevertheless, both our limb--limb retrievals and GCM results show that condensate clouds with different characteristic particle sizes and vertical extents provide reasonable explanations for the observed spectra: high-altitude, optically thick clouds mute the morning limb, whereas lower-altitude clouds of lower opacity preserve the \ce{H2O} bands on the evening limb while producing its short-wavelength slope. The composition of the condensate remains uncertain, however. Joint analysis with the existing MIRI transmission observations (GO 3712, PI: P. E. Cubillos) may help identify the cloud composition, because different condensates produce more distinctive opacity features at mid-infrared wavelengths. Aerosols dominated by photochemical haze can also produce similarly muted spectrum, but would result in thermal inversion that was not detected by the emission spectrum analysis \citep{schlawin2024}. The presence of small mass mixing ratio of haze is not ruled out, though. A joint analysis with HST observations extending into the UV and optical would help constrain the relative contribution of photochemical haze.

%\textcolor{red}{[Add the GCM result showing that condensate clouds composed of \ce{Na2S} and MnS can reproduce the observed limb-resolved transmission spectra and describing their predicted spatial distribution.]} \textcolor{red}{[Add the GCM result showing that photochemical haze cannot reproduce the muted morning-limb spectrum.]}

% \subsection{WASP-69 b at the Aerosol Asymmetry Horizon}

\subsection{WASP-69\,b at the Transition in Aerosol Asymmetry}
\label{sec:asymmetry_horizon}

\begin{figure}[t!]
    \centering
    \includegraphics[width=0.95\linewidth]{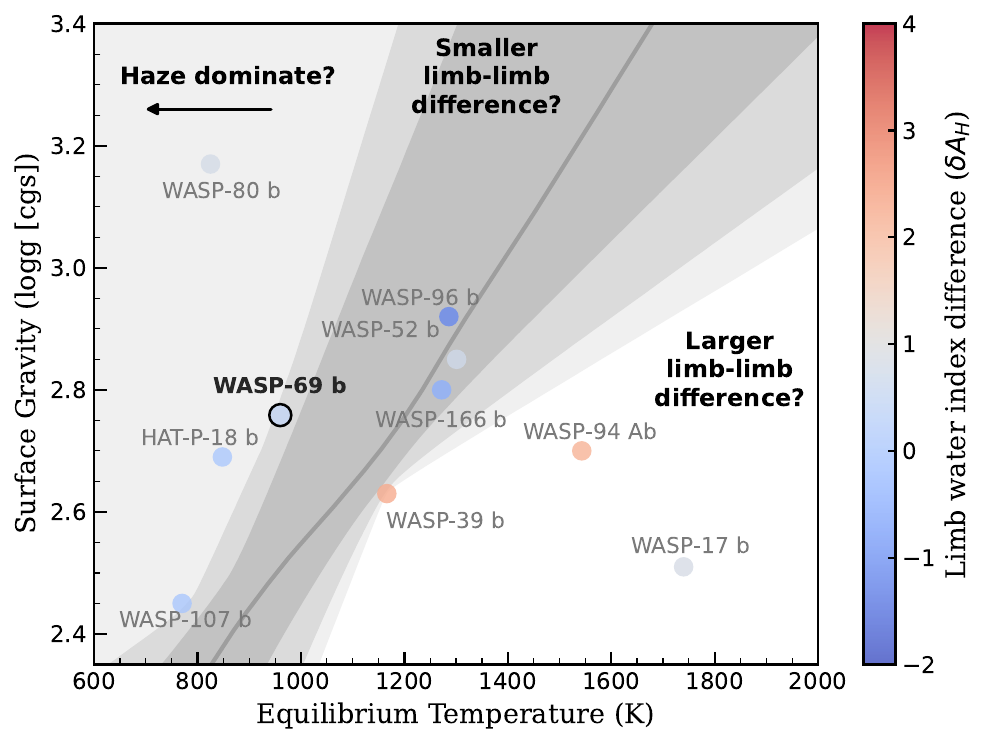}
    \caption{Aerosol asymmetry horizon uniformly recalculated for all nine planets from \citet{fu25} and for WASP-69~b, shown as the black-outlined point. The point color denotes the evening-minus-morning water-index difference in atmospheric scale heights ($\delta A_H$). The gray solid line and dark-to-light gray regions show the median refitted horizon and its 1, 2, and 3$\sigma$ posterior intervals. This figure reproduces Figure 7 of \citet{fu25}, except that the continuum is defined using both the 0.90--1.30 and 1.52--1.80 ~$\micron$ out-of-water bands, rather than only the 0.90--1.30~$\micron$ band adopted by \citet{fu25}.} %The continuum beneath the 1.4~$\micron$ water band is interpolated in wavelength between the 0.90--1.30 and 1.52--1.80~$\micron$ out-of-water bands.
    \label{fig:asymmetry_horizon_local}
\end{figure}

Does longitudinal aerosol heterogeneity vary gradually as planetary and atmospheric conditions change, or does it emerge over a relatively narrow region of parameter space analogous to the L/T transition seen in Brown Dwarfs (see e.g. \citealt{Kirkpatrick2005})? Limb-resolved transmission spectroscopy has the potential to directly probe this question.

% Nevertheless, cloud-regime transitions have also been predicted for hot Jupiters as changes in temperature alter condensate stability, cloud composition, and spatial coverage \citep{Parmentier2016,Gao2020}.

% Using a uniform NIRISS/SOSS analysis of nine giant planets, \citet{fu25} identified an empirical ``asymmetry horizon'' in the $T_{\rm eq}$--$\log g$ plane, suggesting a population-level transition from relatively homogeneous aerosol coverage to stronger morning--evening differences. We place WASP-69\,b within this framework to test whether it lies with the more symmetric cooler planets or the strongly asymmetric systems at higher irradiation.

Using a uniform NIRISS/SOSS analysis of nine giant planets, \citet{fu25} identified an empirical ``asymmetry horizon'' in the $T_{\rm eq}$--$\log g$ plane, suggesting a population-level transition from relatively homogeneous aerosol coverage to stronger morning--evening differences. WASP-69\,b joins the three previously identified planets---WASP-17\,b \citep{fu25}, WASP-39\,b \citep{rustamkulov2023,Espinoza2024}, and WASP-94A\,b \citep{mukherjee2025}---with pronounced longitudinal differences in aerosol opacity. Because WASP-69\,b extends this sample to lower irradiation, it provides meaningful constraints on where this transition of aerosol coverage occurs. We therefore place WASP-69\,b within the \citet{fu25} framework. To summarize, \citet{fu25} quantified the aerosol asymmetry using the evening-minus-morning difference in the amplitude of the 1.4~$\mu$m \ce{H2O} band. This difference was measured relative to the bluer out-of-\ce{H2O} band continuum and normalized by atmospheric scale height. This asymmetry metric was called ``limb water index difference" ($\delta A_H$).

A direct calculation of $\delta A_H$ proposed by \citet{fu25} is biased for WASP-69\,b because the pronounced aerosol-scattering slope on both limbs of WASP-69\,b would alter the blue continuum used to compare with the 1.4~$\mu$m feature. We therefore estimated the continuum from bands on both sides outside of the 1.4~$\mu$m water feature, interpolating the mean transit depths over 0.90--1.30 and 1.52--1.80~$\mu$m. We applied this two-sided continuum definition consistently to WASP-69\,b and to the nine planets from \citet{fu25} before refitting the population trend.

We find WASP-69\,b to be near the transition between relatively homogeneous aerosol coverage and the strongly asymmetric cloudy-morning, clear-evening regime. With the two-sided continuum, WASP-69\,b has an evening-minus-morning water-index difference of
$\delta A_{\rm H}=0.561\pm0.234$ atmospheric scale heights, corresponding to a $2.4\sigma$ limb difference. Its asymmetry is weaker than those of WASP-39\,b and WASP-94A\,b, but larger than the values for WASP-107\,b and HAT-P-18\,b. WASP-69\,b therefore occupies an intermediate regime: its morning limb is sufficiently cloudy to suppress the \ce{H2O} bands, while its evening limb retains short-wavelength aerosol opacity but remains optically thin enough for the 1.4~$\mu$m feature to emerge.

% We refit the sigmoid model of \citet{fu25} to the two-sided-continuum measurements for their nine planets and WASP-69\,b. In this model, $\delta A_H$ changes across a straight boundary in the $T_{\rm eq}$--$\log g$ plane, producing the refitted horizon shown in Figure~\ref{fig:asymmetry_horizon_local}. The sigmoid is favored over a constant-asymmetry model by $\Delta\mathrm{BIC}=38.8$, supporting a broad population trend with irradiation and gravity. However, its poor absolute fit, $\chi_\nu^2=4.20$, shows that these two parameters do not fully describe the transition. Additional variation in circulation, vertical mixing, particle settling, aerosol composition, and nucleation may therefore broaden or shift the apparent horizon \citep{Powell2019,PowellZhang2024,Gao2020}. The horizon should thus be interpreted as a provisional projection of a higher-dimensional transition rather than a sharp universal boundary.

The expanded samples support the population-level trend with equilibrium temperature and surface gravity. We refitted the sigmoid model of \citet{fu25} for their nine planets and WASP-69\,b. In this model, $\delta A_H$ was expected to change across a straight boundary in the $T_{\rm eq}$--$\log g$ plane. This boundary represents the refitted horizon shown in Figure~\ref{fig:asymmetry_horizon_local}. The sigmoid is favored over a constant-asymmetry model by $\Delta\mathrm{BIC}=38.8$, supporting a broad population trend with irradiation and gravity. However, its poor fit quality, $\chi_\nu^2=4.20$, shows that the observed systems cannot be cleanly separated by a single straight diagonal boundary in the $T_{\rm eq}$--$\log g$ plane. Additional variation in circulation, vertical mixing, particle settling, aerosol composition, and nucleation may therefore broaden or shift the apparent horizon \citep{Powell2019,Gao2020, PowellZhang2024}. This could suggest that the asymmetry horizon is a potentially broad, multidimensional transition region rather than binary boundary determined only by $T_{\rm eq}$ and $\log g$.

\section{Conclusion}\label{sec:conclusion}

We find evidence for morning--evening atmospheric asymmetry in WASP-69\,b. The evening limb shows prominent 1.4~$\micron$ \ce{H2O} absorption, whereas the same feature is strongly muted on the morning limb. Aerosols appear to be important on both limbs, but with different characteristic particle sizes and vertical distributions: high-altitude, optically thick aerosols obscure molecular features on the morning limb, whereas lower-opacity aerosols preserve the \ce{H2O} bands on the evening limb. This asymmetry is not readily explained by stellar contamination or high-concentration photochemical haze, while aerosols dominated by condensate clouds remain as the more plausible explanation. Our 3D GCM simulations provide independent support that different cloud opacities and altitudes driven by the temperature difference between the two limbs could reproduce the observed spectrum.

Our analysis also provides constraints on the planet's composition and escaping atmosphere. The evening-limb \ce{H2O} abundance and the limb-limb equilibrium retrieval both favor a stellar-to-superstellar atmospheric metallicity, consistent with the core-accretion scenario in which the envelope is enriched by heavy-element-rich material. In addition, the NIRISS/SOSS spectrum reveals strong metastable-helium absorption and precisely characterizes the outflow morphology.

% By extending the small sample of planets with detected aerosol asymmetry ($N=3$) to lower irradiation ($T_{\rm eq}\lesssim1000~\rm K$), WASP-69\,b offers a new perspective on how heterogeneous cloud distributions emerge across the giant-planet population. 

% WASP-69\,b joins WASP-107\,b as a low-irradiation giant planet with detected aerosol asymmetry ($T_{\rm eq}\lesssim1000~\rm K$), offering a new perspective on how heterogeneous cloud distributions emerge across the giant-planet population.

By extending the small sample of planets with detected aerosol asymmetry ($N=4$) into the 800--1000 K regime, WASP-69\,b offers a new perspective on how heterogeneous cloud distributions emerge across the giant-planet population. Its aerosol distribution suggests that it may lie near the transition between relatively homogeneous aerosol coverage and the strongly asymmetric systems found at higher irradiation. More broadly, our results demonstrate that limb-resolved transmission spectroscopy can reveal aerosol transport and cloud evolution that are hidden in a limb-averaged spectrum.

WASP-69\,b has now been extensively observed with HST and JWST from optical to mid-infrared. A joint panchromatic analysis of WASP-69\,b's HST, NIRISS, NIRSpec, and MIRI transmission spectra, together with its dayside emission spectrum, will be needed for mapping the global transport of aerosols. The broader wavelength coverage will also be essential for determining the atmospheric C/O ratio, testing equilibrium against disequilibrium chemistry, assessing the roles of vertical mixing and photochemistry, and distinguishing among aerosol compositions. Together, these observations have the potential to transform WASP-69\,b into a benchmark warm giant planet to unveil the mystery of how chemistry, circulation, and aerosols interact in three dimensions in cooler regimes.

\section*{ACKNOWLEDGEMENTS}

L.-C.W. thanks Adam Burrows, Patricio Ernesto Cubillos Vallejos, Shreyas Vissapragada, Krishna Kanumalla, and Heather Knutson for insightful discussions.

This work is based on observations made with the NASA/ESA/CSA James Webb Space Telescope. The data were obtained from the Mikulski Archive for Space Telescopes at the Space Telescope Science Institute, which is operated by the Association of Universities for Research in Astronomy, Inc., under NASA contract NAS 5-03127 for JWST.  
These observations are associated with programs \#1201 \& 5924. Support for JWST program GO-5924 was provided by NASA through a grant from the Space Telescope Science Institute, which is operated by the Association of Universities for Research in Astronomy, Inc., under NASA contract NAS 5-26555. 
M.T.M. acknowledges support from the Croucher Postdoctoral Fellowship, funded by the Croucher Foundation.
H.B. is supported by a Science and Technology Facilities Council Studentship [ST/Y509383/1].
S.M. is supported through the 51 Pegasi b fellowship from the Heising-Simons Foundation. M.L-M. is supported by individual
research time under NASA contracts NAS5-26555 and
NAS5-03127 to the Associated Universities for Research
in Astronomy for the operation of the Hubble Space
Telescope and James Webb Space Telescope Science Operations Centers at STScI.
The GCM simulations are performed using Met Office Software and the Monsoon3 system, a collaborative facility supplied under the Joint Weather and Climate Research Programme, a strategic partnership between the Met Office and the Natural Environment Research Council in the UK.

\bibliography{main}{}

@ARTICLE{Allen2024,
       author = {{Allen}, Natalie H. and {Sing}, David K. and {Espinoza}, N{\'e}stor and {O'Steen}, Richard and {Nikolov}, Nikolay K. and {Rustamkulov}, Zafar and {Evans-Soma}, Thomas M. and {Ramos Rosado}, Lakeisha M. and {Alam}, Munazza K. and {L{\'o}pez-Morales}, Mercedes and {Stevenson}, Kevin B. and {Wakeford}, Hannah R. and {May}, Erin M. and {Brahm}, Rafael and {Tala Pinto}, Marcelo},
        title = "{HST SHEL: Enabling Comparative Exoplanetology with HST/STIS}",
      journal = {\aj},
         year = 2024,
        month = sep,
       volume = {168},
       number = {3},
          eid = {111},
        pages = {111},
          doi = {10.3847/1538-3881/ad58e1},
archivePrefix = {arXiv},
       eprint = {2405.20361},
 primaryClass = {astro-ph.EP},
       adsurl = {https://ui.adsabs.harvard.edu/abs/2024AJ....168..111A}
}

@ARTICLE{liu&wang2025,
       author = {{Liu}, Rongrong and {Wang}, Le-Chris and {Rustamkulov}, Zafar and {Sing}, David K.},
        title = "{Unveiling the Atmosphere of the Super-Jupiter HAT-P-14 b with JWST NIRISS and NIRSpec}",
      journal = {\aj},
         year = 2025,
        month = jun,
       volume = {169},
       number = {6},
          eid = {335},
        pages = {335},
          doi = {10.3847/1538-3881/adcba7},
archivePrefix = {arXiv},
       eprint = {2504.08903},
 primaryClass = {astro-ph.EP},
       adsurl = {https://ui.adsabs.harvard.edu/abs/2025AJ....169..335L}
}

@ARTICLE{mang2026,
       author = {{Mang}, James and {Batalha}, Natasha E. and {Morley}, Caroline V. and {Wogan}, Nicholas F. and {Mukherjee}, Sagnick and {Visscher}, Channon and {Marley}, Mark S. and {Fortney}, Jonathan J. and {Chubb}, Katy L. and {Gao}, Peter and {Malsky}, Isaac},
        title = "{PICASO 4.0: Clouds and Photochemistry in Climate Models of Brown Dwarfs and Exoplanets}",
      journal = {\apj},
         year = 2026,
        month = mar,
       volume = {1000},
       number = {1},
          eid = {98},
        pages = {98},
          doi = {10.3847/1538-4357/ae47ff},
archivePrefix = {arXiv},
       eprint = {2602.22468},
 primaryClass = {astro-ph.EP},
       adsurl = {https://ui.adsabs.harvard.edu/abs/2026ApJ..1000...98M}
}

@ARTICLE{batalha19,
       author = {{Batalha}, Natasha E. and {Marley}, Mark S. and {Lewis}, Nikole K. and {Fortney}, Jonathan J.},
        title = "{Exoplanet Reflected-light Spectroscopy with PICASO}",
      journal = {\apj},
         year = 2019,
        month = jun,
       volume = {878},
       number = {1},
          eid = {70},
        pages = {70},
          doi = {10.3847/1538-4357/ab1b51},
archivePrefix = {arXiv},
       eprint = {1904.09355},
 primaryClass = {astro-ph.EP},
       adsurl = {https://ui.adsabs.harvard.edu/abs/2019ApJ...878...70B}
}

@ARTICLE{Mukherjee2023,
       author = {{Mukherjee}, Sagnick and {Batalha}, Natasha E. and {Fortney}, Jonathan J. and {Marley}, Mark S.},
        title = "{PICASO 3.0: A One-dimensional Climate Model for Giant Planets and Brown Dwarfs}",
      journal = {\apj},
         year = 2023,
        month = jan,
       volume = {942},
       number = {2},
          eid = {71},
        pages = {71},
          doi = {10.3847/1538-4357/ac9f48},
archivePrefix = {arXiv},
       eprint = {2208.07836},
 primaryClass = {astro-ph.EP},
       adsurl = {https://ui.adsabs.harvard.edu/abs/2023ApJ...942...71M}
}

@ARTICLE{schmidt2025,
       author = {{Schmidt}, Stephen P. and {MacDonald}, Ryan J. and {Tsai}, Shang-Min and {Radica}, Michael and {Wang}, Le-Chris and {Ahrer}, Eva-Maria and {Bell}, Taylor J. and {Fisher}, Chloe and {Thorngren}, Daniel P. and {Wogan}, Nicholas and {May}, Erin M. and {Ferrari}, Piero and {Bennett}, Katherine A. and {Rustamkulov}, Zafar and {L{\'o}pez-Morales}, Mercedes and {Sing}, David K.},
        title = "{A Comprehensive Reanalysis of K2-18 b's JWST NIRISS+NIRSpec Transmission Spectrum}",
      journal = {arXiv e-prints},
         year = 2025,
        month = jan,
          eid = {arXiv:2501.18477},
        pages = {arXiv:2501.18477},
          doi = {10.48550/arXiv.2501.18477},
archivePrefix = {arXiv},
       eprint = {2501.18477},
 primaryClass = {astro-ph.EP},
       adsurl = {https://ui.adsabs.harvard.edu/abs/2025arXiv250118477S}
}

@ARTICLE{rustamkulov2022,
       author = {{Rustamkulov}, Zafar and {Sing}, David K. and {Liu}, Rongrong and {Wang}, Ashley},
        title = "{Analysis of a JWST NIRSpec Lab Time Series: Characterizing Systematics, Recovering Exoplanet Transit Spectroscopy, and Constraining a Noise Floor}",
      journal = {\apjl},
         year = 2022,
        month = mar,
       volume = {928},
       number = {1},
          eid = {L7},
        pages = {L7},
          doi = {10.3847/2041-8213/ac5b6f},
archivePrefix = {arXiv},
       eprint = {2203.04173},
 primaryClass = {astro-ph.EP},
       adsurl = {https://ui.adsabs.harvard.edu/abs/2022ApJ...928L...7R}
}

@ARTICLE{rustamkulov2023,
       author = {{Rustamkulov}, Z. and {Sing}, D.~K. and {Mukherjee}, S. and {May}, E.~M. and {Kirk}, J. and {Schlawin}, E. and {Line}, M.~R. and {Piaulet}, C. and {Carter}, A.~L. and {Batalha}, N.~E. and {Goyal}, J.~M. and {L{\'o}pez-Morales}, M. and {Lothringer}, J.~D. and {MacDonald}, R.~J. and {Moran}, S.~E. and {Stevenson}, K.~B. and {Wakeford}, H.~R. and {Espinoza}, N. and {Bean}, J.~L. and {Batalha}, N.~M. and {Benneke}, B. and {Berta-Thompson}, Z.~K. and {Crossfield}, I.~J.~M. and {Gao}, P. and {Kreidberg}, L. and {Powell}, D.~K. and {Cubillos}, P.~E. and {Gibson}, N.~P. and {Leconte}, J. and {Molaverdikhani}, K. and {Nikolov}, N.~K. and {Parmentier}, V. and {Roy}, P. and {Taylor}, J. and {Turner}, J.~D. and {Wheatley}, P.~J. and {Aggarwal}, K. and {Ahrer}, E. and {Alam}, M.~K. and {Alderson}, L. and {Allen}, N.~H. and {Banerjee}, A. and {Barat}, S. and {Barrado}, D. and {Barstow}, J.~K. and {Bell}, T.~J. and {Blecic}, J. and {Brande}, J. and {Casewell}, S. and {Changeat}, Q. and {Chubb}, K.~L. and {Crouzet}, N. and {Daylan}, T. and {Decin}, L. and {D{\'e}sert}, J. and {Mikal-Evans}, T. and {Feinstein}, A.~D. and {Flagg}, L. and {Fortney}, J.~J. and {Harrington}, J. and {Heng}, K. and {Hong}, Y. and {Hu}, R. and {Iro}, N. and {Kataria}, T. and {Kempton}, E.~M. -R. and {Krick}, J. and {Lendl}, M. and {Lillo-Box}, J. and {Louca}, A. and {Lustig-Yaeger}, J. and {Mancini}, L. and {Mansfield}, M. and {Mayne}, N.~J. and {Miguel}, Y. and {Morello}, G. and {Ohno}, K. and {Palle}, E. and {Petit dit de la Roche}, D.~J.~M. and {Rackham}, B.~V. and {Radica}, M. and {Ramos-Rosado}, L. and {Redfield}, S. and {Rogers}, L.~K. and {Shkolnik}, E.~L. and {Southworth}, J. and {Teske}, J. and {Tremblin}, P. and {Tucker}, G.~S. and {Venot}, O. and {Waalkes}, W.~C. and {Welbanks}, L. and {Zhang}, X. and {Zieba}, S.},
        title = "{Early Release Science of the exoplanet WASP-39b with JWST NIRSpec PRISM}",
      journal = {\nat},
         year = 2023,
        month = feb,
       volume = {614},
       number = {7949},
        pages = {659-663},
          doi = {10.1038/s41586-022-05677-y},
archivePrefix = {arXiv},
       eprint = {2211.10487},
 primaryClass = {astro-ph.EP},
       adsurl = {https://ui.adsabs.harvard.edu/abs/2023Natur.614..659R}
}

@ARTICLE{bell2022,
       author = {{Bell}, Taylor and {Ahrer}, Eva-Maria and {Brande}, Jonathan and {Carter}, Aarynn and {Feinstein}, Adina and {Caloca}, Giannina and {Mansfield}, Megan and {Zieba}, Sebastian and {Piaulet}, Caroline and {Benneke}, Bj{\"o}rn and {Filippazzo}, Joseph and {May}, Erin and {Roy}, Pierre-Alexis and {Kreidberg}, Laura and {Stevenson}, Kevin},
        title = "{Eureka!: An End-to-End Pipeline for JWST Time-Series Observations}",
      journal = {The Journal of Open Source Software},
         year = 2022,
        month = nov,
       volume = {7},
       number = {79},
          eid = {4503},
        pages = {4503},
          doi = {10.21105/joss.04503},
archivePrefix = {arXiv},
       eprint = {2207.03585},
 primaryClass = {astro-ph.IM},
       adsurl = {https://ui.adsabs.harvard.edu/abs/2022JOSS....7.4503B}
}

@software{buchner16,
       author = {{Buchner}, Johannes},
        title = "{PyMultiNest: Python interface for MultiNest}",
 howpublished = {Astrophysics Source Code Library, record ascl:1606.005},
         year = 2016,
        month = jun,
          eid = {ascl:1606.005},
archivePrefix = {ascl},
       eprint = {1606.005},
       adsurl = {https://ui.adsabs.harvard.edu/abs/2016ascl.soft06005B}
}

@ARTICLE{visscher2006,
       author = {{Visscher}, Channon and {Lodders}, Katharina and {Fegley}, Jr., Bruce},
        title = "{Atmospheric Chemistry in Giant Planets, Brown Dwarfs, and Low-Mass Dwarf Stars. II. Sulfur and Phosphorus}",
      journal = {\apj},
         year = 2006,
        month = sep,
       volume = {648},
       number = {2},
        pages = {1181-1195},
          doi = {10.1086/506245},
archivePrefix = {arXiv},
       eprint = {astro-ph/0511136},
 primaryClass = {astro-ph},
       adsurl = {https://ui.adsabs.harvard.edu/abs/2006ApJ...648.1181V}
}

@ARTICLE{ackerman2001,
       author = {{Ackerman}, Andrew S. and {Marley}, Mark S.},
        title = "{Precipitating Condensation Clouds in Substellar Atmospheres}",
      journal = {\apj},
         year = 2001,
        month = aug,
       volume = {556},
       number = {2},
        pages = {872-884},
          doi = {10.1086/321540},
archivePrefix = {arXiv},
       eprint = {astro-ph/0103423},
 primaryClass = {astro-ph},
       adsurl = {https://ui.adsabs.harvard.edu/abs/2001ApJ...556..872A}
}

@ARTICLE{Powell_and_Zhang_2024,
       author = {{Powell}, Diana and {Zhang}, Xi},
        title = "{Two-dimensional Models of Microphysical Clouds on Hot Jupiters. I. Cloud Properties}",
      journal = {\apj},
         year = 2024,
        month = jul,
       volume = {969},
       number = {1},
          eid = {5},
        pages = {5},
          doi = {10.3847/1538-4357/ad3de4},
archivePrefix = {arXiv},
       eprint = {2404.08759},
 primaryClass = {astro-ph.EP},
       adsurl = {https://ui.adsabs.harvard.edu/abs/2024ApJ...969....5P}
}

@ARTICLE{morley2012,
       author = {{Morley}, Caroline V. and {Fortney}, Jonathan J. and {Marley}, Mark S. and {Visscher}, Channon and {Saumon}, Didier and {Leggett}, S.~K.},
        title = "{Neglected Clouds in T and Y Dwarf Atmospheres}",
      journal = {\apj},
         year = 2012,
        month = sep,
       volume = {756},
       number = {2},
          eid = {172},
        pages = {172},
          doi = {10.1088/0004-637X/756/2/172},
archivePrefix = {arXiv},
       eprint = {1206.4313},
 primaryClass = {astro-ph.SR},
       adsurl = {https://ui.adsabs.harvard.edu/abs/2012ApJ...756..172M}
}

@ARTICLE{fu25,
       author = {{Fu}, Guangwei and {Mukherjee}, Sagnick and {Stevenson}, Kevin B. and {Sing}, David K. and {Ashtari}, Reza and {Mayne}, Nathan and {Lothringer}, Joshua D. and {Zamyatina}, Maria and {Schmidt}, Stephen P. and {Gasc{\'o}n}, Carlos and {Allen}, Natalie H. and {Bennett}, Katherine A. and {L{\'o}pez-Morales}, Mercedes},
        title = "{Overcast Mornings and Clear Evenings in Hot Jupiter Exoplanet Atmospheres}",
      journal = {\apjl},
         year = 2025,
        month = aug,
       volume = {989},
       number = {1},
          eid = {L17},
        pages = {L17},
          doi = {10.3847/2041-8213/adf20f},
archivePrefix = {arXiv},
       eprint = {2507.15854},
 primaryClass = {astro-ph.EP},
       adsurl = {https://ui.adsabs.harvard.edu/abs/2025ApJ...989L..17F}
}

@ARTICLE{moses2013,
       author = {{Moses}, J.~I. and {Line}, M.~R. and {Visscher}, C. and {Richardson}, M.~R. and {Nettelmann}, N. and {Fortney}, J.~J. and {Barman}, T.~S. and {Stevenson}, K.~B. and {Madhusudhan}, N.},
        title = "{Compositional Diversity in the Atmospheres of Hot Neptunes, with Application to GJ 436b}",
      journal = {\apj},
         year = 2013,
        month = nov,
       volume = {777},
       number = {1},
          eid = {34},
        pages = {34},
          doi = {10.1088/0004-637X/777/1/34},
archivePrefix = {arXiv},
       eprint = {1306.5178},
 primaryClass = {astro-ph.EP},
       adsurl = {https://ui.adsabs.harvard.edu/abs/2013ApJ...777...34M}
}

@ARTICLE{Mukherjee2025chem,
       author = {{Mukherjee}, Sagnick and {Fortney}, Jonathan J. and {Wogan}, Nicholas F. and {Sing}, David K. and {Ohno}, Kazumasa},
        title = "{Effects of Planetary Parameters on Disequilibrium Chemistry in Irradiated Planetary Atmospheres: From Gas Giants to Sub-Neptunes}",
      journal = {\apj},
         year = 2025,
        month = jun,
       volume = {985},
       number = {2},
          eid = {209},
        pages = {209},
          doi = {10.3847/1538-4357/adc7b3},
archivePrefix = {arXiv},
       eprint = {2410.17169},
 primaryClass = {astro-ph.EP},
       adsurl = {https://ui.adsabs.harvard.edu/abs/2025ApJ...985..209M}
}

@ARTICLE{Guillot2010,
       author = {{Guillot}, T.},
        title = "{On the radiative equilibrium of irradiated planetary atmospheres}",
      journal = {\aap},
         year = 2010,
        month = sep,
       volume = {520},
          eid = {A27},
        pages = {A27},
          doi = {10.1051/0004-6361/200913396},
archivePrefix = {arXiv},
       eprint = {1006.4702},
 primaryClass = {astro-ph.EP},
       adsurl = {https://ui.adsabs.harvard.edu/abs/2010A&A...520A..27G}
}

@ARTICLE{batalha2026,
       author = {{Batalha}, Natasha E. and {Rooney}, Caoimhe M. and {Visscher}, Channon and {Moran}, Sarah E. and {Marley}, Mark S. and {Sengupta}, Aditya R. and {Kiefer}, Sven and {Lodge}, Matt G. and {Mang}, James and {Morley}, Caroline V. and {Mukherjee}, Sagnick and {Fortney}, Jonathan J. and {Gao}, Peter and {Lewis}, Nikole K. and {Mayorga}, L.~C. and {Pearce}, Logan A. and {Wakeford}, Hannah R.},
        title = "{Condensation Clouds in Substellar Atmospheres with Virga}",
      journal = {\aj},
         year = 2026,
        month = feb,
       volume = {171},
       number = {2},
          eid = {98},
        pages = {98},
          doi = {10.3847/1538-3881/ae29e5},
archivePrefix = {arXiv},
       eprint = {2508.15102},
 primaryClass = {astro-ph.EP},
       adsurl = {https://ui.adsabs.harvard.edu/abs/2026AJ....171...98B}
}

@ARTICLE{emcee,
       author = {{Foreman-Mackey}, Daniel and {Hogg}, David W. and {Lang}, Dustin and {Goodman}, Jonathan},
        title = "{emcee: The MCMC Hammer}",
      journal = {\pasp},
         year = 2013,
        month = mar,
       volume = {125},
       number = {925},
        pages = {306},
          doi = {10.1086/670067},
archivePrefix = {arXiv},
       eprint = {1202.3665},
 primaryClass = {astro-ph.IM},
       adsurl = {https://ui.adsabs.harvard.edu/abs/2013PASP..125..306F}
}

@ARTICLE{Christie_2022,
       author = {{Christie}, D.~A. and {Mayne}, N.~J. and {Gillard}, R.~M. and {Manners}, J. and {H{\'e}brard}, E. and {Lines}, S. and {Kohary}, K.},
        title = "{The impact of phase equilibrium cloud models on GCM simulations of GJ 1214b}",
      journal = {\mnras},
         year = 2022,
        month = nov,
       volume = {517},
       number = {1},
        pages = {1407-1421},
          doi = {10.1093/mnras/stac2763},
archivePrefix = {arXiv},
       eprint = {2209.12205},
 primaryClass = {astro-ph.EP},
       adsurl = {https://ui.adsabs.harvard.edu/abs/2022MNRAS.517.1407C}
}

@ARTICLE{mukherjee2025,
       author = {{Mukherjee}, Sagnick and {Sing}, David K. and {Fu}, Guangwei and {Stevenson}, Kevin B. and {Schmidt}, Stephen P. and {Baskett}, Harry and {Mak}, Mei Ting and {McCreery}, Patrick and {Allen}, Natalie H. and {Bennett}, Katherine A. and {Christie}, Duncan A. and {Gasc{\'o}n}, Carlos and {Goyal}, Jayesh and {H{\'e}brard}, {\'E}ric and {Lothringer}, Joshua D. and {L{\'o}pez-Morales}, Mercedes and {Lustig-Yaeger}, Jacob and {May}, Erin M. and {Mayorga}, L.~C. and {Mayne}, Nathan and {Ramos Rosado}, Lakeisha M. and {Reggiani}, Henrique and {Rustamkulov}, Zafar and {Schlaufman}, Kevin C. and {Sotzen}, Kristin S. and {Thorngren}, Daniel and {Wang}, Le-Chris and {Zamyatina}, Maria},
        title = "{Cloudy mornings and clear evenings on a gas giant exoplanet}",
      journal = {Science},
         year = 2026,
        month = may,
       volume = {392},
       number = {6800},
        pages = {858-862},
          doi = {10.1126/science.adx5903},
archivePrefix = {arXiv},
       eprint = {2505.10910},
 primaryClass = {astro-ph.EP},
       adsurl = {https://ui.adsabs.harvard.edu/abs/2026Sci...392..858M}
}

@ARTICLE{LCWang2026,
       author = {{Wang}, Le-Chris and {Rustamkulov}, Zafar and {Sing}, David K. and {Lothringer}, Joshua and {McCreery}, Patrick and {Thorngren}, Daniel and {Alam}, Munazza K.},
        title = "{A Comprehensive Analysis of the Panchromatic Transmission Spectrum of the Hot-Saturn WASP-96 b: Nondetection of Haze, Possible Sodium Limb Asymmetry, Stellar Characterization, and Formation History}",
      journal = {\aj},
         year = 2026,
        month = mar,
       volume = {171},
       number = {3},
          eid = {147},
        pages = {147},
          doi = {10.3847/1538-3881/ae231b},
archivePrefix = {arXiv},
       eprint = {2511.16771},
 primaryClass = {astro-ph.EP},
       adsurl = {https://ui.adsabs.harvard.edu/abs/2026AJ....171..147W}
}

@ARTICLE{PR2024,
       author = {{Petit dit de la Roche}, D.~J.~M. and {Chakraborty}, H. and {Lendl}, M. and {Kitzmann}, D. and {Pietrow}, A.~G.~M. and {Akinsanmi}, B. and {Boffin}, H.~M.~J. and {Cubillos}, Patricio E. and {Deline}, A. and {Ehrenreich}, D. and {Fossati}, L. and {Sedaghati}, E.},
        title = "{Detection of faculae in the transit and transmission spectrum of WASP-69b}",
      journal = {\aap},
         year = 2024,
        month = dec,
       volume = {692},
          eid = {A83},
        pages = {A83},
          doi = {10.1051/0004-6361/202451740},
archivePrefix = {arXiv},
       eprint = {2410.18663},
 primaryClass = {astro-ph.EP},
       adsurl = {https://ui.adsabs.harvard.edu/abs/2024A&A...692A..83P}
}

@software{Bushouse2023,
       author = {{Bushouse}, Howard and {Eisenhamer}, Jonathan and {Dencheva}, Nadia and {Davies}, James and {Greenfield}, Perry and {Morrison}, Jane and {Hodge}, Phil and {Simon}, Bernie and {Grumm}, David and {Droettboom}, Michael and {Slavich}, Edward and {Sosey}, Megan and {Pauly}, Tyler and {Miller}, Todd and {Jedrzejewski}, Robert and {Hack}, Warren and {Davis}, David and {Crawford}, Steven and {Law}, David and {Gordon}, Karl and {Regan}, Michael and {Cara}, Mihai and {MacDonald}, Ken and {Bradley}, Larry and {Shanahan}, Clare and {Jamieson}, William and {Teodoro}, Mairan and {Williams}, Thomas},
        title = "{JWST Calibration Pipeline}",
         year = 2023,
        month = jul,
          eid = {10.5281/zenodo.8140011},
          doi = {10.5281/zenodo.8140011},
      version = {1.11.2},
    publisher = {Zenodo},
       adsurl = {https://ui.adsabs.harvard.edu/abs/2023zndo...8140011B}
}

@ARTICLE{kreidberg2015,
       author = {{Kreidberg}, Laura},
        title = "{batman: BAsic Transit Model cAlculatioN in Python}",
      journal = {\pasp},
         year = 2015,
        month = nov,
       volume = {127},
       number = {957},
        pages = {1161},
          doi = {10.1086/683602},
archivePrefix = {arXiv},
       eprint = {1507.08285},
 primaryClass = {astro-ph.EP},
       adsurl = {https://ui.adsabs.harvard.edu/abs/2015PASP..127.1161K}
}

@ARTICLE{Anderson2014,
       author = {{Anderson}, D.~R. and {Collier Cameron}, A. and {Delrez}, L. and {Doyle}, A.~P. and {Faedi}, F. and {Fumel}, A. and {Gillon}, M. and {G{\'o}mez Maqueo Chew}, Y. and {Hellier}, C. and {Jehin}, E. and {Lendl}, M. and {Maxted}, P.~F.~L. and {Pepe}, F. and {Pollacco}, D. and {Queloz}, D. and {S{\'e}gransan}, D. and {Skillen}, I. and {Smalley}, B. and {Smith}, A.~M.~S. and {Southworth}, J. and {Triaud}, A.~H.~M.~J. and {Turner}, O.~D. and {Udry}, S. and {West}, R.~G.},
        title = "{Three newly discovered sub-Jupiter-mass planets: WASP-69b and WASP-84b transit active K dwarfs and WASP-70Ab transits the evolved primary of a G4+K3 binary}",
      journal = {\mnras},
         year = 2014,
        month = dec,
       volume = {445},
       number = {2},
        pages = {1114-1129},
          doi = {10.1093/mnras/stu1737},
archivePrefix = {arXiv},
       eprint = {1310.5654},
 primaryClass = {astro-ph.EP},
       adsurl = {https://ui.adsabs.harvard.edu/abs/2014MNRAS.445.1114A}
}

@ARTICLE{Magic2015,
       author = {{Magic}, Z. and {Chiavassa}, A. and {Collet}, R. and {Asplund}, M.},
        title = "{The Stagger-grid: A grid of 3D stellar atmosphere models. IV. Limb darkening coefficients}",
      journal = {\aap},
         year = 2015,
        month = jan,
       volume = {573},
          eid = {A90},
        pages = {A90},
          doi = {10.1051/0004-6361/201423804},
archivePrefix = {arXiv},
       eprint = {1403.3487},
 primaryClass = {astro-ph.SR},
       adsurl = {https://ui.adsabs.harvard.edu/abs/2015A&A...573A..90M}
}

@ARTICLE{catwoman2020,
       author = {{Jones}, Kathryn and {Espinoza}, N{\'e}stor},
        title = "{catwoman: A transit modelling Python package for asymmetric light curves}",
      journal = {The Journal of Open Source Software},
         year = 2020,
        month = nov,
       volume = {5},
       number = {55},
          eid = {2382},
        pages = {2382},
          doi = {10.21105/joss.02382},
archivePrefix = {arXiv},
       eprint = {2106.15643},
 primaryClass = {astro-ph.IM},
       adsurl = {https://ui.adsabs.harvard.edu/abs/2020JOSS....5.2382J}
}

@ARTICLE{catwoman2021,
       author = {{Espinoza}, N{\'e}stor and {Jones}, Kathryn},
        title = "{Constraining Mornings and Evenings on Distant Worlds: A new Semianalytical Approach and Prospects with Transmission Spectroscopy}",
      journal = {\aj},
         year = 2021,
        month = oct,
       volume = {162},
       number = {4},
          eid = {165},
        pages = {165},
          doi = {10.3847/1538-3881/ac134d},
archivePrefix = {arXiv},
       eprint = {2106.15687},
 primaryClass = {astro-ph.EP},
       adsurl = {https://ui.adsabs.harvard.edu/abs/2021AJ....162..165E}
}

@article{Nortmann2018,
  author  = {Nortmann, Lisa and Pall{\'e}, Enric and Salz, Michael and Sanz-Forcada, Jorge and Nagel, Evangelos and Alonso-Floriano, F. Javier and Czesla, Stefan and Yan, Fei and Chen, Guo and Snellen, Ignas A. G. and Zechmeister, Mathias and Schmitt, J{\"u}rgen H. M. M. and L{\'o}pez-Puertas, Manuel and Casasayas-Barris, N{\'u}ria and Bauer, Florian F. and Amado, Pedro J. and Caballero, Jos{\'e} A. and Dreizler, Stefan and Henning, Thomas and Lamp{\'o}n, Manuel and Montes, David and Molaverdikhani, Karan and Quirrenbach, Andreas and Reiners, Ansgar and Ribas, Ignasi and S{\'a}nchez-L{\'o}pez, Alejandro and Schneider, P. Christian and Zapatero Osorio, Mar{\'i}a R.},
  title   = {Ground-based detection of an extended helium atmosphere in the Saturn-mass exoplanet WASP-69b},
  journal = {Science},
  year    = {2018},
  volume  = {362},
  pages   = {1388--1391},
  doi     = {10.1126/science.aat5348}
}

@article{Vissapragada2020,
  author  = {Vissapragada, Shreyas and Knutson, Heather A. and Jovanovic, Nemanja and Harada, Caleb K. and Oklop{\v c}i{\'c}, Antonija and Eriksen, James and Mawet, Dimitri and Millar-Blanchaer, Maxwell A. and Tinyanont, Samaporn and Vasisht, Gautam},
  title   = {Constraints on Metastable Helium in the Atmospheres of WASP-69b and WASP-52b with Ultranarrowband Photometry},
  journal = {AJ},
  year    = {2020},
  volume  = {159},
  number  = {6},
  pages   = {278},
  doi     = {10.3847/1538-3881/ab8e34}
}

@article{Allart2023,
  author  = {Allart, Romain and Lem{\'e}e-Joliecoeur, Pierre-Benoit and Jaziri, Amina Y. and Bourrier, Vincent and Lovis, Christophe and Ehrenreich, David and Bonfils, Xavier and Cadieux, Charles and Carmona, Andr{\'e}s and Cook, Neil J. and Delisle, Jean-Baptiste and Doyon, Ren{\'e} and Moutou, Claire and Piaulet-Ghorayeb, Caroline and Vandal, Thomas and Artigau, {\'E}tienne},
  title   = {Homogeneous search for helium in the atmosphere of 11 gas giant exoplanets with SPIRou},
  journal = {A\&A},
  year    = {2023},
  volume  = {677},
  pages   = {A164},
  doi     = {10.1051/0004-6361/202245832}
}

@article{Levine2024,
  author  = {Levine, W. Garrett and Vissapragada, Shreyas and Feinstein, Adina D. and King, George W. and Hernandez, Aleck and Corrales, L{\'i}a and Greklek-McKeon, Michael and Knutson, Heather A.},
  title   = {Exoplanet Aeronomy: A Case Study of WASP-69 b's Variable Thermosphere},
  journal = {AJ},
  year    = {2024},
  volume  = {168},
  number  = {2},
  pages   = {65},
  doi     = {10.3847/1538-3881/ad5354}
}

@article{Guilluy2024,
  author  = {Guilluy, G. and D'Arpa, M. C. and Bonomo, A. S. and Spinelli, R. and Biassoni, F. and Fossati, L. and Maggio, A. and Giacobbe, P. and Lanza, A. F. and Sozzetti, A. and Borsa, F. and Rainer, M. and Micela, G. and Affer, L. and Andreuzzi, G. and Bignamini, A. and Boschin, W. and Carleo, I. and Cecconi, S. and Desidera, S. and Fardella, V. and Ghedina, A. and Mantovan, G. and Mancini, L. and Nascimbeni, V. and Knapic, C. and Pedani, M. and Petralia, A. and Pino, L. and Scandariato, D. and Sicilia, D. and Stangret, M. and Zingales, T.},
  title   = {The GAPS Programme at TNG. LIV. A He I survey of close-in giant planets hosted by M-K dwarf stars with GIANO-B},
  journal = {A\&A},
  year    = {2024},
  volume  = {686},
  pages   = {A83},
  doi     = {10.1051/0004-6361/202348997}
}

@article{Masson2024,
  author  = {Masson, A. and Vinatier, S. and B{\'e}zard, B. and L{\'o}pez-Puertas, M. and Lamp{\'o}n, M. and Debras, F. and Carmona, A. and Klein, B. and Artigau, {\'E}. and Dethier, W. and Pelletier, S. and Hood, T. and Allart, R. and Bourrier, V. and Cadieux, C. and Charnay, B. and Cowan, N. B. and Cook, N. J. and Delfosse, X. and Donati, J.-F. and Gu, P.-G. and H{\'e}brard, G. and Martioli, E. and Moutou, C. and Venot, O. and Wyttenbach, A.},
  title   = {Probing atmospheric escape through metastable He I triplet lines in 15 exoplanets observed with SPIRou},
  journal = {A\&A},
  year    = {2024},
  volume  = {688},
  pages   = {A179},
  doi     = {10.1051/0004-6361/202449608}
}

@article{Allart2025,
  author  = {Allart, Romain and Carteret, Yann and Bourrier, Vincent and Mignon, Lucile and Baron, Fr{\'e}d{\'e}rique and Cadieux, Charles and Carmona, Andr{\'e}s and Lovis, Christophe and Chakraborty, Hritam and Delgado-Mena, Elisa and Artigau, {\'E}tienne and others},
  title   = {NIRPS detection of delayed atmospheric escape from the warm and misaligned Saturn-mass exoplanet WASP-69 b},
  journal = {A\&A},
  year    = {2025},
  volume  = {700},
  pages   = {A7},
  doi     = {10.1051/0004-6361/202452525}
}

@ARTICLE{Tyler2024,
       author = {{Tyler}, Dakotah and {Petigura}, Erik A. and {Oklop{\v{c}}i{\'c}}, Antonija and {David}, Trevor J.},
        title = "{WASP-69b's Escaping Envelope Is Confined to a Tail Extending at Least 7 R$_{p}$}",
      journal = {\apj},
         year = 2024,
        month = jan,
       volume = {960},
       number = {2},
          eid = {123},
        pages = {123},
          doi = {10.3847/1538-4357/ad11d0},
archivePrefix = {arXiv},
       eprint = {2312.02381},
 primaryClass = {astro-ph.EP},
       adsurl = {https://ui.adsabs.harvard.edu/abs/2024ApJ...960..123T}
}

@ARTICLE{Wood14,
       author = {{Wood}, Nigel and {Staniforth}, Andrew and {White}, Andy and {Allen}, Thomas and {Diamantakis}, Michail and {Gross}, Markus and {Melvin}, Thomas and {Smith}, Chris and {Vosper}, Simon and {Zerroukat}, Mohamed and {Thuburn}, John},
        title = "{An inherently mass-conserving semi-implicit semi-Lagrangian discretization of the deep-atmosphere global non-hydrostatic equations}",
      journal = {Quarterly Journal of the Royal Meteorological Society},
         year = 2014,
        month = jul,
       volume = {140},
       number = {682},
        pages = {1505-1520},
          doi = {10.1002/qj.2235},
       adsurl = {https://ui.adsabs.harvard.edu/abs/2014QJRMS.140.1505W}
}

@ARTICLE{Mayne14a,
       author = {{Mayne}, N.~J. and {Baraffe}, I. and {Acreman}, D.~M. and {Smith}, C. and {Wood}, N. and {Amundsen}, D.~S. and {Thuburn}, J. and {Jackson}, D.~R.},
        title = "{Using the UM dynamical cores to reproduce idealised 3-D flows}",
      journal = {Geoscientific Model Development},
         year = 2014,
        month = dec,
       volume = {7},
       number = {6},
        pages = {3059-3087},
          doi = {10.5194/gmd-7-3059-2014},
archivePrefix = {arXiv},
       eprint = {1310.6041},
 primaryClass = {astro-ph.EP},
       adsurl = {https://ui.adsabs.harvard.edu/abs/2014GMD.....7.3059M}
}

@ARTICLE{Mayne14b,
       author = {{Mayne}, Nathan J. and {Baraffe}, Isabelle and {Acreman}, David M. and {Smith}, Chris and {Browning}, Matthew K. and {Sk{\r{a}}lid Amundsen}, David and {Wood}, Nigel and {Thuburn}, John and {Jackson}, David R.},
        title = "{The unified model, a fully-compressible, non-hydrostatic, deep atmosphere global circulation model, applied to hot Jupiters. ENDGame for a HD 209458b test case}",
      journal = {\aap},
         year = 2014,
        month = jan,
       volume = {561},
          eid = {A1},
        pages = {A1},
          doi = {10.1051/0004-6361/201322174},
       adsurl = {https://ui.adsabs.harvard.edu/abs/2014A&A...561A...1M}
}

@ARTICLE{Mayne17,
       author = {{Mayne}, Nathan J. and {Debras}, Florian and {Baraffe}, Isabelle and {Thuburn}, John and {Amundsen}, David S. and {Acreman}, David M. and {Smith}, Chris and {Browning}, Matthew K. and {Manners}, James and {Wood}, Nigel},
        title = "{Results from a set of three-dimensional numerical experiments of a hot Jupiter atmosphere}",
      journal = {\aap},
         year = 2017,
        month = aug,
       volume = {604},
          eid = {A79},
        pages = {A79},
          doi = {10.1051/0004-6361/201730465},
archivePrefix = {arXiv},
       eprint = {1704.00539},
 primaryClass = {astro-ph.EP},
       adsurl = {https://ui.adsabs.harvard.edu/abs/2017A&A...604A..79M}
}

@ARTICLE{Christie21,
       author = {{Christie}, D.~A. and {Mayne}, N.~J. and {Lines}, S. and {Parmentier}, V. and {Manners}, J. and {Boutle}, I. and {Drummond}, B. and {Mikal-Evans}, T. and {Sing}, D.~K. and {Kohary}, K.},
        title = "{The impact of mixing treatments on cloud modelling in 3D simulations of hot Jupiters}",
      journal = {\mnras},
         year = 2021,
        month = sep,
       volume = {506},
       number = {3},
        pages = {4500-4515},
          doi = {10.1093/mnras/stab2027},
archivePrefix = {arXiv},
       eprint = {2107.05732},
 primaryClass = {astro-ph.EP},
       adsurl = {https://ui.adsabs.harvard.edu/abs/2021MNRAS.506.4500C}
}

@ARTICLE{Zamyatina23,
       author = {{Zamyatina}, Maria and {H{\'e}brard}, Eric and {Drummond}, Benjamin and {Mayne}, Nathan J. and {Manners}, James and {Christie}, Duncan A. and {Tremblin}, Pascal and {Sing}, David K. and {Kohary}, Krisztian},
        title = "{Observability of signatures of transport-induced chemistry in clear atmospheres of hot gas giant exoplanets}",
      journal = {\mnras},
         year = 2023,
        month = feb,
       volume = {519},
       number = {2},
        pages = {3129-3153},
          doi = {10.1093/mnras/stac3432},
archivePrefix = {arXiv},
       eprint = {2211.09071},
 primaryClass = {astro-ph.EP},
       adsurl = {https://ui.adsabs.harvard.edu/abs/2023MNRAS.519.3129Z}
}

@ARTICLE{Zamyatina24,
       author = {{Zamyatina}, Maria and {Christie}, Duncan A. and {H{\'e}brard}, Eric and {Mayne}, Nathan J. and {Radica}, Michael and {Taylor}, Jake and {Baskett}, Harry and {Moore}, Ben and {Lils}, Craig and {Sergeev}, Denis and {Ahrer}, Eva-Maria and {Manners}, James and {Kohary}, Krisztian and {Feinstein}, Adina D.},
        title = "{Quenching-driven equatorial depletion and limb asymmetries in hot Jupiter atmospheres: WASP-96b example}",
      journal = {arXiv e-prints},
         year = 2024,
        month = feb,
          eid = {arXiv:2402.14535},
        pages = {arXiv:2402.14535},
          doi = {10.48550/arXiv.2402.14535},
archivePrefix = {arXiv},
       eprint = {2402.14535},
 primaryClass = {astro-ph.EP},
       adsurl = {https://ui.adsabs.harvard.edu/abs/2024arXiv240214535Z}
}

@ARTICLE{Mak25,
       author = {{Mak}, Mei Ting and {Sergeev}, Denis E. and {Mayne}, Nathan J. and {Zamyatina}, Maria and {Steinrueck}, Maria E. and {Manners}, James and {H{\'e}brard}, {\'E}ric and {Sing}, David K. and {Kohary}, Krisztian},
        title = "{The impact of different haze types on the atmospheres and observations of hot Jupiters: 3D simulations of HD 189733b, HD 209458b, and WASP-39b}",
      journal = {\mnras},
         year = 2025,
        month = sep,
       volume = {542},
       number = {3},
        pages = {1873-1900},
          doi = {10.1093/mnras/staf1250},
archivePrefix = {arXiv},
       eprint = {2507.20366},
 primaryClass = {astro-ph.EP},
       adsurl = {https://ui.adsabs.harvard.edu/abs/2025MNRAS.542.1873M}
}

@ARTICLE{Edwards_and_Slingo_1996,
       author = {{Edwards}, J. M. and {Slingo}, A.},
        title = "{Studies with a flexible new radiation code. I: Choosing a configuration for a large-scale mode}",
      journal = {Royal Meteorological Society},
         year = 1996,
        month = nov,
       volume = {122},
       number = {A},
        pages = {689-719},
          doi = {10.1089/ast.2015.1422},
archivePrefix = {arXiv},
       eprint = {1610.04515},
 primaryClass = {astro-ph.EP},
       adsurl = {https://ui.adsabs.harvard.edu/abs/2016AsBio..16..873A}
}

@ARTICLE{Christie24,
       author = {{Christie}, D.~A. and {Mayne}, N.~J. and {Zamyatina}, M. and {Baskett}, H. and {Evans-Soma}, T.~M. and {Wood}, N. and {Kohary}, K.},
        title = "{Longitudinal filtering, sponge layers, and equatorial jet formation in a general circulation model of gaseous exoplanets}",
      journal = {\mnras},
         year = 2024,
        month = aug,
       volume = {532},
       number = {3},
        pages = {3001-3019},
          doi = {10.1093/mnras/stae1408},
archivePrefix = {arXiv},
       eprint = {2406.02231},
 primaryClass = {astro-ph.EP},
       adsurl = {https://ui.adsabs.harvard.edu/abs/2024MNRAS.532.3001C}
}

@ARTICLE{Drummond20,
       author = {{Drummond}, Benjamin and {H{\'e}brard}, Eric and {Mayne}, Nathan J. and {Venot}, Olivia and {Ridgway}, Robert J. and {Changeat}, Quentin and {Tsai}, Shang-Min and {Manners}, James and {Tremblin}, Pascal and {Abraham}, Nathan Luke and {Sing}, David and {Kohary}, Krisztian},
        title = "{Implications of three-dimensional chemical transport in hot Jupiter atmospheres: Results from a consistently coupled chemistry-radiation-hydrodynamics model}",
      journal = {\aap},
         year = 2020,
        month = apr,
       volume = {636},
          eid = {A68},
        pages = {A68},
          doi = {10.1051/0004-6361/201937153},
archivePrefix = {arXiv},
       eprint = {2001.11444},
 primaryClass = {astro-ph.EP},
       adsurl = {https://ui.adsabs.harvard.edu/abs/2020A&A...636A..68D}
}

@ARTICLE{Venot19,
       author = {{Venot}, O. and {Bounaceur}, R. and {Dobrijevic}, M. and {H{\'e}brard}, E. and {Cavali{\'e}}, T. and {Tremblin}, P. and {Drummond}, B. and {Charnay}, B.},
        title = "{Reduced chemical scheme for modelling warm to hot hydrogen-dominated atmospheres}",
      journal = {\aap},
         year = 2019,
        month = apr,
       volume = {624},
          eid = {A58},
        pages = {A58},
          doi = {10.1051/0004-6361/201834861},
archivePrefix = {arXiv},
       eprint = {1902.04939},
 primaryClass = {astro-ph.EP},
       adsurl = {https://ui.adsabs.harvard.edu/abs/2019A&A...624A..58V}
}

@ARTICLE{gascon25,
       author = {{Gasc{\'o}n}, Carlos and {L{\'o}pez-Morales}, Mercedes and {Vissapragada}, Shreyas and {MacLeod}, Morgan and {Wakeford}, Hannah R. and {Grant}, David and {Ribas}, Ignasi and {Anglada-Escud{\'e}}, Guillem},
        title = "{Modeling Tails of Escaping Gas in Exoplanet Atmospheres with Harmonica}",
      journal = {\apjl},
         year = 2025,
        month = oct,
       volume = {991},
       number = {2},
          eid = {L47},
        pages = {L47},
          doi = {10.3847/2041-8213/adfb77},
archivePrefix = {arXiv},
       eprint = {2508.14846},
 primaryClass = {astro-ph.EP},
       adsurl = {https://ui.adsabs.harvard.edu/abs/2025ApJ...991L..47G}
}

@ARTICLE{foreman13,
       author = {{Foreman-Mackey}, Daniel and {Hogg}, David W. and {Lang}, Dustin and {Goodman}, Jonathan},
        title = "{emcee: The MCMC Hammer}",
      journal = {\pasp},
         year = 2013,
        month = mar,
       volume = {125},
       number = {925},
        pages = {306},
          doi = {10.1086/670067},
archivePrefix = {arXiv},
       eprint = {1202.3665},
 primaryClass = {astro-ph.IM},
       adsurl = {https://ui.adsabs.harvard.edu/abs/2013PASP..125..306F}
}

@article{grant22,
  title={Transmission strings: a technique for spatially mapping exoplanet atmospheres around their terminators},
  author={Grant, David and Wakeford, Hannah R},
  journal={Monthly Notices of the Royal Astronomical Society},
  volume={519},
  number={4},
  pages={5114--5127},
  year={2023},
  publisher={Oxford University Press}
}

@ARTICLE{thorngren2026,
       author = {{Thorngren}, Daniel P. and {Sing}, David K. and {Mukherjee}, Sagnick},
        title = "{Bayesian Model Comparison and Significance: Widespread Errors and How to Correct Them}",
      journal = {\apjs},
         year = 2026,
        month = mar,
       volume = {283},
       number = {1},
          eid = {10},
        pages = {10},
          doi = {10.3847/1538-4365/ae0e71},
archivePrefix = {arXiv},
       eprint = {2510.00169},
 primaryClass = {astro-ph.EP},
       adsurl = {https://ui.adsabs.harvard.edu/abs/2026ApJS..283...10T}
}

@ARTICLE{mccreery2025,
       author = {{McCreery}, Patrick and {Dos Santos}, Leonardo A. and {Espinoza}, N{\'e}stor and {Allart}, Romain and {Kirk}, James},
        title = "{Tracing the Winds: A Uniform Interpretation of Helium Escape in Exoplanets from Archival Spectroscopic Observations}",
      journal = {\apj},
         year = 2025,
        month = feb,
       volume = {980},
       number = {1},
          eid = {125},
        pages = {125},
          doi = {10.3847/1538-4357/ada6b9},
archivePrefix = {arXiv},
       eprint = {2501.03998},
 primaryClass = {astro-ph.EP},
       adsurl = {https://ui.adsabs.harvard.edu/abs/2025ApJ...980..125M}
}

@ARTICLE{dossantos22,
       author = {{Dos Santos}, Leonardo A. and {Vidotto}, Aline A. and {Vissapragada}, Shreyas and {Alam}, Munazza K. and {Allart}, Romain and {Bourrier}, Vincent and {Kirk}, James and {Seidel}, Julia V. and {Ehrenreich}, David},
        title = "{p-winds: An open-source Python code to model planetary outflows and upper atmospheres}",
      journal = {\aap},
         year = 2022,
        month = mar,
       volume = {659},
          eid = {A62},
        pages = {A62},
          doi = {10.1051/0004-6361/202142038},
archivePrefix = {arXiv},
       eprint = {2111.11370},
 primaryClass = {astro-ph.EP},
       adsurl = {https://ui.adsabs.harvard.edu/abs/2022A&A...659A..62D}
}

@ARTICLE{musclesI,
       author = {{France}, Kevin and {Loyd}, R.~O. Parke and {Youngblood}, Allison and {Brown}, Alexander and {Schneider}, P. Christian and {Hawley}, Suzanne L. and {Froning}, Cynthia S. and {Linsky}, Jeffrey L. and {Roberge}, Aki and {Buccino}, Andrea P. and {Davenport}, James R.~A. and {Fontenla}, Juan M. and {Kaltenegger}, Lisa and {Kowalski}, Adam F. and {Mauas}, Pablo J.~D. and {Miguel}, Yamila and {Redfield}, Seth and {Rugheimer}, Sarah and {Tian}, Feng and {Vieytes}, Mariela C. and {Walkowicz}, Lucianne M. and {Weisenburger}, Kolby L.},
        title = "{The MUSCLES Treasury Survey. I. Motivation and Overview}",
      journal = {\apj},
         year = 2016,
        month = apr,
       volume = {820},
       number = {2},
          eid = {89},
        pages = {89},
          doi = {10.3847/0004-637X/820/2/89},
archivePrefix = {arXiv},
       eprint = {1602.09142},
 primaryClass = {astro-ph.SR},
       adsurl = {https://ui.adsabs.harvard.edu/abs/2016ApJ...820...89F}
}

@ARTICLE{musclesII,
       author = {{Youngblood}, Allison and {France}, Kevin and {Loyd}, R.~O. Parke and {Linsky}, Jeffrey L. and {Redfield}, Seth and {Schneider}, P. Christian and {Wood}, Brian E. and {Brown}, Alexander and {Froning}, Cynthia and {Miguel}, Yamila and {Rugheimer}, Sarah and {Walkowicz}, Lucianne},
        title = "{The MUSCLES Treasury Survey. II. Intrinsic LY{\ensuremath{\alpha}} and Extreme Ultraviolet Spectra of K and M Dwarfs with Exoplanets*}",
      journal = {\apj},
         year = 2016,
        month = jun,
       volume = {824},
       number = {2},
          eid = {101},
        pages = {101},
          doi = {10.3847/0004-637X/824/2/101},
archivePrefix = {arXiv},
       eprint = {1604.01032},
 primaryClass = {astro-ph.SR},
       adsurl = {https://ui.adsabs.harvard.edu/abs/2016ApJ...824..101Y}
}

@ARTICLE{musclesIII,
       author = {{Loyd}, R.~O.~P. and {France}, Kevin and {Youngblood}, Allison and {Schneider}, Christian and {Brown}, Alexander and {Hu}, Renyu and {Linsky}, Jeffrey and {Froning}, Cynthia S. and {Redfield}, Seth and {Rugheimer}, Sarah and {Tian}, Feng},
        title = "{The MUSCLES Treasury Survey. III. X-Ray to Infrared Spectra of 11 M and K Stars Hosting Planets}",
      journal = {\apj},
         year = 2016,
        month = jun,
       volume = {824},
       number = {2},
          eid = {102},
        pages = {102},
          doi = {10.3847/0004-637X/824/2/102},
archivePrefix = {arXiv},
       eprint = {1604.04776},
 primaryClass = {astro-ph.SR},
       adsurl = {https://ui.adsabs.harvard.edu/abs/2016ApJ...824..102L}
}

@ARTICLE{musclesIV,
       author = {{Wilson}, David J. and {Froning}, Cynthia S. and {Duvvuri}, Girish M. and {France}, Kevin and {Youngblood}, Allison and {Schneider}, P. Christian and {Berta-Thompson}, Zachory and {Brown}, Alexander and {Buccino}, Andrea P. and {Hawley}, Suzanne and {Irwin}, Jonathan and {Kaltenegger}, Lisa and {Kowalski}, Adam and {Linsky}, Jeffrey and {Loyd}, R.~O. Parke and {Miguel}, Yamila and {Pineda}, J. Sebastian and {Redfield}, Seth and {Roberge}, Aki and {Rugheimer}, Sarah and {Tian}, Feng and {Vieytes}, Mariela},
        title = "{The Mega-MUSCLES Spectral Energy Distribution of TRAPPIST-1}",
      journal = {\apj},
         year = 2021,
        month = apr,
       volume = {911},
       number = {1},
          eid = {18},
        pages = {18},
          doi = {10.3847/1538-4357/abe771},
archivePrefix = {arXiv},
       eprint = {2102.11415},
 primaryClass = {astro-ph.SR},
       adsurl = {https://ui.adsabs.harvard.edu/abs/2021ApJ...911...18W}
}

@ARTICLE{musclesV,
       author = {{Behr}, Patrick R. and {France}, Kevin and {Brown}, Alexander and {Duvvuri}, Girish and {Bean}, Jacob L. and {Berta-Thompson}, Zachory and {Froning}, Cynthia and {Miguel}, Yamila and {Pineda}, J. Sebastian and {Wilson}, David J. and {Youngblood}, Allison},
        title = "{The MUSCLES Extension for Atmospheric Transmission Spectroscopy: UV and X-Ray Host-star Observations for JWST ERS \& GTO Targets}",
      journal = {\aj},
         year = 2023,
        month = jul,
       volume = {166},
       number = {1},
          eid = {35},
        pages = {35},
          doi = {10.3847/1538-3881/acdb70},
archivePrefix = {arXiv},
       eprint = {2306.05322},
 primaryClass = {astro-ph.SR},
       adsurl = {https://ui.adsabs.harvard.edu/abs/2023AJ....166...35B}
}

@ARTICLE{Salz2016a,
       author = {{Salz}, M. and {Schneider}, P.~C. and {Czesla}, S. and {Schmitt}, J.~H.~M.~M.},
        title = "{Energy-limited escape revised. The transition from strong planetary winds to stable thermospheres}",
      journal = {\aap},
         year = 2016,
        month = jan,
       volume = {585},
          eid = {L2},
        pages = {L2},
          doi = {10.1051/0004-6361/201527042},
archivePrefix = {arXiv},
       eprint = {1511.09348},
 primaryClass = {astro-ph.EP},
       adsurl = {https://ui.adsabs.harvard.edu/abs/2016A\&A...585L...2S}
}

@ARTICLE{Rumenskikh2022,
       author = {{Rumenskikh}, M.~S. and {Shaikhislamov}, I.~F. and {Khodachenko}, M.~L. and {Lammer}, H. and {Miroshnichenko}, I.~B. and {Berezutsky}, A.~G. and {Fossati}, L.},
        title = "{Global 3D Simulation of the Upper Atmosphere of HD189733b and Absorption in Metastable He I and Ly{\ensuremath{\alpha}} Lines}",
      journal = {\apj},
         year = 2022,
        month = mar,
       volume = {927},
       number = {2},
          eid = {238},
        pages = {238},
          doi = {10.3847/1538-4357/ac441d},
archivePrefix = {arXiv},
       eprint = {2205.01341},
 primaryClass = {astro-ph.EP},
       adsurl = {https://ui.adsabs.harvard.edu/abs/2022ApJ...927..238R}
}

@ARTICLE{Yan2024,
       author = {{Yan}, Dongdong and {Guo}, Jianheng and {Seon}, Kwang-il and {L{\'o}pez-Puertas}, Manuel and {Czesla}, Stefan and {Lamp{\'o}n}, Manuel},
        title = "{A possibly solar metallicity atmosphere escaping from HAT-P-32b revealed by H{\ensuremath{\alpha}} and He absorption}",
      journal = {\aap},
         year = 2024,
        month = jun,
       volume = {686},
          eid = {A208},
        pages = {A208},
          doi = {10.1051/0004-6361/202348210},
archivePrefix = {arXiv},
       eprint = {2403.17325},
 primaryClass = {astro-ph.EP},
       adsurl = {https://ui.adsabs.harvard.edu/abs/2024A\&A...686A.208Y}
}

@ARTICLE{schlawin2024,
       author = {{Schlawin}, Everett and {Mukherjee}, Sagnick and {Ohno}, Kazumasa and {Bell}, Taylor J. and {Beatty}, Thomas G. and {Greene}, Thomas P. and {Line}, Michael and {Challener}, Ryan C. and {Parmentier}, Vivien and {Fortney}, Jonathan J. and {Rauscher}, Emily and {Wiser}, Lindsey and {Welbanks}, Luis and {Murphy}, Matthew and {Edelman}, Isaac and {Batalha}, Natasha and {Moran}, Sarah E. and {Mehta}, Nishil and {Rieke}, Marcia},
        title = "{Multiple Clues for Dayside Aerosols and Temperature Gradients in WASP-69 b from a Panchromatic JWST Emission Spectrum}",
      journal = {\aj},
         year = 2024,
        month = sep,
       volume = {168},
       number = {3},
          eid = {104},
        pages = {104},
          doi = {10.3847/1538-3881/ad58e0},
archivePrefix = {arXiv},
       eprint = {2406.15543},
 primaryClass = {astro-ph.EP},
       adsurl = {https://ui.adsabs.harvard.edu/abs/2024AJ....168..104S}
}

@ARTICLE{skymapper,
       author = {{Onken}, Christopher A. and {Wolf}, Christian and {Bessell}, Michael S. and {Chang}, Seo-Won and {Luvaul}, Lance C. and {Tonry}, John L. and {White}, Marc C. and {Da Costa}, Gary S.},
        title = "{SkyMapper Southern Survey: Data release 4}",
      journal = {\pasa},
         year = 2024,
        month = oct,
       volume = {41},
          eid = {e061},
        pages = {e061},
          doi = {10.1017/pasa.2024.53},
archivePrefix = {arXiv},
       eprint = {2402.02015},
 primaryClass = {astro-ph.CO},
       adsurl = {https://ui.adsabs.harvard.edu/abs/2024PASA...41...61O}
}

@MISC{mor15,
   author = {{Morton}, T.~D.},
    title = "{isochrones: Stellar model grid package}",
howpublished = {Astrophysics Source Code Library},
     year = 2015,
    month = mar,
      url = {https://isochrones.readthedocs.io/en/latest/},
archivePrefix = "ascl",
   eprint = {1503.010},
   adsurl = {https://ui.adsabs.harvard.edu/abs/2015ascl.soft03010M}
}

@ARTICLE{pax11,
       author = {{Paxton}, Bill and {Bildsten}, Lars and {Dotter}, Aaron and
         {Herwig}, Falk and {Lesaffre}, Pierre and {Timmes}, Frank},
        title = "{Modules for Experiments in Stellar Astrophysics (MESA)}",
      journal = {\apjs},
         year = 2011,
        month = jan,
       volume = {192},
       number = {1},
          eid = {3},
        pages = {3},
          doi = {10.1088/0067-0049/192/1/3},
archivePrefix = {arXiv},
       eprint = {1009.1622},
 primaryClass = {astro-ph.SR},
       adsurl = {https://ui.adsabs.harvard.edu/abs/2011ApJS..192....3P}
}

@ARTICLE{pax13,
       author = {{Paxton}, Bill and {Cantiello}, Matteo and {Arras}, Phil and
         {Bildsten}, Lars and {Brown}, Edward F. and {Dotter}, Aaron and
         {Mankovich}, Christopher and {Montgomery}, M.~H. and {Stello}, Dennis and
         {Timmes}, F.~X. and {Townsend}, Richard},
        title = "{Modules for Experiments in Stellar Astrophysics (MESA): Planets, Oscillations, Rotation, and Massive Stars}",
      journal = {\apjs},
         year = 2013,
        month = sep,
       volume = {208},
       number = {1},
          eid = {4},
        pages = {4},
          doi = {10.1088/0067-0049/208/1/4},
archivePrefix = {arXiv},
       eprint = {1301.0319},
 primaryClass = {astro-ph.SR},
       adsurl = {https://ui.adsabs.harvard.edu/abs/2013ApJS..208....4P}
}

@ARTICLE{pax18,
       author = {{Paxton}, Bill and {Schwab}, Josiah and {Bauer}, Evan B. and
         {Bildsten}, Lars and {Blinnikov}, Sergei and {Duffell}, Paul and
         {Farmer}, R. and {Goldberg}, Jared A. and {Marchant}, Pablo and
         {Sorokina}, Elena and {Thoul}, Anne and {Townsend}, Richard H.~D. and
         {Timmes}, F.~X.},
        title = "{Modules for Experiments in Stellar Astrophysics (MESA): Convective Boundaries, Element Diffusion, and Massive Star Explosions}",
      journal = {\apjs},
         year = 2018,
        month = feb,
       volume = {234},
       number = {2},
          eid = {34},
        pages = {34},
          doi = {10.3847/1538-4365/aaa5a8},
archivePrefix = {arXiv},
       eprint = {1710.08424},
 primaryClass = {astro-ph.SR},
       adsurl = {https://ui.adsabs.harvard.edu/abs/2018ApJS..234...34P}
}

@ARTICLE{pax19,
       author = {{Paxton}, Bill and {Smolec}, R. and {Schwab}, Josiah and {Gautschy}, A. and {Bildsten}, Lars and {Cantiello}, Matteo and {Dotter}, Aaron and {Farmer}, R. and {Goldberg}, Jared A. and {Jermyn}, Adam S. and {Kanbur}, S.~M. and {Marchant}, Pablo and {Thoul}, Anne and {Townsend}, Richard H.~D. and {Wolf}, William M. and {Zhang}, Michael and {Timmes}, F.~X.},
        title = "{Modules for Experiments in Stellar Astrophysics (MESA): Pulsating Variable Stars, Rotation, Convective Boundaries, and Energy Conservation}",
      journal = {\apjs},
         year = 2019,
        month = jul,
       volume = {243},
       number = {1},
          eid = {10},
        pages = {10},
          doi = {10.3847/1538-4365/ab2241},
archivePrefix = {arXiv},
       eprint = {1903.01426},
 primaryClass = {astro-ph.SR},
       adsurl = {https://ui.adsabs.harvard.edu/abs/2019ApJS..243...10P}
}

@ARTICLE{jer23,
       author = {{Jermyn}, Adam S. and {Bauer}, Evan B. and {Schwab}, Josiah and {Farmer}, R. and {Ball}, Warrick H. and {Bellinger}, Earl P. and {Dotter}, Aaron and {Joyce}, Meridith and {Marchant}, Pablo and {Mombarg}, Joey S.~G. and {Wolf}, William M. and {Sunny Wong}, Tin Long and {Cinquegrana}, Giulia C. and {Farrell}, Eoin and {Smolec}, R. and {Thoul}, Anne and {Cantiello}, Matteo and {Herwig}, Falk and {Toloza}, Odette and {Bildsten}, Lars and {Townsend}, Richard H.~D. and {Timmes}, F.~X.},
        title = "{Modules for Experiments in Stellar Astrophysics (MESA): Time-dependent Convection, Energy Conservation, Automatic Differentiation, and Infrastructure}",
      journal = {\apjs},
         year = 2023,
        month = mar,
       volume = {265},
       number = {1},
          eid = {15},
        pages = {15},
          doi = {10.3847/1538-4365/acae8d},
archivePrefix = {arXiv},
       eprint = {2208.03651},
 primaryClass = {astro-ph.SR},
       adsurl = {https://ui.adsabs.harvard.edu/abs/2023ApJS..265...15J}
}

@ARTICLE{dot16,
       author = {{Dotter}, Aaron},
        title = "{MESA Isochrones and Stellar Tracks (MIST) 0: Methods for the Construction of Stellar Isochrones}",
      journal = {\apjs},
         year = 2016,
        month = jan,
       volume = {222},
       number = {1},
          eid = {8},
        pages = {8},
          doi = {10.3847/0067-0049/222/1/8},
archivePrefix = {arXiv},
       eprint = {1601.05144},
 primaryClass = {astro-ph.SR},
       adsurl = {https://ui.adsabs.harvard.edu/abs/2016ApJS..222....8D}
}

@ARTICLE{cho16,
       author = {{Choi}, Jieun and {Dotter}, Aaron and {Conroy}, Charlie and
         {Cantiello}, Matteo and {Paxton}, Bill and {Johnson}, Benjamin D.},
        title = "{Mesa Isochrones and Stellar Tracks (MIST). I. Solar-scaled Models}",
      journal = {\apj},
         year = 2016,
        month = jun,
       volume = {823},
       number = {2},
          eid = {102},
        pages = {102},
          doi = {10.3847/0004-637X/823/2/102},
archivePrefix = {arXiv},
       eprint = {1604.08592},
 primaryClass = {astro-ph.SR},
       adsurl = {https://ui.adsabs.harvard.edu/abs/2016ApJ...823..102C}
}

@ARTICLE{fer08,
       author = {{Feroz}, F. and {Hobson}, M.~P.},
        title = "{Multimodal nested sampling: an efficient and robust alternative to Markov Chain Monte Carlo methods for astronomical data analyses}",
      journal = {\mnras},
         year = 2008,
        month = feb,
       volume = {384},
       number = {2},
        pages = {449-463},
          doi = {10.1111/j.1365-2966.2007.12353.x},
archivePrefix = {arXiv},
       eprint = {0704.3704},
 primaryClass = {astro-ph},
       adsurl = {https://ui.adsabs.harvard.edu/abs/2008MNRAS.384..449F}
}

@ARTICLE{fer09,
       author = {{Feroz}, F. and {Hobson}, M.~P. and {Bridges}, M.},
        title = "{MULTINEST: an efficient and robust Bayesian inference tool for cosmology and particle physics}",
      journal = {\mnras},
         year = 2009,
        month = oct,
       volume = {398},
       number = {4},
        pages = {1601-1614},
          doi = {10.1111/j.1365-2966.2009.14548.x},
archivePrefix = {arXiv},
       eprint = {0809.3437},
 primaryClass = {astro-ph},
       adsurl = {https://ui.adsabs.harvard.edu/abs/2009MNRAS.398.1601F}
}

@ARTICLE{fer19,
       author = {{Feroz}, Farhan and {Hobson}, Michael P. and {Cameron}, Ewan and
         {Pettitt}, Anthony N.},
        title = "{Importance Nested Sampling and the MultiNest Algorithm}",
      journal = {The Open Journal of Astrophysics},
         year = 2019,
        month = nov,
       volume = {2},
       number = {1},
          eid = {10},
        pages = {10},
          doi = {10.21105/astro.1306.2144},
archivePrefix = {arXiv},
       eprint = {1306.2144},
 primaryClass = {astro-ph.IM},
       adsurl = {https://ui.adsabs.harvard.edu/abs/2019OJAp....2E..10F}
}

@ARTICLE{Reggiani22,
       author = {{Reggiani}, Henrique and {Ji}, Alexander P. and {Schlaufman}, Kevin C. and {Frebel}, Anna and {Necib}, Lina and {Nelson}, Tyler and {Hawkins}, Keith and {Galarza}, Jhon Yana},
        title = "{The Chemical Composition of Extreme-velocity Stars}",
      journal = {\aj},
         year = 2022,
        month = jun,
       volume = {163},
       number = {6},
          eid = {252},
        pages = {252},
          doi = {10.3847/1538-3881/ac62d9},
archivePrefix = {arXiv},
       eprint = {2203.16364},
 primaryClass = {astro-ph.SR},
       adsurl = {https://ui.adsabs.harvard.edu/abs/2022AJ....163..252R}
}

@ARTICLE{Reggiani24,
       author = {{Reggiani}, Henrique and {Galarza}, Jhon Yana and {Schlaufman}, Kevin C. and {Sing}, David K. and {Healy}, Brian F. and {McWilliam}, Andrew and {Lothringer}, Joshua D. and {Pueyo}, Laurent},
        title = "{Insight into the Formation of {\ensuremath{\beta}} Pic b through the Composition of Its Parent Protoplanetary Disk as Revealed by the {\ensuremath{\beta}} Pic Moving Group Member HD 181327}",
      journal = {\aj},
         year = 2024,
        month = jan,
       volume = {167},
       number = {1},
          eid = {45},
        pages = {45},
          doi = {10.3847/1538-3881/ad0f93},
archivePrefix = {arXiv},
       eprint = {2311.12210},
 primaryClass = {astro-ph.SR},
       adsurl = {https://ui.adsabs.harvard.edu/abs/2024AJ....167...45R}
}

@ARTICLE{Hamer22,
       author = {{Hamer}, Jacob H. and {Schlaufman}, Kevin C.},
        title = "{Evidence for the Late Arrival of Hot Jupiters in Systems with High Host-star Obliquities}",
      journal = {\aj},
         year = 2022,
        month = jul,
       volume = {164},
       number = {1},
          eid = {26},
        pages = {26},
          doi = {10.3847/1538-3881/ac69ef},
archivePrefix = {arXiv},
       eprint = {2205.00040},
 primaryClass = {astro-ph.EP},
       adsurl = {https://ui.adsabs.harvard.edu/abs/2022AJ....164...26H}
}

@Article{Gaia_mission2016,
  author        = {{Gaia Collaboration} and {Prusti}, T. and {de Bruijne}, J.~H.~J. and {Brown}, A.~G.~A. and {Vallenari}, A. and {Babusiaux}, C. and {Bailer-Jones}, C.~A.~L. and {Bastian}, U. and {Biermann}, M. and {Evans}, D.~W. and {Eyer}, L. and {Jansen}, F. and {Jordi}, C. and {Klioner}, S.~A. and {Lammers}, U. and {Lindegren}, L. and {Luri}, X. and {Mignard}, F. and {Milligan}, D.~J. and {Panem}, C. and {Poinsignon}, V. and {Pourbaix}, D. and {Randich}, S. and {Sarri}, G. and {Sartoretti}, P. and {Siddiqui}, H.~I. and {Soubiran}, C. and {Valette}, V. and {van Leeuwen}, F. and {Walton}, N.~A. and {Aerts}, C. and {Arenou}, F. and {Cropper}, M. and {Drimmel}, R. and {H{\o}g}, E. and {Katz}, D. and {Lattanzi}, M.~G. and {O'Mullane}, W. and {Grebel}, E.~K. and {Holland}, A.~D. and {Huc}, C. and {Passot}, X. and {Bramante}, L. and {Cacciari}, C. and {Casta{\~n}eda}, J. and {Chaoul}, L. and {Cheek}, N. and {De Angeli}, F. and {Fabricius}, C. and {Guerra}, R. and {Hern{\'a}ndez}, J. and {Jean-Antoine-Piccolo}, A. and {Masana}, E. and {Messineo}, R. and {Mowlavi}, N. and {Nienartowicz}, K. and {Ord{\'o}{\~n}ez-Blanco}, D. and {Panuzzo}, P. and {Portell}, J. and {Richards}, P.~J. and {Riello}, M. and {Seabroke}, G.~M. and {Tanga}, P. and {Th{\'e}venin}, F. and {Torra}, J. and {Els}, S.~G. and {Gracia-Abril}, G. and {Comoretto}, G. and {Garcia-Reinaldos}, M. and {Lock}, T. and {Mercier}, E. and {Altmann}, M. and {Andrae}, R. and {Astraatmadja}, T.~L. and {Bellas-Velidis}, I. and {Benson}, K. and {Berthier}, J. and {Blomme}, R. and {Busso}, G. and {Carry}, B. and {Cellino}, A. and {Clementini}, G. and {Cowell}, S. and {Creevey}, O. and {Cuypers}, J. and {Davidson}, M. and {De Ridder}, J. and {de Torres}, A. and {Delchambre}, L. and {Dell'Oro}, A. and {Ducourant}, C. and {Fr{\'e}mat}, Y. and {Garc{\'\i}a-Torres}, M. and {Gosset}, E. and {Halbwachs}, J. -L. and {Hambly}, N.~C. and {Harrison}, D.~L. and {Hauser}, M. and {Hestroffer}, D. and {Hodgkin}, S.~T. and {Huckle}, H.~E. and {Hutton}, A. and {Jasniewicz}, G. and {Jordan}, S. and {Kontizas}, M. and {Korn}, A.~J. and {Lanzafame}, A.~C. and {Manteiga}, M. and {Moitinho}, A. and {Muinonen}, K. and {Osinde}, J. and {Pancino}, E. and {Pauwels}, T. and {Petit}, J. -M. and {Recio-Blanco}, A. and {Robin}, A.~C. and {Sarro}, L.~M. and {Siopis}, C. and {Smith}, M. and {Smith}, K.~W. and {Sozzetti}, A. and {Thuillot}, W. and {van Reeven}, W. and {Viala}, Y. and {Abbas}, U. and {Abreu Aramburu}, A. and {Accart}, S. and {Aguado}, J.~J. and {Allan}, P.~M. and {Allasia}, W. and {Altavilla}, G. and {{\'A}lvarez}, M.~A. and {Alves}, J. and {Anderson}, R.~I. and {Andrei}, A.~H. and {Anglada Varela}, E. and {Antiche}, E. and {Antoja}, T. and {Ant{\'o}n}, S. and {Arcay}, B. and {Atzei}, A. and {Ayache}, L. and {Bach}, N. and {Baker}, S.~G. and {Balaguer-N{\'u}{\~n}ez}, L. and {Barache}, C. and {Barata}, C. and {Barbier}, A. and {Barblan}, F. and {Baroni}, M. and {Barrado y Navascu{\'e}s}, D. and {Barros}, M. and {Barstow}, M.~A. and {Becciani}, U. and {Bellazzini}, M. and {Bellei}, G. and {Bello Garc{\'\i}a}, A. and {Belokurov}, V. and {Bendjoya}, P. and {Berihuete}, A. and {Bianchi}, L. and {Bienaym{\'e}}, O. and {Billebaud}, F. and {Blagorodnova}, N. and {Blanco-Cuaresma}, S. and {Boch}, T. and {Bombrun}, A. and {Borrachero}, R. and {Bouquillon}, S. and {Bourda}, G. and {Bouy}, H. and {Bragaglia}, A. and {Breddels}, M.~A. and {Brouillet}, N. and {Br{\"u}semeister}, T. and {Bucciarelli}, B. and {Budnik}, F. and {Burgess}, P. and {Burgon}, R. and {Burlacu}, A. and {Busonero}, D. and {Buzzi}, R. and {Caffau}, E. and {Cambras}, J. and {Campbell}, H. and {Cancelliere}, R. and {Cantat-Gaudin}, T. and {Carlucci}, T. and {Carrasco}, J.~M. and {Castellani}, M. and {Charlot}, P. and {Charnas}, J. and {Charvet}, P. and {Chassat}, F. and {Chiavassa}, A. and {Clotet}, M. and {Cocozza}, G. and {Collins}, R.~S. and {Collins}, P. and {Costigan}, G. and {Crifo}, F. and {Cross}, N.~J.~G. and {Crosta}, M. and {Crowley}, C. and {Dafonte}, C. and {Damerdji}, Y. and {Dapergolas}, A. and {David}, P. and {David}, M. and {De Cat}, P. and {de Felice}, F. and {de Laverny}, P. and {De Luise}, F. and {De March}, R. and {de Martino}, D. and {de Souza}, R. and {Debosscher}, J. and {del Pozo}, E. and {Delbo}, M. and {Delgado}, A. and {Delgado}, H.~E. and {di Marco}, F. and {Di Matteo}, P. and {Diakite}, S. and {Distefano}, E. and {Dolding}, C. and {Dos Anjos}, S. and {Drazinos}, P. and {Dur{\'a}n}, J. and {Dzigan}, Y. and {Ecale}, E. and {Edvardsson}, B. and {Enke}, H. and {Erdmann}, M. and {Escolar}, D. and {Espina}, M. and {Evans}, N.~W. and {Eynard Bontemps}, G. and {Fabre}, C. and {Fabrizio}, M. and {Faigler}, S. and {Falc{\~a}o}, A.~J. and {Farr{\`a}s Casas}, M. and {Faye}, F. and {Federici}, L. and {Fedorets}, G. and {Fern{\'a}ndez-Hern{\'a}ndez}, J. and {Fernique}, P. and {Fienga}, A. and {Figueras}, F. and {Filippi}, F. and {Findeisen}, K. and {Fonti}, A. and {Fouesneau}, M. and {Fraile}, E. and {Fraser}, M. and {Fuchs}, J. and {Furnell}, R. and {Gai}, M. and {Galleti}, S. and {Galluccio}, L. and {Garabato}, D. and {Garc{\'\i}a-Sedano}, F. and {Gar{\'e}}, P. and {Garofalo}, A. and {Garralda}, N. and {Gavras}, P. and {Gerssen}, J. and {Geyer}, R. and {Gilmore}, G. and {Girona}, S. and {Giuffrida}, G. and {Gomes}, M. and {Gonz{\'a}lez-Marcos}, A. and {Gonz{\'a}lez-N{\'u}{\~n}ez}, J. and {Gonz{\'a}lez-Vidal}, J.~J. and {Granvik}, M. and {Guerrier}, A. and {Guillout}, P. and {Guiraud}, J. and {G{\'u}rpide}, A. and {Guti{\'e}rrez-S{\'a}nchez}, R. and {Guy}, L.~P. and {Haigron}, R. and {Hatzidimitriou}, D. and {Haywood}, M. and {Heiter}, U. and {Helmi}, A. and {Hobbs}, D. and {Hofmann}, W. and {Holl}, B. and {Holland}, G. and {Hunt}, J.~A.~S. and {Hypki}, A. and {Icardi}, V. and {Irwin}, M. and {Jevardat de Fombelle}, G. and {Jofr{\'e}}, P. and {Jonker}, P.~G. and {Jorissen}, A. and {Julbe}, F. and {Karampelas}, A. and {Kochoska}, A. and {Kohley}, R. and {Kolenberg}, K. and {Kontizas}, E. and {Koposov}, S.~E. and {Kordopatis}, G. and {Koubsky}, P. and {Kowalczyk}, A. and {Krone-Martins}, A. and {Kudryashova}, M. and {Kull}, I. and {Bachchan}, R.~K. and {Lacoste-Seris}, F. and {Lanza}, A.~F. and {Lavigne}, J. -B. and {Le Poncin-Lafitte}, C. and {Lebreton}, Y. and {Lebzelter}, T. and {Leccia}, S. and {Leclerc}, N. and {Lecoeur-Taibi}, I. and {Lemaitre}, V. and {Lenhardt}, H. and {Leroux}, F. and {Liao}, S. and {Licata}, E. and {Lindstr{\o}m}, H.~E.~P. and {Lister}, T.~A. and {Livanou}, E. and {Lobel}, A. and {L{\"o}ffler}, W. and {L{\'o}pez}, M. and {Lopez-Lozano}, A. and {Lorenz}, D. and {Loureiro}, T. and {MacDonald}, I. and {Magalh{\~a}es Fernandes}, T. and {Managau}, S. and {Mann}, R.~G. and {Mantelet}, G. and {Marchal}, O. and {Marchant}, J.~M. and {Marconi}, M. and {Marie}, J. and {Marinoni}, S. and {Marrese}, P.~M. and {Marschalk{\'o}}, G. and {Marshall}, D.~J. and {Mart{\'\i}n-Fleitas}, J.~M. and {Martino}, M. and {Mary}, N. and {Matijevi{\v{c}}}, G. and {Mazeh}, T. and {McMillan}, P.~J. and {Messina}, S. and {Mestre}, A. and {Michalik}, D. and {Millar}, N.~R. and {Miranda}, B.~M.~H. and {Molina}, D. and {Molinaro}, R. and {Molinaro}, M. and {Moln{\'a}r}, L. and {Moniez}, M. and {Montegriffo}, P. and {Monteiro}, D. and {Mor}, R. and {Mora}, A. and {Morbidelli}, R. and {Morel}, T. and {Morgenthaler}, S. and {Morley}, T. and {Morris}, D. and {Mulone}, A.~F. and {Muraveva}, T. and {Musella}, I. and {Narbonne}, J. and {Nelemans}, G. and {Nicastro}, L. and {Noval}, L. and {Ord{\'e}novic}, C. and {Ordieres-Mer{\'e}}, J. and {Osborne}, P. and {Pagani}, C. and {Pagano}, I. and {Pailler}, F. and {Palacin}, H. and {Palaversa}, L. and {Parsons}, P. and {Paulsen}, T. and {Pecoraro}, M. and {Pedrosa}, R. and {Pentik{\"a}inen}, H. and {Pereira}, J. and {Pichon}, B. and {Piersimoni}, A.~M. and {Pineau}, F. -X. and {Plachy}, E. and {Plum}, G. and {Poujoulet}, E. and {Pr{\v{s}}a}, A. and {Pulone}, L. and {Ragaini}, S. and {Rago}, S. and {Rambaux}, N. and {Ramos-Lerate}, M. and {Ranalli}, P. and {Rauw}, G. and {Read}, A. and {Regibo}, S. and {Renk}, F. and {Reyl{\'e}}, C. and {Ribeiro}, R.~A. and {Rimoldini}, L. and {Ripepi}, V. and {Riva}, A. and {Rixon}, G. and {Roelens}, M. and {Romero-G{\'o}mez}, M. and {Rowell}, N. and {Royer}, F. and {Rudolph}, A. and {Ruiz-Dern}, L. and {Sadowski}, G. and {Sagrist{\`a} Sell{\'e}s}, T. and {Sahlmann}, J. and {Salgado}, J. and {Salguero}, E. and {Sarasso}, M. and {Savietto}, H. and {Schnorhk}, A. and {Schultheis}, M. and {Sciacca}, E. and {Segol}, M. and {Segovia}, J.~C. and {Segransan}, D. and {Serpell}, E. and {Shih}, I. -C. and {Smareglia}, R. and {Smart}, R.~L. and {Smith}, C. and {Solano}, E. and {Solitro}, F. and {Sordo}, R. and {Soria Nieto}, S. and {Souchay}, J. and {Spagna}, A. and {Spoto}, F. and {Stampa}, U. and {Steele}, I.~A. and {Steidelm{\"u}ller}, H. and {Stephenson}, C.~A. and {Stoev}, H. and {Suess}, F.~F. and {S{\"u}veges}, M. and {Surdej}, J. and {Szabados}, L. and {Szegedi-Elek}, E. and {Tapiador}, D. and {Taris}, F. and {Tauran}, G. and {Taylor}, M.~B. and {Teixeira}, R. and {Terrett}, D. and {Tingley}, B. and {Trager}, S.~C. and {Turon}, C. and {Ulla}, A. and {Utrilla}, E. and {Valentini}, G. and {van Elteren}, A. and {Van Hemelryck}, E. and {van Leeuwen}, M. and {Varadi}, M. and {Vecchiato}, A. and {Veljanoski}, J. and {Via}, T. and {Vicente}, D. and {Vogt}, S. and {Voss}, H. and {Votruba}, V. and {Voutsinas}, S. and {Walmsley}, G. and {Weiler}, M. and {Weingrill}, K. and {Werner}, D. and {Wevers}, T. and {Whitehead}, G. and {Wyrzykowski}, {\L}. and {Yoldas}, A. and {{\v{Z}}erjal}, M. and {Zucker}, S. and {Zurbach}, C. and {Zwitter}, T. and {Alecu}, A. and {Allen}, M. and {Allende Prieto}, C. and {Amorim}, A. and {Anglada-Escud{\'e}}, G. and {Arsenijevic}, V. and {Azaz}, S. and {Balm}, P. and {Beck}, M. and {Bernstein}, H. -H. and {Bigot}, L. and {Bijaoui}, A. and {Blasco}, C. and {Bonfigli}, M. and {Bono}, G. and {Boudreault}, S. and {Bressan}, A. and {Brown}, S. and {Brunet}, P. -M. and {Bunclark}, P. and {Buonanno}, R. and {Butkevich}, A.~G. and {Carret}, C. and {Carrion}, C. and {Chemin}, L. and {Ch{\'e}reau}, F. and {Corcione}, L. and {Darmigny}, E. and {de Boer}, K.~S. and {de Teodoro}, P. and {de Zeeuw}, P.~T. and {Delle Luche}, C. and {Domingues}, C.~D. and {Dubath}, P. and {Fodor}, F. and {Fr{\'e}zouls}, B. and {Fries}, A. and {Fustes}, D. and {Fyfe}, D. and {Gallardo}, E. and {Gallegos}, J. and {Gardiol}, D. and {Gebran}, M. and {Gomboc}, A. and {G{\'o}mez}, A. and {Grux}, E. and {Gueguen}, A. and {Heyrovsky}, A. and {Hoar}, J. and {Iannicola}, G. and {Isasi Parache}, Y. and {Janotto}, A. -M. and {Joliet}, E. and {Jonckheere}, A. and {Keil}, R. and {Kim}, D. -W. and {Klagyivik}, P. and {Klar}, J. and {Knude}, J. and {Kochukhov}, O. and {Kolka}, I. and {Kos}, J. and {Kutka}, A. and {Lainey}, V. and {LeBouquin}, D. and {Liu}, C. and {Loreggia}, D. and {Makarov}, V.~V. and {Marseille}, M.~G. and {Martayan}, C. and {Martinez-Rubi}, O. and {Massart}, B. and {Meynadier}, F. and {Mignot}, S. and {Munari}, U. and {Nguyen}, A. -T. and {Nordlander}, T. and {Ocvirk}, P. and {O'Flaherty}, K.~S. and {Olias Sanz}, A. and {Ortiz}, P. and {Osorio}, J. and {Oszkiewicz}, D. and {Ouzounis}, A. and {Palmer}, M. and {Park}, P. and {Pasquato}, E. and {Peltzer}, C. and {Peralta}, J. and {P{\'e}turaud}, F. and {Pieniluoma}, T. and {Pigozzi}, E. and {Poels}, J. and {Prat}, G. and {Prod'homme}, T. and {Raison}, F. and {Rebordao}, J.~M. and {Risquez}, D. and {Rocca-Volmerange}, B. and {Rosen}, S. and {Ruiz-Fuertes}, M.~I. and {Russo}, F. and {Sembay}, S. and {Serraller Vizcaino}, I. and {Short}, A. and {Siebert}, A. and {Silva}, H. and {Sinachopoulos}, D. and {Slezak}, E. and {Soffel}, M. and {Sosnowska}, D. and {Strai{\v{z}}ys}, V. and {ter Linden}, M. and {Terrell}, D. and {Theil}, S. and {Tiede}, C. and {Troisi}, L. and {Tsalmantza}, P. and {Tur}, D. and {Vaccari}, M. and {Vachier}, F. and {Valles}, P. and {Van Hamme}, W. and {Veltz}, L. and {Virtanen}, J. and {Wallut}, J. -M. and {Wichmann}, R. and {Wilkinson}, M.~I. and {Ziaeepour}, H. and {Zschocke}, S.},
  title         = {{The Gaia mission}},
  doi           = {10.1051/0004-6361/201629272},
  eid           = {A1},
  eprint        = {1609.04153},
  pages         = {A1},
  volume        = {595},
  adsurl        = {https://ui.adsabs.harvard.edu/abs/2016A&A...595A...1G},
  archiveprefix = {arXiv},
  journal       = {\aap},
  month         = nov,
  primaryclass  = {astro-ph.IM},
  year          = {2016},
}

@Article{GaiaEDR3Validation,
  author        = {{Fabricius}, C. and {Luri}, X. and {Arenou}, F. and {Babusiaux}, C. and {Helmi}, A. and {Muraveva}, T. and {Reyl{\'e}}, C. and {Spoto}, F. and {Vallenari}, A. and {Antoja}, T. and {Balbinot}, E. and {Barache}, C. and {Bauchet}, N. and {Bragaglia}, A. and {Busonero}, D. and {Cantat-Gaudin}, T. and {Carrasco}, J.~M. and {Diakit{\'e}}, S. and {Fabrizio}, M. and {Figueras}, F. and {Garcia-Gutierrez}, A. and {Garofalo}, A. and {Jordi}, C. and {Kervella}, P. and {Khanna}, S. and {Leclerc}, N. and {Licata}, E. and {Lambert}, S. and {Marrese}, P.~M. and {Masip}, A. and {Ramos}, P. and {Robichon}, N. and {Robin}, A.~C. and {Romero-G{\'o}mez}, M. and {Rubele}, S. and {Weiler}, M.},
  title         = {{Gaia Early Data Release 3. Catalogue validation}},
  doi           = {10.1051/0004-6361/202039834},
  eid           = {A5},
  eprint        = {2012.06242},
  pages         = {A5},
  volume        = {649},
  adsurl        = {https://ui.adsabs.harvard.edu/abs/2021A&A...649A...5F},
  archiveprefix = {arXiv},
  journal       = {\aap},
  month         = may,
  primaryclass  = {astro-ph.GA},
  year          = {2021},
}

@Article{EDR3photometry,
  author        = {{Riello}, M. and {De Angeli}, F. and {Evans}, D.~W. and {Montegriffo}, P. and {Carrasco}, J.~M. and {Busso}, G. and {Palaversa}, L. and {Burgess}, P.~W. and {Diener}, C. and {Davidson}, M. and {Rowell}, N. and {Fabricius}, C. and {Jordi}, C. and {Bellazzini}, M. and {Pancino}, E. and {Harrison}, D.~L. and {Cacciari}, C. and {van Leeuwen}, F. and {Hambly}, N.~C. and {Hodgkin}, S.~T. and {Osborne}, P.~J. and {Altavilla}, G. and {Barstow}, M.~A. and {Brown}, A.~G.~A. and {Castellani}, M. and {Cowell}, S. and {De Luise}, F. and {Gilmore}, G. and {Giuffrida}, G. and {Hidalgo}, S. and {Holland}, G. and {Marinoni}, S. and {Pagani}, C. and {Piersimoni}, A.~M. and {Pulone}, L. and {Ragaini}, S. and {Rainer}, M. and {Richards}, P.~J. and {Sanna}, N. and {Walton}, N.~A. and {Weiler}, M. and {Yoldas}, A.},
  title         = {{Gaia Early Data Release 3. Photometric content and validation}},
  doi           = {10.1051/0004-6361/202039587},
  eid           = {A3},
  eprint        = {2012.01916},
  pages         = {A3},
  volume        = {649},
  adsurl        = {https://ui.adsabs.harvard.edu/abs/2021A&A...649A...3R},
  archiveprefix = {arXiv},
  journal       = {\aap},
  month         = may,
  primaryclass  = {astro-ph.IM},
  year          = {2021},
}

@ARTICLE{row21,
       author = {{Rowell}, N. and {Davidson}, M. and {Lindegren}, L. and {van Leeuwen}, F. and {Casta{\~n}eda}, J. and {Fabricius}, C. and {Bastian}, U. and {Hambly}, N.~C. and {Hern{\'a}ndez}, J. and {Bombrun}, A. and {Evans}, D.~W. and {De Angeli}, F. and {Riello}, M. and {Busonero}, D. and {Crowley}, C. and {Mora}, A. and {Lammers}, U. and {Gracia}, G. and {Portell}, J. and {Biermann}, M. and {Brown}, A.~G.~A.},
        title = "{Gaia Early Data Release 3. Modelling and calibration of Gaia's point and line spread functions}",
      journal = {\aap},
         year = 2021,
        month = may,
       volume = {649},
          eid = {A11},
        pages = {A11},
          doi = {10.1051/0004-6361/202039448},
archivePrefix = {arXiv},
       eprint = {2012.02069},
 primaryClass = {astro-ph.IM},
       adsurl = {https://ui.adsabs.harvard.edu/abs/2021A&A...649A..11R}
}

@ARTICLE{tor21,
       author = {{Torra}, F. and {Casta{\~n}eda}, J. and {Fabricius}, C. and {Lindegren}, L. and {Clotet}, M. and {Gonz{\'a}lez-Vidal}, J.~J. and {Bartolom{\'e}}, S. and {Bastian}, U. and {Bernet}, M. and {Biermann}, M. and {Garralda}, N. and {G{\'u}rpide}, A. and {Lammers}, U. and {Portell}, J. and {Torra}, J.},
        title = "{Gaia Early Data Release 3. Building the Gaia DR3 source list - Cross-match of Gaia observations}",
      journal = {\aap},
         year = 2021,
        month = may,
       volume = {649},
          eid = {A10},
        pages = {A10},
          doi = {10.1051/0004-6361/202039637},
archivePrefix = {arXiv},
       eprint = {2012.06420},
 primaryClass = {astro-ph.IM},
       adsurl = {https://ui.adsabs.harvard.edu/abs/2021A&A...649A..10T}
}

@ARTICLE{skr06,
       author = {{Skrutskie}, M.~F. and {Cutri}, R.~M. and {Stiening}, R. and {Weinberg}, M.~D. and {Schneider}, S. and {Carpenter}, J.~M. and {Beichman}, C. and {Capps}, R. and {Chester}, T. and {Elias}, J. and {Huchra}, J. and {Liebert}, J. and {Lonsdale}, C. and {Monet}, D.~G. and {Price}, S. and {Seitzer}, P. and {Jarrett}, T. and {Kirkpatrick}, J.~D. and {Gizis}, J.~E. and {Howard}, E. and {Evans}, T. and {Fowler}, J. and {Fullmer}, L. and {Hurt}, R. and {Light}, R. and {Kopan}, E.~L. and {Marsh}, K.~A. and {McCallon}, H.~L. and {Tam}, R. and {Van Dyk}, S. and {Wheelock}, S.},
        title = "{The Two Micron All Sky Survey (2MASS)}",
      journal = {\aj},
         year = 2006,
        month = feb,
       volume = {131},
       number = {2},
        pages = {1163-1183},
          doi = {10.1086/498708},
       adsurl = {https://ui.adsabs.harvard.edu/abs/2006AJ....131.1163S}
}

@ARTICLE{wri10,
       author = {{Wright}, Edward L. and {Eisenhardt}, Peter R.~M. and {Mainzer}, Amy K. and {Ressler}, Michael E. and {Cutri}, Roc M. and {Jarrett}, Thomas and {Kirkpatrick}, J. Davy and {Padgett}, Deborah and {McMillan}, Robert S. and {Skrutskie}, Michael and {Stanford}, S.~A. and {Cohen}, Martin and {Walker}, Russell G. and {Mather}, John C. and {Leisawitz}, David and {Gautier}, Thomas N., III and {McLean}, Ian and {Benford}, Dominic and {Lonsdale}, Carol J. and {Blain}, Andrew and {Mendez}, Bryan and {Irace}, William R. and {Duval}, Valerie and {Liu}, Fengchuan and {Royer}, Don and {Heinrichsen}, Ingolf and {Howard}, Joan and {Shannon}, Mark and {Kendall}, Martha and {Walsh}, Amy L. and {Larsen}, Mark and {Cardon}, Joel G. and {Schick}, Scott and {Schwalm}, Mark and {Abid}, Mohamed and {Fabinsky}, Beth and {Naes}, Larry and {Tsai}, Chao-Wei},
        title = "{The Wide-field Infrared Survey Explorer (WISE): Mission Description and Initial On-orbit Performance}",
      journal = {\aj},
         year = 2010,
        month = dec,
       volume = {140},
       number = {6},
        pages = {1868-1881},
          doi = {10.1088/0004-6256/140/6/1868},
archivePrefix = {arXiv},
       eprint = {1008.0031},
 primaryClass = {astro-ph.IM},
       adsurl = {https://ui.adsabs.harvard.edu/abs/2010AJ....140.1868W}
}

@ARTICLE{lin21a,
       author = {{Lindegren}, L. and {Bastian}, U. and {Biermann}, M. and {Bombrun}, A. and {de Torres}, A. and {Gerlach}, E. and {Geyer}, R. and {Hern{\'a}ndez}, J. and {Hilger}, T. and {Hobbs}, D. and {Klioner}, S.~A. and {Lammers}, U. and {McMillan}, P.~J. and {Ramos-Lerate}, M. and {Steidelm{\"u}ller}, H. and {Stephenson}, C.~A. and {van Leeuwen}, F.},
        title = "{Gaia Early Data Release 3. Parallax bias versus magnitude, colour, and position}",
      journal = {\aap},
         year = 2021,
        month = may,
       volume = {649},
          eid = {A4},
        pages = {A4},
          doi = {10.1051/0004-6361/202039653},
archivePrefix = {arXiv},
       eprint = {2012.01742},
 primaryClass = {astro-ph.IM},
       adsurl = {https://ui.adsabs.harvard.edu/abs/2021A&A...649A...4L}
}

@ARTICLE{lin21b,
       author = {{Lindegren}, L. and {Klioner}, S.~A. and {Hern{\'a}ndez}, J. and {Bombrun}, A. and {Ramos-Lerate}, M. and {Steidelm{\"u}ller}, H. and {Bastian}, U. and {Biermann}, M. and {de Torres}, A. and {Gerlach}, E. and {Geyer}, R. and {Hilger}, T. and {Hobbs}, D. and {Lammers}, U. and {McMillan}, P.~J. and {Stephenson}, C.~A. and {Casta{\~n}eda}, J. and {Davidson}, M. and {Fabricius}, C. and {Gracia-Abril}, G. and {Portell}, J. and {Rowell}, N. and {Teyssier}, D. and {Torra}, F. and {Bartolom{\'e}}, S. and {Clotet}, M. and {Garralda}, N. and {Gonz{\'a}lez-Vidal}, J.~J. and {Torra}, J. and {Abbas}, U. and {Altmann}, M. and {Anglada Varela}, E. and {Balaguer-N{\'u}{\~n}ez}, L. and {Balog}, Z. and {Barache}, C. and {Becciani}, U. and {Bernet}, M. and {Bertone}, S. and {Bianchi}, L. and {Bouquillon}, S. and {Brown}, A.~G.~A. and {Bucciarelli}, B. and {Busonero}, D. and {Butkevich}, A.~G. and {Buzzi}, R. and {Cancelliere}, R. and {Carlucci}, T. and {Charlot}, P. and {Cioni}, M. -R.~L. and {Crosta}, M. and {Crowley}, C. and {del Peloso}, E.~F. and {del Pozo}, E. and {Drimmel}, R. and {Esquej}, P. and {Fienga}, A. and {Fraile}, E. and {Gai}, M. and {Garcia-Reinaldos}, M. and {Guerra}, R. and {Hambly}, N.~C. and {Hauser}, M. and {Jan{\ss}en}, K. and {Jordan}, S. and {Kostrzewa-Rutkowska}, Z. and {Lattanzi}, M.~G. and {Liao}, S. and {Licata}, E. and {Lister}, T.~A. and {L{\"o}ffler}, W. and {Marchant}, J.~M. and {Masip}, A. and {Mignard}, F. and {Mints}, A. and {Molina}, D. and {Mora}, A. and {Morbidelli}, R. and {Murphy}, C.~P. and {Pagani}, C. and {Panuzzo}, P. and {Pe{\~n}alosa Esteller}, X. and {Poggio}, E. and {Re Fiorentin}, P. and {Riva}, A. and {Sagrist{\`a} Sell{\'e}s}, A. and {Sanchez Gimenez}, V. and {Sarasso}, M. and {Sciacca}, E. and {Siddiqui}, H.~I. and {Smart}, R.~L. and {Souami}, D. and {Spagna}, A. and {Steele}, I.~A. and {Taris}, F. and {Utrilla}, E. and {van Reeven}, W. and {Vecchiato}, A.},
        title = "{Gaia Early Data Release 3. The astrometric solution}",
      journal = {\aap},
         year = 2021,
        month = may,
       volume = {649},
          eid = {A2},
        pages = {A2},
          doi = {10.1051/0004-6361/202039709},
archivePrefix = {arXiv},
       eprint = {2012.03380},
 primaryClass = {astro-ph.IM},
       adsurl = {https://ui.adsabs.harvard.edu/abs/2021A&A...649A...2L}
}

@ARTICLE{bai21,
       author = {{Bailer-Jones}, C.~A.~L. and {Rybizki}, J. and {Fouesneau}, M. and {Demleitner}, M. and {Andrae}, R.},
        title = "{Estimating Distances from Parallaxes. V. Geometric and Photogeometric Distances to 1.47 Billion Stars in Gaia Early Data Release 3}",
      journal = {\aj},
         year = 2021,
        month = mar,
       volume = {161},
       number = {3},
          eid = {147},
        pages = {147},
          doi = {10.3847/1538-3881/abd806},
archivePrefix = {arXiv},
       eprint = {2012.05220},
 primaryClass = {astro-ph.SR},
       adsurl = {https://ui.adsabs.harvard.edu/abs/2021AJ....161..147B}
}

@ARTICLE{lal22,
       author = {{Lallement}, R. and {Vergely}, J.~L. and {Babusiaux}, C. and {Cox}, N.~L.~J.},
        title = "{Updated Gaia-2MASS 3D maps of Galactic interstellar dust}",
      journal = {\aap},
         year = 2022,
        month = may,
       volume = {661},
          eid = {A147},
        pages = {A147},
          doi = {10.1051/0004-6361/202142846},
archivePrefix = {arXiv},
       eprint = {2203.01627},
 primaryClass = {astro-ph.GA},
       adsurl = {https://ui.adsabs.harvard.edu/abs/2022A&A...661A.147L}
}

@ARTICLE{ver22,
       author = {{Vergely}, J.~L. and {Lallement}, R. and {Cox}, N.~L.~J.},
        title = "{Three-dimensional extinction maps: Inverting inter-calibrated extinction catalogues}",
      journal = {\aap},
         year = 2022,
        month = aug,
       volume = {664},
          eid = {A174},
        pages = {A174},
          doi = {10.1051/0004-6361/202243319},
archivePrefix = {arXiv},
       eprint = {2205.09087},
 primaryClass = {astro-ph.GA},
       adsurl = {https://ui.adsabs.harvard.edu/abs/2022A&A...664A.174V}
}

@ARTICLE{GaiaDR32023,
       author = {{Gaia Collaboration} and {Vallenari}, A. and {Brown}, A.~G.~A. and {Prusti}, T. and {de Bruijne}, J.~H.~J. and {Arenou}, F. and {Babusiaux}, C. and {Biermann}, M. and {Creevey}, O.~L. and {Ducourant}, C. and {Evans}, D.~W. and {Eyer}, L. and {Guerra}, R. and {Hutton}, A. and {Jordi}, C. and {Klioner}, S.~A. and {Lammers}, U.~L. and {Lindegren}, L. and {Luri}, X. and {Mignard}, F. and {Panem}, C. and {Pourbaix}, D. and {Randich}, S. and {Sartoretti}, P. and {Soubiran}, C. and {Tanga}, P. and {Walton}, N.~A. and {Bailer-Jones}, C.~A.~L. and {Bastian}, U. and {Drimmel}, R. and {Jansen}, F. and {Katz}, D. and {Lattanzi}, M.~G. and {van Leeuwen}, F. and {Bakker}, J. and {Cacciari}, C. and {Casta{\~n}eda}, J. and {De Angeli}, F. and {Fabricius}, C. and {Fouesneau}, M. and {Fr{\'e}mat}, Y. and {Galluccio}, L. and {Guerrier}, A. and {Heiter}, U. and {Masana}, E. and {Messineo}, R. and {Mowlavi}, N. and {Nicolas}, C. and {Nienartowicz}, K. and {Pailler}, F. and {Panuzzo}, P. and {Riclet}, F. and {Roux}, W. and {Seabroke}, G.~M. and {Sordo}, R. and {Th{\'e}venin}, F. and {Gracia-Abril}, G. and {Portell}, J. and {Teyssier}, D. and {Altmann}, M. and {Andrae}, R. and {Audard}, M. and {Bellas-Velidis}, I. and {Benson}, K. and {Berthier}, J. and {Blomme}, R. and {Burgess}, P.~W. and {Busonero}, D. and {Busso}, G. and {C{\'a}novas}, H. and {Carry}, B. and {Cellino}, A. and {Cheek}, N. and {Clementini}, G. and {Damerdji}, Y. and {Davidson}, M. and {de Teodoro}, P. and {Nu{\~n}ez Campos}, M. and {Delchambre}, L. and {Dell'Oro}, A. and {Esquej}, P. and {Fern{\'a}ndez-Hern{\'a}ndez}, J. and {Fraile}, E. and {Garabato}, D. and {Garc{\'\i}a-Lario}, P. and {Gosset}, E. and {Haigron}, R. and {Halbwachs}, J.-L. and {Hambly}, N.~C. and {Harrison}, D.~L. and {Hern{\'a}ndez}, J. and {Hestroffer}, D. and {Hodgkin}, S.~T. and {Holl}, B. and {Jan{\ss}en}, K. and {Jevardat de Fombelle}, G. and {Jordan}, S. and {Krone-Martins}, A. and {Lanzafame}, A.~C. and {L{\"o}ffler}, W. and {Marchal}, O. and {Marrese}, P.~M. and {Moitinho}, A. and {Muinonen}, K. and {Osborne}, P. and {Pancino}, E. and {Pauwels}, T. and {Recio-Blanco}, A. and {Reyl{\'e}}, C. and {Riello}, M. and {Rimoldini}, L. and {Roegiers}, T. and {Rybizki}, J. and {Sarro}, L.~M. and {Siopis}, C. and {Smith}, M. and {Sozzetti}, A. and {Utrilla}, E. and {van Leeuwen}, M. and {Abbas}, U. and {{\'A}brah{\'a}m}, P. and {Abreu Aramburu}, A. and {Aerts}, C. and {Aguado}, J.~J. and {Ajaj}, M. and {Aldea-Montero}, F. and {Altavilla}, G. and {{\'A}lvarez}, M.~A. and {Alves}, J. and {Anders}, F. and {Anderson}, R.~I. and {Anglada Varela}, E. and {Antoja}, T. and {Baines}, D. and {Baker}, S.~G. and {Balaguer-N{\'u}{\~n}ez}, L. and {Balbinot}, E. and {Balog}, Z. and {Barache}, C. and {Barbato}, D. and {Barros}, M. and {Barstow}, M.~A. and {Bartolom{\'e}}, S. and {Bassilana}, J.-L. and {Bauchet}, N. and {Becciani}, U. and {Bellazzini}, M. and {Berihuete}, A. and {Bernet}, M. and {Bertone}, S. and {Bianchi}, L. and {Binnenfeld}, A. and {Blanco-Cuaresma}, S. and {Blazere}, A. and {Boch}, T. and {Bombrun}, A. and {Bossini}, D. and {Bouquillon}, S. and {Bragaglia}, A. and {Bramante}, L. and {Breedt}, E. and {Bressan}, A. and {Brouillet}, N. and {Brugaletta}, E. and {Bucciarelli}, B. and {Burlacu}, A. and {Butkevich}, A.~G. and {Buzzi}, R. and {Caffau}, E. and {Cancelliere}, R. and {Cantat-Gaudin}, T. and {Carballo}, R. and {Carlucci}, T. and {Carnerero}, M.~I. and {Carrasco}, J.~M. and {Casamiquela}, L. and {Castellani}, M. and {Castro-Ginard}, A. and {Chaoul}, L. and {Charlot}, P. and {Chemin}, L. and {Chiaramida}, V. and {Chiavassa}, A. and {Chornay}, N. and {Comoretto}, G. and {Contursi}, G. and {Cooper}, W.~J. and {Cornez}, T. and {Cowell}, S. and {Crifo}, F. and {Cropper}, M. and {Crosta}, M. and {Crowley}, C. and {Dafonte}, C. and {Dapergolas}, A. and {David}, M. and {David}, P. and {de Laverny}, P. and {De Luise}, F. and {De March}, R.},
        title = "{Gaia Data Release 3. Summary of the content and survey properties}",
      journal = {\aap},
         year = 2023,
        month = jun,
       volume = {674},
          eid = {A1},
        pages = {A1},
          doi = {10.1051/0004-6361/202243940},
archivePrefix = {arXiv},
       eprint = {2208.00211},
 primaryClass = {astro-ph.GA},
       adsurl = {https://ui.adsabs.harvard.edu/abs/2023A&A...674A...1G}
}

@ARTICLE{Murphy2024,
       author = {{Murphy}, Matthew M. and {Beatty}, Thomas G. and {Schlawin}, Everett and {Bell}, Taylor J. and {Line}, Michael R. and {Greene}, Thomas P. and {Parmentier}, Vivien and {Rauscher}, Emily and {Welbanks}, Luis and {Fortney}, Jonathan J. and {Rieke}, Marcia},
        title = "{Evidence for morning-to-evening limb asymmetry on the cool low-density exoplanet WASP-107 b}",
      journal = {Nature Astronomy},
         year = 2024,
        month = dec,
       volume = {8},
        pages = {1562-1574},
          doi = {10.1038/s41550-024-02367-9},
archivePrefix = {arXiv},
       eprint = {2406.09863},
 primaryClass = {astro-ph.EP}
}

@ARTICLE{murphy24T0,
       author = {{Murphy}, Matthew M. and {Beatty}, Thomas G. and {Apai}, D{\'a}niel},
        title = "{An Analytic Characterization of the Limb Asymmetry{\textemdash}Transit Time Degeneracy}",
      journal = {\apj},
         year = 2024,
        month = oct,
       volume = {974},
       number = {2},
          eid = {179},
        pages = {179},
          doi = {10.3847/1538-4357/ad7114},
archivePrefix = {arXiv},
       eprint = {2407.17564},
 primaryClass = {astro-ph.EP},
       adsurl = {https://ui.adsabs.harvard.edu/abs/2024ApJ...974..179M}
}

@ARTICLE{Espinoza2024,
       author = {{Espinoza}, N{\'e}stor and {Steinrueck}, Maria E. and {Kirk}, James and {MacDonald}, Ryan J. and {Savel}, Arjun B. and {Arnold}, Kenneth and {Kempton}, Eliza M. -R. and {Murphy}, Matthew M. and {Carone}, Ludmila and {Zamyatina}, Maria and {Lewis}, David A. and {Samra}, Dominic and {Kiefer}, Sven and {Rauscher}, Emily and {Christie}, Duncan and {Mayne}, Nathan and {Helling}, Christiane and {Rustamkulov}, Zafar and {Parmentier}, Vivien and {May}, Erin M. and others},
        title = "{Inhomogeneous terminators on the exoplanet WASP-39 b}",
      journal = {\nat},
         year = 2024,
        month = aug,
       volume = {632},
       number = {8027},
        pages = {1017-1020},
          doi = {10.1038/s41586-024-07768-4},
archivePrefix = {arXiv},
       eprint = {2407.10294},
 primaryClass = {astro-ph.EP}
}

@ARTICLE{Murgas2020,
       author = {{Murgas}, F. and {Chen}, G. and {Nortmann}, L. and {Pall{\'e}}, E. and {Nowak}, G.},
        title = "{The GTC exoplanet transit spectroscopy survey. XI. Possible detection of Rayleigh scattering in the atmosphere of the Saturn-mass planet WASP-69b}",
      journal = {\aap},
         year = 2020,
        month = sep,
       volume = {641},
          eid = {A158},
        pages = {A158},
          doi = {10.1051/0004-6361/202038161},
archivePrefix = {arXiv},
       eprint = {2007.02741},
 primaryClass = {astro-ph.EP}
}

@ARTICLE{Estrela2021,
       author = {{Estrela}, Raissa and {Swain}, Mark R. and {Roudier}, Ga{\"e}l and {Mugnai}, Lorenzo V. and {Mousis}, Olivier and {Tinetti}, Giovanna and {Yurchenko}, Sergei N. and {Tennyson}, Jonathan},
        title = "{Detection of Aerosols at Microbar Pressures in an Exoplanet Atmosphere}",
      journal = {\aj},
         year = 2021,
        month = sep,
       volume = {162},
       number = {3},
          eid = {91},
        pages = {91},
          doi = {10.3847/1538-3881/ac0c7c},
archivePrefix = {arXiv},
       eprint = {2106.10292},
 primaryClass = {astro-ph.EP}
}

@ARTICLE{Ouyang2023,
       author = {{Ouyang}, Qinglin and {Wang}, Wei and {Zhai}, Meng and {Chen}, Guo and {Rojo}, Patricio and {Liu}, Yujuan and {Zhao}, Fei and {Huang}, Jia-Sheng and {Zhao}, Gang},
        title = "{Tentative detection of titanium oxide in the atmosphere of WASP-69 b with a 4m ground-based telescope}",
      journal = {\mnras},
         year = 2023,
        month = jun,
       volume = {521},
       number = {4},
        pages = {5860-5879},
          doi = {10.1093/mnras/stad893},
archivePrefix = {arXiv},
       eprint = {2303.13202},
 primaryClass = {astro-ph.EP}
}

@ARTICLE{Guilluy2022,
       author = {{Guilluy}, G. and {Giacobbe}, P. and {Carleo}, I. and {Cubillos}, P. E. and {Sozzetti}, A. and {Bonomo}, A. S. and {Brogi}, M. and {Gandhi}, S. and {Fossati}, L. and {Nascimbeni}, V. and {Turrini}, D. and {Schisano}, E. and {Borsa}, F. and {Lanza}, A. F. and {Mancini}, L. and {Maggio}, A. and {Malavolta}, L. and {Micela}, G. and {Pino}, L. and {Rainer}, M. and others},
        title = "{The GAPS Programme at TNG. XXXVIII. Five molecules in the atmosphere of the warm giant planet WASP-69b detected at high spectral resolution}",
      journal = {\aap},
         year = 2022,
        month = sep,
       volume = {665},
          eid = {A104},
        pages = {A104},
          doi = {10.1051/0004-6361/202243854},
archivePrefix = {arXiv},
       eprint = {2207.09760},
 primaryClass = {astro-ph.EP}
}

@ARTICLE{Gao2020,
       author = {{Gao}, Peter and {Thorngren}, Daniel P. and {Lee}, Elspeth K. H. and {Fortney}, Jonathan J. and {Morley}, Caroline V. and {Wakeford}, Hannah R. and {Powell}, Diana and {Stevenson}, Kevin B. and {Zhang}, Xi},
        title = "{Aerosol composition of hot giant exoplanets dominated by silicates and hydrocarbon hazes}",
      journal = {Nature Astronomy},
         year = 2020,
        month = oct,
       volume = {4},
        pages = {951-956},
          doi = {10.1038/s41550-020-1114-3},
archivePrefix = {arXiv},
       eprint = {2005.11939},
 primaryClass = {astro-ph.EP}
}

@ARTICLE{Powell2019,
       author = {{Powell}, Diana and {Louden}, Tom and {Kreidberg}, Laura and {Zhang}, Xi and {Gao}, Peter and {Parmentier}, Vivien},
        title = "{Transit Signatures of Inhomogeneous Clouds on Hot Jupiters: Insights from Microphysical Cloud Modeling}",
      journal = {\apj},
         year = 2019,
        month = dec,
       volume = {887},
       number = {2},
          eid = {170},
        pages = {170},
          doi = {10.3847/1538-4357/ab55d9},
archivePrefix = {arXiv},
       eprint = {1910.07527},
 primaryClass = {astro-ph.EP}
}

@ARTICLE{Khalafinejad2021,
       author = {{Khalafinejad}, S. and {Molaverdikhani}, K. and {Blecic}, J. and {Mallonn}, M. and {Nortmann}, L. and {Caballero}, J. A. and {Rahmati}, H. and {Kaminski}, A. and {Sadegi}, S. and {Nagel}, E. and {Carone}, L. and {Amado}, P. J. and {Azzaro}, M. and {Bauer}, F. F. and {Casasayas-Barris}, N. and {Czesla}, S. and {von Essen}, C. and {Fossati}, L. and {G{\"u}del}, M. and {Henning}, Th. and {L{\'o}pez-Puertas}, M. and {Lendl}, M. and {L{\"u}ftinger}, T. and {Montes}, D. and {Oshagh}, M. and {Pall{\'e}}, E. and {Quirrenbach}, A. and {Reffert}, S. and {Reiners}, A. and {Ribas}, I. and {Stock}, S. and {Yan}, F. and {Zapatero Osorio}, M. R. and {Zechmeister}, M.},
        title = "{Probing the atmosphere of WASP-69 b with low- and high-resolution transmission spectroscopy}",
      journal = {\aap},
         year = 2021,
        month = dec,
       volume = {656},
          eid = {A142},
        pages = {A142},
          doi = {10.1051/0004-6361/202141191},
archivePrefix = {arXiv},
       eprint = {2109.06335},
 primaryClass = {astro-ph.EP}
}

@ARTICLE{PowellZhang2024,
       author = {{Powell}, Diana and {Zhang}, Xi},
        title = "{Two-dimensional Models of Microphysical Clouds on Hot Jupiters. I. Cloud Properties}",
      journal = {\apj},
         year = 2024,
        month = jul,
       volume = {969},
       number = {1},
          eid = {5},
        pages = {5},
          doi = {10.3847/1538-4357/ad3de4},
archivePrefix = {arXiv},
       eprint = {2404.08759},
 primaryClass = {astro-ph.EP},
       adsurl = {https://ui.adsabs.harvard.edu/abs/2024ApJ...969....5P}
}

@ARTICLE{dynesty,
       author = {{Speagle}, Joshua S.},
        title = "{DYNESTY: a dynamic nested sampling package for estimating Bayesian posteriors and evidences}",
      journal = {\mnras},
         year = 2020,
        month = apr,
       volume = {493},
       number = {3},
        pages = {3132-3158},
          doi = {10.1093/mnras/staa278},
archivePrefix = {arXiv},
       eprint = {1904.02180},
 primaryClass = {astro-ph.IM},
       adsurl = {https://ui.adsabs.harvard.edu/abs/2020MNRAS.493.3132S}
}

@ARTICLE{Kirkpatrick2005,
       author = {{Kirkpatrick}, J. Davy},
        title = "{New Spectral Types L and T}",
      journal = {\araa},
         year = 2005,
        month = sep,
       volume = {43},
       number = {1},
        pages = {195-245},
          doi = {10.1146/annurev.astro.42.053102.134017},
       adsurl = {https://ui.adsabs.harvard.edu/abs/2005ARA&A..43..195K}
}

@ARTICLE{Radica2026,
       author = {{Radica}, Michael and {Taylor}, Jake and {Rotman}, Yoav and {Blecic}, Jasmina and {Welbanks}, Luis and {Ahrer}, Eva-Maria and {Christie}, Duncan and {Coulombe}, Louis-Philippe and {Lowry}, Gillis and {Murphy}, Matthew M. and {Feinstein}, Adina D. and {Lafreni{\`e}re}, David and {MacDonald}, Ryan J. and {Mayne}, Nathan J. and {Tsai}, Shang-Min and {Zamyatina}, Maria},
        title = "{Supersolar Metallicity and Tentative Evidence for Photochemistry on WASP-96 b from JWST and Ground-based VLT Transmission Spectroscopy}",
      journal = {\aj},
         year = 2026,
        month = may,
       volume = {171},
       number = {5},
          eid = {314},
        pages = {314},
          doi = {10.3847/1538-3881/ae5b9f},
archivePrefix = {arXiv},
       eprint = {2604.05049},
 primaryClass = {astro-ph.EP},
       adsurl = {https://ui.adsabs.harvard.edu/abs/2026AJ....171..314R}
}

@ARTICLE{Mak2026,
       author = {{Mak}, Mei Ting and {Komacek}, Thaddeus D. and {Mayne}, Nathan J. and {Sing}, David K.},
        title = "{Flow-Driven Limb-Asymmetry of Haze Distribution Part I: An Analytical Framework for Predicting the Size Distribution of Photochemical Hazes Across the Two Limbs of hot-Jupiters}",
      journal = {\mnras},
         year = 2026,
        month = jul,
          doi = {10.1093/mnras/stag1376},
archivePrefix = {arXiv},
       eprint = {2607.14845},
 primaryClass = {astro-ph.EP},
       adsurl = {https://ui.adsabs.harvard.edu/abs/2026MNRAS.tmp.1288M}
}

@ARTICLE{mps2,
       author = {{Kostogryz}, N. and {Shapiro}, A.~I. and {Witzke}, V. and {Grant}, D. and {Wakeford}, H.~R. and {Stevenson}, K.~B. and {Solanki}, S.~K. and {Gizon}, L.},
        title = "{MPS-ATLAS Library of Stellar Model Atmospheres and Spectra}",
      journal = {Research Notes of the American Astronomical Society},
         year = 2023,
        month = mar,
       volume = {7},
       number = {3},
          eid = {39},
        pages = {39},
          doi = {10.3847/2515-5172/acc180},
archivePrefix = {arXiv},
       eprint = {2303.02685},
 primaryClass = {astro-ph.SR},
       adsurl = {https://ui.adsabs.harvard.edu/abs/2023RNAAS...7...39K}
}

@ARTICLE{schmidt2026,
       author = {{Schmidt}, Stephen P. and {May}, Erin M. and {Lothringer}, Joshua D. and {McCreery}, Patrick and {Mak}, Mei Ting and {Pope}, Myles and {Baskett}, Harry and {Mukherjee}, Sagnick and {Sing}, David K. and {Bennett}, Katherine A. and {Egan}, Arika and {Fu}, Guangwei and {Thorngren}, Daniel P. and {Christie}, Duncan A. and {Gasc{\'o}n}, Carlos and {Wang}, Le-Chris and {Ramos Rosado}, Lakeisha M. and {Mayne}, Nathan J. and {Allen}, Natalie H. and {Rustamkulov}, Zafar and {L{\'o}pez-Morales}, Mercedes and {Schlaufman}, Kevin C.},
        title = "{Mitigating Charge Migration in JWST NIRISS Reveals That KELT-7 b is a Metal-enriched Ultra-hot Jupiter Orbiting a Young Metal-rich Star}",
      journal = {arXiv e-prints},
         year = 2026,
        month = jul,
          eid = {arXiv:2607.06708},
        pages = {arXiv:2607.06708},
          doi = {10.48550/arXiv.2607.06708},
archivePrefix = {arXiv},
       eprint = {2607.06708},
 primaryClass = {astro-ph.EP},
       adsurl = {https://ui.adsabs.harvard.edu/abs/2026arXiv260706708S}
}

@ARTICLE{Roth2024,
       author = {{Roth}, Alexander and {Parmentier}, Vivien and {Hammond}, Mark},
        title = "{Hot Jupiter diversity and the onset of TiO/VO revealed by a large grid of non-grey global circulation models}",
      journal = {\mnras},
         year = 2024,
        month = jun,
       volume = {531},
       number = {1},
        pages = {1056-1083},
          doi = {10.1093/mnras/stae984},
archivePrefix = {arXiv},
       eprint = {2404.09626},
 primaryClass = {astro-ph.EP},
       adsurl = {https://ui.adsabs.harvard.edu/abs/2024MNRAS.531.1056R}
}

@misc{Cthulu,
	author = {{MacDonald}, Ryan J and {Agrawal}, Arnav},
	title = {{G}it{H}ub - {M}artian{C}olonist/{C}thulhu: {A} {P}ython package to calculate cross sections for substellar atmospheres --- github.com},
	howpublished = {\url{https://github.com/MartianColonist/Cthulhu}},
	year = {2024}
}
\bibliographystyle{aasjournalv7}



\end{document}